\documentclass[a4paper,11pt]{article}
\pdfoutput=1 

\usepackage{jheppub} 
\usepackage[T1]{fontenc} 
\usepackage[table]{xcolor}
\usepackage[caption=false]{subfig}
\usepackage{epsfig}
\usepackage{colortbl}
\usepackage[T1]{fontenc} 
\usepackage[compat=1.1.0]{tikz-feynman}
\usepackage{adjustbox}
\usepackage{multirow}
\usepackage{slashed}
\usepackage{lipsum}
\usepackage{float}
\usepackage{chngcntr}
\usepackage{bigints}
\usepackage[normalem]{ulem}
\usepackage{amsmath}
\usepackage{textcomp}
\usepackage{cleveref}
\usepackage{lscape}
\usepackage[table]{xcolor}  
\usepackage{colortbl}       
\usepackage{array}

\usepackage{float}
\usepackage{pstricks}
\usepackage[toc,page]{appendix}
\usepackage{pifont}
\usepackage{braket}
\usepackage{ragged2e}
\usepackage{slashed}
\usepackage{rotating}
\usepackage{tablefootnote}

\usepackage{hyperref}
\usepackage{tikz}
\usepackage{tikz-feynman}
\usetikzlibrary{arrows}
\usepackage{caption} 
\usepackage{orcidlink}
\crefname{figure}{fig\,.}{figs\,.} 
\crefname{equation}{eq\,.}{eqs\,.} 
\newcommand{\stkout}[1]{\ifmmode\text{\sout{\ensuremath{#1}}}\else\sout{#1}\fi}
\definecolor{calpolypomonagreen}{rgb}{0.12, 0.3, 0.17}
\newcommand{\bea}{\begin{eqnarray}}
	\newcommand{\eea}{\end{eqnarray}}

\usepackage{xcolor}
   
\allowdisplaybreaks

\title{Study of Form Factors and Observables in $B_c^- \rightarrow D_{s}^{*-}\ell^+\ell^-$ and $B_c^- \rightarrow D_{s}^{*-}\nu\bar{\nu}$ decays}

\author[a]{Utsab Dey\orcidlink{0000-0002-6228-741X},}
\author[a]{Soumitra Nandi\orcidlink{0000-0001-6567-0302}.}

\affiliation[a]{Department of Physics, Indian Institute of Technology Guwahati,\\North Guwahati, Assam-781039, India,}

\emailAdd{utsab\_dey@iitg.ac.in}
\emailAdd{soumitra.nandi@iitg.ac.in}

\abstract{We investigate the decays $B_c^- \rightarrow D_{s}^{(*)-}\ell^+\ell^-$ and $B_c^- \rightarrow D_{s}^{(*)-}\nu\bar{\nu}$ within the Standard Model (SM), employing perturbative QCD (pQCD) form factors that are sensitive to the wave functions of $B_c$ and $D_{s}^{(*)}$ mesons. We determine the shape parameters of these mesons and the $B_c \to D_s^{(*)}$ form factors at $q^2 = 0$ from available lattice QCD inputs for $B_s \to D_s^{(*)}$ and $B_c \to D_s$ transitions.
To obtain the $q^2$ dependence of the $B_c \to D_s^*$ form factors, we employ heavy-quark spin symmetry (HQSS) and an appropriate parametrization scheme over the allowed $q^2$ region. Based on these inputs, we present predictions for branching ratios and lepton-flavour-sensitive observables. Furthermore, we perform a detailed angular analysis of the cascade decay $B_c^- \to D_s^{*-}(\to D_s^- \pi^0)\,\ell^+\ell^-$, providing Standard Model predictions for several angular observables.}


\begin{document}
	
		\maketitle
	\flushbottom

	\section{Introduction}
	\label{section:Introduction}
	
The study of semileptonic and leptonic decays of $B$ mesons is crucial for determining the Cabibbo-Kobayashi-Maskawa (CKM) matrix elements and probing possible scenarios of New Physics (NP). Among the key observables, lepton flavor universality (LFU) ratios play a central role, as they test a fundamental prediction of the Standard Model (SM) that interactions involving different charged-lepton flavors are identical up to mass effects.

The $B_c$ meson has emerged as an important subject in flavor physics. First observed by the CDF Collaboration at the Tevatron in 1998 through the semileptonic decay $B_c \to J/\psi(\mu^+\mu^-)\ell^+X$~\cite{CDF:1998ihx}, it opened new avenues for experimental studies. With the LHCb experiment expected to produce about $5\times10^{10}$ $B_c$ mesons annually~\cite{PepeAltarelli:2008yyl}, precision measurements of its decay properties are now feasible. Unlike other heavy mesons, $B_c$ decays exclusively via weak interactions, as strong and electromagnetic annihilation channels are forbidden. This unique feature results in a rich spectrum of decay modes, whose systematic study provides critical insights into weak interaction dynamics within the Standard Model and potential signatures of New Physics.

Rare transitions such as $b \to s\ell^+\ell^-$ and $b \to d\ell^+\ell^-$ are key probes of New Physics (NP) in heavy-flavor phenomenology. As Flavor-Changing Neutral Current (FCNC) processes, they are forbidden at tree level in the Standard Model (SM) and occur only through loop diagrams, primarily via $Z$- and photon-penguin contributions and $W^+W^-$ box diagrams. Their sensitivity to short-distance effects makes them powerful tools for NP searches. Experimental studies began with $B \to K^{(*)}\ell^+\ell^-$ decays, first observed by CDF in 1998~\cite{CDF:1999uew}, and have since been extensively measured by BELLE~\cite{Belle:2001oey,Belle:2003ivt,Belle:2009zue,Belle:2016fev,BELLE:2019xld,Belle:2019oag}, BABAR~\cite{BaBar:2003szi,BaBar:2008jdv,BaBar:2012mrf}, CMS~\cite{CMS:2015bcy}, and LHCb~\cite{LHCb:2012juf,LHCb:2013ghj,LHCb:2014vgu,LHCb:2017avl,LHCb:2021trn}, driving intense theoretical and experimental interest.

In recent works, we have analyzed semileptonic, nonleptonic, and rare decay modes of the $B_c$ meson to $S$- and $P$-wave charmonium states~\cite{Dey:2025xdx}, as well as to $D^{(*)}$ meson final states~\cite{Dey:2025xjg}. Building on this foundation, the present study focuses on the rare channels $B_c^- \rightarrow D_s^{(*)-}\,\ell^+\ell^- \quad \text{and} \quad B_c^- \rightarrow D_s^{(*)-}\,\nu\bar{\nu}, \; (\ell = e,\mu,\tau)$. These modes are sensitive to short-distance electroweak dynamics through penguin and box diagrams, analogous to $b \to s (d)$ transitions, but in a heavy-heavy quark system. Their study provides complementary information to the well-explored $b \to s \ell^+\ell^-$ transitions and offers an additional probe of lepton flavor universality (LFU). Our aim is to extract the $q^{2}$-dependence of the $B_{c}\rightarrow D_{s}^{*}$ form factors using suitable parametrization methods, and subsequently to present predictions for the branching ratios for different lepton modes. We also perform a comprehensive angular analysis, including observables such as the forward–backward asymmetry, the longitudinal and transverse polarization fractions of the $D_{s}^{*}$ meson, as well as a set of form-factor–independent clean observables within the SM framework. 

Comparing observables in $B_c \to D_s^{(*)}$ and $B_c \to D^{(*)}$ decays~\cite{Dey:2025xjg} provides a powerful test of SU(3)-flavor symmetry and its breaking. Such comparisons are not merely academic, they are essential for understanding nonperturbative QCD effects that govern heavy meson dynamics. SU(3) breaking directly impacts hadronic form factors and wave functions, which are critical inputs for precise theoretical predictions. Constraining these effects improves the reliability of SM calculations and enhances sensitivity to potential New Physics contributions. With the anticipated large $B_c$ sample at LHCb, these rare channels offer an unprecedented opportunity for high-precision measurements, making them a key probe for both SM validation and NP searches.

The contents of the paper are organised as follows: In section \ref{section:Theoretical background} we describe analytic expressions of the various physical observables we intend to predict and analyze in this work, along-with brief discussions on the relevant form factors in modified pQCD framework. In addition we also discuss the form of the LCDAs of the participating mesons. In section \ref{section:Extraction of LCDA parameters} we extract the LCDA shape parameters of the participating mesons, and present predictions of the relevant form factors at $q^{2}=0$, calculated using the extracted parameters as inputs. In section \ref{section:Extrapolation to full physical range} we extract information of $B_{c}\rightarrow D_{s}^{*}$ form factors over the full physical $q^{2}$ region, using some suitable symmetry relations and appropriate form factor parametrization. In section \ref{section:Prediction of physical observables}, we present our prediction of some physical observable, involving branching ratios and a number of angular observables along-with observables like forward backward asymmetry and longitudinal and transverse polarization fractions. Finally in section \ref{section:Summary and conclusions} we briefly summarize our work.

	\section{Theoretical background}
	\label{section:Theoretical background}

	The very first thing to be clearly addressed is the theoretical background for the various aspects of the analysis in this work, particularly the physical observables. This section introduces the readers with discussions and explicit analytic expressions for the same. In subsection \ref{subsection:physical observables}, we briefly introduce the physical observables we intend to predict in this work. In subsection \ref{subsection:form factors} we discuss the form factor definitions in pQCD, and finally in subsection \ref{subsection:LCDAs} we discuss the distribution amplitudes of the mesons participating in the processes studied in this work.
	
		\subsection{Physical observables}
		\label{subsection:physical observables}
		
		In this subsection we briefly discuss about the various physical observables that we will be predicting in this work. 
		\subsubsection{Decay widths and branching fractions}
		\label{subsubsection:decay width theory}
		
		Contrary to the $B_{c}\rightarrow D^{(*)}$ modes, where both charged-current and neutral-current processes are allowed, the semileptonic charged-current channel is forbidden in the $B_{c}\rightarrow D_{s}^{(*)}$ modes. This is because in SM, the charged-current interaction couples the $b$ quark only to $c$ or $u$, and therefore cannot produce the $\bar{c}s$ configuration required to form a $D_{s}^{(*)}$ meson, resulting in tree-level $B_{c}^{-}\rightarrow D_{s}^{(*)-}\,\ell^{-}\bar{\nu}_{\ell}$ modes being forbidden. Consequently, for the $B_{c}\rightarrow D_{s}^{(*)}$ transitions, only the rare FCNC processes $b\rightarrow s\,\ell^{+}\ell^{-}$ and $b\rightarrow s\,\nu\bar{\nu}$ are allowed and arise at loop level via penguin and box diagrams.

			\begin{itemize}

				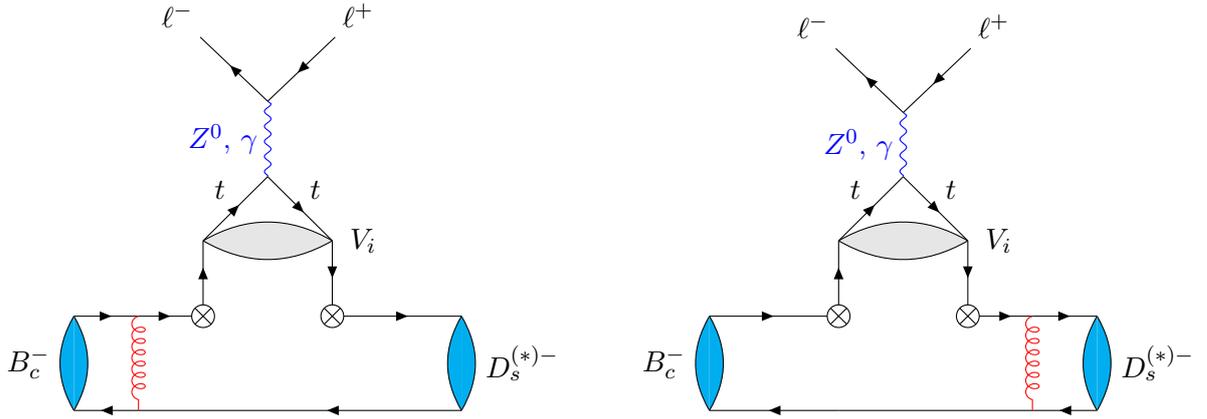
\begin{figure}[t!]
					\begin{tikzpicture}
						\begin{feynman}
							\vertex[crossed dot](a){};
							\vertex[crossed dot][right=1.7cm of a](a1){};
							\vertex[left=0.85cm of a](a2);
							\vertex[left=0.85cm of a2](a4);
							\vertex[right=1.7cm of a1](a3);
							\vertex[above=1.0cm of a](d);
							\vertex[right=1.7cm of d](d1);
							\vertex[above right=0.85cm and 0.85cm of d](c);
							\vertex[above=1.0cm of c](c1);
							\vertex[above left=0.85cm and 0.85cm of c1](c2){$\ell^{-}$};
							\vertex[above right=0.85cm and 0.85cm of c1](c3){$\ell^{+}$};
							\vertex[below=1.25cm of a](b);
							\vertex[left=0.85cm of b](b1);
							\vertex[left=0.85cm of b1](b2);
							\vertex[right=3.4cm of b](b3);
							\vertex[right=0.1cm of d1](d2){$V_{i}$};
							\diagram*{(a4)--[arrow size=1pt,fermion](a2)--[arrow size=1pt,fermion](a)--[arrow size=1pt,fermion](d)--[arrow size=1pt,fermion,edge label=$t$](c)--[arrow size=1pt,fermion,edge label=$t$](d1)--[arrow size=1pt,fermion](a1)--[arrow size=1pt,fermion](a3),(b3)--[arrow size=1pt,fermion](b)--[arrow size=1pt,plain](b1)--[arrow size=1pt,fermion](b2),(c)--[style=blue,photon,edge label={$Z^{0}$, $\gamma$}](c1),(c1)--[arrow size=1pt,fermion](c2),(c3)--[arrow size=1pt,fermion](c1),(a2)--[style=red,gluon](b1),(d)--[fill=gray!20,bend left,plain](d1),(d)--[fill=gray!20,bend right,plain](d1),(a4)--[fill=cyan,bend left,plain](b2),(a4)--[fill=cyan,bend right,plain,edge label'=\(B_{c}^{-}\)](b2),(a3)--[fill=cyan,bend left,plain,edge label={\(D_{s}^{(*)-}\)}](b3),(a3)--[fill=cyan,bend right,plain](b3)};
						\end{feynman}
					\end{tikzpicture}
					\qquad
					\begin{tikzpicture}
						\begin{feynman}
							\vertex[crossed dot](a){};
							\vertex[crossed dot][right=1.7cm of a](a1){};
							\vertex[left=1.7cm of a](a2);
							\vertex[right=0.85cm of a1](a3);
							\vertex[right=0.85cm of a3](a4);
							\vertex[above=1.0cm of a](d);
							\vertex[right=1.7cm of d](d1);
							\vertex[above right=0.85cm and 0.85cm of d](c);
							\vertex[above=0.85cm of c](c1);
							\vertex[above left=0.85cm and 0.85cm of c1](c2){$\ell^{-}$};
							\vertex[above right=0.85cm and 0.85cm of c1](c3){$\ell^{+}$};
							\vertex[below=1.25cm of a](b);
							\vertex[left=1.7cm of b](b1);
							\vertex[right=2.55cm of b](b2);
							\vertex[right=0.85cm of b2](b3);
							\vertex[right=0.1cm of d1](d2){$V_{i}$};
							\diagram*{(a2)--[arrow size=1pt,fermion](a)--[arrow size=1pt,fermion](d)--[arrow size=1pt,fermion,edge label=$t$](c)--[arrow size=1pt,fermion,edge label=$t$](d1)--[arrow size=1pt,fermion](a1)--[arrow size=1pt,fermion](a3)--[arrow size=1pt,fermion](a4),(b3)--[arrow size=1pt,fermion](b2)--[plain](b)--[arrow size=1pt,fermion](b1),(c)--[style=blue,photon,edge label={$Z^{0}$, $\gamma$}](c1),(c1)--[arrow size=1pt,fermion](c2),(c3)--[arrow size=1pt,fermion](c1),(a3)--[style=red,gluon](b2),(d)--[fill=gray!20,bend left,plain](d1),(d)--[fill=gray!20,bend right,plain](d1),(a2)--[fill=cyan,bend left,plain](b1),(a2)--[fill=cyan,bend right,plain,edge label'=\(B_{c}^{-}\)](b1),(a4)--[fill=cyan,bend left,plain,edge label={\(D_{s}^{(*)-}\)}](b3),(a4)--[fill=cyan,bend right,plain](b3)};
						\end{feynman}
					\end{tikzpicture}
					\caption{$Z^{0}$ and $\gamma$ penguin diagrams in effective theory for $B_{c}^{-}\rightarrow D_{s}^{(*)-}\ell^{+}\ell^{-}$ channel with $\ell=(e,\mu,\tau)$. $V_{i}$ denotes the intermediate resonance states $\rho$, $\omega$, $\phi$, $J/\psi$ and $\psi(2S)$.}
					\label{fig:FCNC feynman diagram BcDStar}
				\end{figure}
				\begin{figure}[htb!]
					\begin{tikzpicture}
						\begin{feynman}
							\vertex[crossed dot](a){};
							\vertex[crossed dot][right=1.7cm of a](a1){};
							\vertex[right=1.5cm of a1](a2);
							\vertex[left=1.7cm of a](a3);
							\vertex[right=0.85cm of a3](a4);
							\vertex[crossed dot][above=1.0cm of a](b){};
							\vertex[crossed dot][right=1.7cm of b](b1){};
							\vertex[above right=0.85cm and 0.85cm of b1](b2){$\ell^{-}$};
							\vertex[above left=0.85cm and 0.85cm of b](b3){$\ell^{+}$};
							\vertex[below left=1.3cm and 0.85cm of a](c);
							\vertex[right=4.05cm of c](c1);
							\vertex[left=0.85cm of c](c2);
							\diagram*{(a3)--[arrow size=1pt,fermion](a4)--[plain](a)--[arrow size=1pt,fermion, edge label=$t$](a1)--[arrow size=1pt,fermion](a2),(c1)--[arrow size=1pt,fermion](c)--[arrow size=1pt,fermion](c2),(b3)--[arrow size=1pt,fermion](b)--[arrow size=1pt,fermion, edge label=$\nu_{l}$](b1)--[arrow size=1pt,fermion](b2),(a4)--[style=red,gluon](c),(a3)--[fill=cyan,bend left,plain](c2),(a3)--[fill=cyan,bend right,plain,edge label'=\(B_{c}^{-}\)](c2),(a2)--[fill=cyan,bend left,plain,edge label={\(D_{s}^{(*)-}\)}](c1),(a2)--[fill=cyan,bend right,plain](c1)};
						\end{feynman}
					\end{tikzpicture}	
					\qquad
					\begin{tikzpicture}
						\begin{feynman}
							\vertex[crossed dot](a){};
							\vertex[crossed dot][right=1.7cm of a](a1){};
							\vertex[right=1.7cm of a1](a2);
							\vertex[left=1.7cm of a](a3);
							\vertex[right=0.85cm of a1](a4);
							\vertex[crossed dot][above=1.0cm of a](b){};
							\vertex[crossed dot][right=1.7cm of b](b1){};
							\vertex[above right=0.85cm and 0.85cm of b1](b2){$\ell^{-}$};
							\vertex[above left=0.85cm and 0.85cm of b](b3){$\ell^{+}$};
							\vertex[below right=1.3cm and 0.85cm of a1](c);
							\vertex[right=0.85cm of c](c1);
							\vertex[left=4.25cm of c](c2);
							\diagram*{(a3)--[arrow size=1pt,fermion](a)--[arrow size=1pt,fermion, edge label=$t$](a1)--[plain](a4)--[arrow size=1pt,fermion](a2),(c1)--[arrow size=1pt,fermion](c)--[arrow size=1pt,fermion](c2),(b3)--[arrow size=1pt,fermion](b)--[arrow size=1pt,fermion, edge label=$\nu_{l}$](b1)--[arrow size=1pt,fermion](b2),(a4)--[style=red,gluon](c),(a3)--[fill=cyan,bend left,plain](c2),(a3)--[fill=cyan,bend right,plain,edge label'=\(B_{c}^{-}\)](c2),(a2)--[fill=cyan,bend left,plain,edge label={\(D_{s}^{(*)-}\)}](c1),(a2)--[fill=cyan,bend right,plain](c1)};
						\end{feynman}
					\end{tikzpicture}	
					\caption{$W^{+}W^{-}$ box diagrams in effective theory for $B_{c}^{-}\rightarrow D_{s}^{(*)-} \ell^{+}\ell^{-}$ channel with $\ell=(e,\mu,\tau)$.}
					\label{fig:FCNC feynman diagram BcDStar2}
				\end{figure}
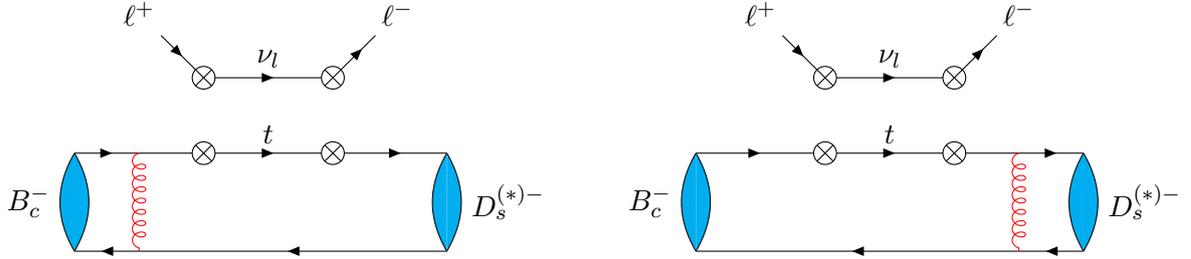
\item For the decay mode $B_{c}^{-}\rightarrow D_{s}^{(*)-}~\ell^{+}\ell^{-}$ governed by $b\rightarrow s$ quark level transition, and shown in figures \ref{fig:FCNC feynman diagram BcDStar} and \ref{fig:FCNC feynman diagram BcDStar2}. The most general effective Hamiltonian representing the above transition can be written as \cite{Altmannshofer:2008dz}
\begin{equation}
	\mathcal{H}_{eff}=-\frac{4G_{F}}{\sqrt{2}}\left(\lambda_{t}\mathcal{H}_{eff}^{t}+\lambda_{u}\mathcal{H}_{eff}^{u}\right),
	\label{eqn:effective hamiltonian}
\end{equation}
where $\lambda_{i}=V_{ib}V_{is}^{*}$ represents the CKM combination, and 
\begin{equation}
	H_{eff}^{t}=C_{1}O_{1}^{c}+C_{2}O_{2}^{c}+\sum_{i=3}^{6}C_{i}O_{i}+\sum_{i=7}^{10}\left(C_{i}O_{i}+C_{i}^{'}O_{i}^{'}\right),
\end{equation}	 
and
\begin{equation}
	H_{eff}^{u}=C_{1}\left(O_{1}^{c}-O_{1}^{u}\right)+C_{2}\left(O_{2}^{c}-O_{2}^{u}\right),
\end{equation}
in the SM framework, $O_{i}\equiv O_{i}(\mu)$ represent the four Fermi operators, $C_{i}\equiv C_{i}(\mu)$ represent the corresponding Wilson coefficients, and $\mu$ being the re-normalization scale. The operators $O_{1,2}^{c,u}$ are the current-current operators, $O_{3-6}$ the QCD penguin operators, $O_{7,8}$ the electromagnetic and chromomagnetic operators, and $O_{9,10}$ the semileptonic operators respectively. The expressions for the decay width are pretty much complicated and involved to be mentioned here. Hence, we refrain from discussing it in detail in this work and refer the readers to the references \cite{Ali:1999mm,Wang:2012ab} for their convenience.

\item For the decay mode $B_{c}^{-}\rightarrow D_{s}^{*-}\nu\bar{\nu}$ the effective Hamiltonian is
\begin{equation}\label{eq:Heffbtosll}
\begin{split}
\mathcal{H}_{eff}^{b\rightarrow s \nu\bar{\nu}}=\frac{G_{F}}{\sqrt{2}}\frac{\alpha_{EM}}{2\pi \sin^{2}(\theta_{W})}V_{tb}V_{ts}^{*}\eta_{X}X(x_{t})\left[\bar{s}\gamma^{\mu}(1-\gamma_{5})b\right]\left[\bar{\nu}\gamma_{\mu}(1-\gamma_{5})\nu\right],
\end{split}
\end{equation}
where $\theta_{W}$ is the Weinberg angle with $\sin^{2}(\theta_{W})=0.231$. $V_{tb}$ and $V_{ts}$ are the CKM matrix elements. The function $X(x_{t})$ has been taken from \cite{Buchalla:1995vs}, and $\eta_{X}\approx 1$ represents the QCD correction factor. The differential decay width is expressed as \cite{Barakat:2001ef,Wang:2012ab,PhysRevD.90.094018}
\begin{equation}
\resizebox{0.95\hsize}{!}{$
					\begin{split}
						\frac{d\Gamma(B_{c}\rightarrow D_{s}^{*}\nu\bar{\nu})}{dq^{2}}=&\frac{G_{F}^{2}\alpha_{EM}^{2}}{2^{10}\pi^{5}m_{B_{c}}^{3}}\cdot\Biggl|\frac{X(x_{t})}{\sin^{2}(\theta_{W})}\Biggr|^{2}\cdot \eta_{X}^{2}\cdot |V_{tb}V_{ts}^{*}|^{2}\sqrt{\lambda(q^{2})}
						\Biggl\{8\lambda(q^{2})q^{2}\frac{V(q^{2})^{2}}{(m_{B_{c}}+m_{D_{s}^{*}})^{2}}\\[1em]&+\frac{\lambda(q^{2})^{2}}{m_{D_{s}^{*}}^{2}}\cdot\frac{A_{2}(q^{2})^{2}}{(m_{B_{c}}+m_{D_{s}^{*}})^{2}}+\frac{1}{m_{D_{s}^{*}}^{2}}(m_{B_{c}}+m_{D_{s}^{*}})^{2}(\lambda(q^{2})+12m_{D_{s}^{*}}^{2}q^{2})\cdot A_{1}(q^{2})^{2}\\[1em]&-\frac{2\lambda(q^{2})}{m_{D_{s}^{*}}^{2}}(m_{B_{c}}^{2}-m_{D_{s}^{*}}^{2}-q^{2})\cdot Re[A_{1}(q^{2})^{*}A_{2}(q^{2})]\Biggr\},
					\end{split}$
				}
				\end{equation}	
				where $\lambda(q^{2})$, the phase space factor is expressed as 
				\begin{equation}
					\label{eqn:phasefactor}
					\lambda(q^{2})=(m_{B_{c}}^{2}+m_{D_{s}^{*}}^{2}-q^{2})^{2}-4m_{B_{c}}^{2}m_{D_{s}^{*}}^{2}.
				\end{equation}
			\end{itemize}

			\subsubsection{Angular analysis of rare $B_{c}^{-}\rightarrow D_{s}^{*-}\ell^{+}\ell^{-}$ channel}
			\label{subsubsection:Angular analysis theory}
			
			\begin{figure}[t!]
				\centering
				\includegraphics[width=9.0cm]{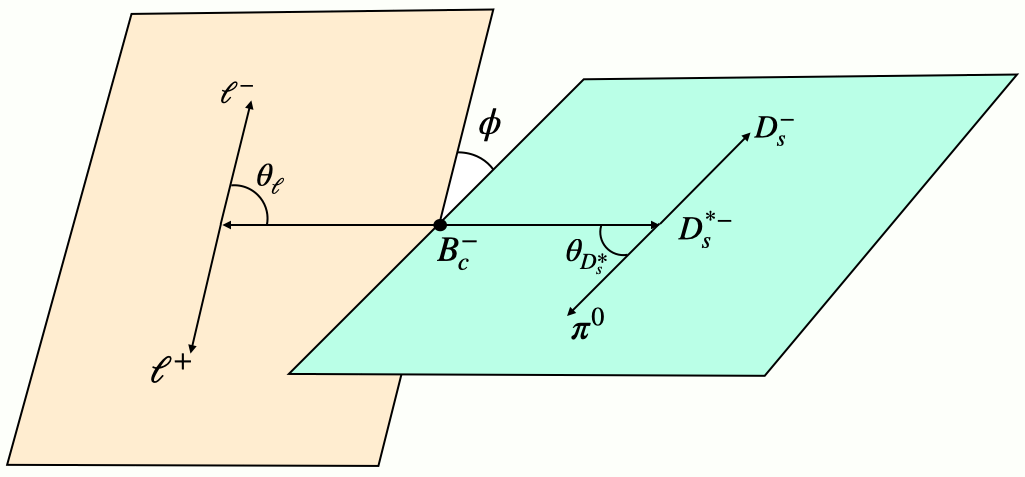}
				\caption{Kinematics of $B_{c}^{-}\rightarrow D_{s}^{*-}(\rightarrow D_{s}^{-}\pi^{0}) \ell^{+}\ell^{-}$ four body decay.}
				\label{fig: Angular observables decay planes}
			\end{figure}

Apart from the branching fractions, there are a number of other observables that are important in the present scenario of phenomenology. These primarily involve the angular observables, and a number of observables derived from them. These observables, being derived from $b\rightarrow s \ell^{+}\ell^{-}$ transitions and governed via FCNC, are forbidden at tree-level in the SM and are sensitive to loop-level contributions, like the penguin and box diagrams shown in figures \ref{fig:FCNC feynman diagram BcDStar} and \ref{fig:FCNC feynman diagram BcDStar2}, thus making these channels interesting as probes to look for possible NP scenarios.  With the effective Hamiltonian in eq.~\eqref{eqn:effective hamiltonian}, a full angular decay rate  distribution of $B_{c}^{-}\rightarrow D_{s}^{*-}(\rightarrow D_{s}^{-}\pi^{0}) \ell^{+}\ell^{-}$ can be obtained, and can be expressed as
\begin{equation}
\frac{d^{4}\Gamma}{dq^{2}d\cos\theta_{D_{s}^{*}}d\cos\theta_{\ell}d\phi}=\frac{9}{32\pi}\sum_{i}I_{i}(q^{2})f_{i}(\theta_{D_{s}^{*}},\theta_{\ell},\phi),
\label{eqn:differential decay width angular}
\end{equation}
with the angles $\theta_{D_{s}^{*}}$, $\theta_{\ell}$ and $\phi$ has been shown in figure \ref{fig: Angular observables decay planes}, where
\begin{itemize}
\item $\theta_{D_{s}^{*}}$ denotes the angle made by $\pi$ in the centre of mass system of $D_{s}^{-}$ and $\pi^{0}$ with respect to the direction of flight of $D_{s}^{*-}$,
\item $\theta_{\ell}$ denotes the angle made by $\ell^{-}$ in the center of mass system of $\ell^{+}$ and $\ell^{-}$ with respect to the direction of flight of the lepton pair, and
\item $\phi$ denotes the angle between the decay planes formed by the $(D_{s}^{-}\pi^{0})$ and $(\ell^{+},\ell^{-})$ pairs.
\end{itemize}
Revisiting eq.\eqref{eqn:differential decay width angular}, we can see that the differential decay width has been expressed as a function of four kinematic variables. However, there is one more variable that does not appear in the expression. It is the invariant mass squared of the $D_{s}^{-}\pi^{0}$ pair, namely $p_{D_s\pi}^2 ( = (p_{D_s}+p_\pi)^2)$, is omitted from our final phase space because we have employed the narrow width approximation (NWA) for the intermediate $D_{s}^{*}$ resonance. To explain this in more details, if we consider $s$ to be the invariant mass of the $D_{s}^{-}\pi^{0}$ pair, propagator of the intermediate $D_{s}^{*}$ resonance can be expressed as
\begin{equation}
    \frac{1}{(p_{D_s\pi}^2-m_{D_{s}^{*}}^{2})+i~\Gamma_{D_{s}^{*}}m_{D_{s}^{*}}},
\end{equation}
where $m_{D_{s}^{*}}$ and $\Gamma_{D_{s}^{*}}$ represent the mass and total width of $D_{s}^{*}$ meson, respectively. The squared amplitude can then be represented as 
\begin{equation}
\begin{split}
    &\left[\frac{1}{(p_{D_s\pi}^2-m_{D_{s}^{*}}^{2})+i~\Gamma_{D_{s}^{*}}m_{D_{s}^{*}}}\right]\left[\frac{1}{(p_{D_s\pi}^2-m_{D_{s}^{*}}^{2})+i~\Gamma_{D_{s}^{*}}m_{D_{s}^{*}}}\right]^{*},\\
    =&\frac{1}{(p_{D_s\pi}^2-m_{D_{s}^{*}}^{2})^{2}+(\Gamma_{D_{s}^{*}}m_{D_{s}^{*}})^{2}}.
    \end{split}
\end{equation}
Now, $\Gamma_{D_{s}^{*}}<1.9~\text{MeV}$ and $m_{D_{s}^{*}}=2.112~\text{GeV}$, which gives 
\begin{equation}
    \frac{\Gamma_{D_{s}^{*}}}{m_{D_{s}^{*}}}=\frac{1.9\times 10^{-3}~\text{GeV}}{2.112~\text{GeV}}=9\times 10^{-4}.
\end{equation}
We can see $\frac{\Gamma_{D_{s}^{*}}}{m_{D_{s}^{*}}}<<1$. Hence, following the discussion in \cite{Altmannshofer:2008dz}, we can use NWA to replace the squared $D_{s}^{*}$ propagator with
\begin{equation}
    \frac{1}{(p_{D_s\pi}^2-m_{D_{s}^{*}}^{2})^{2}+(\Gamma_{D_{s}^{*}}m_{D_{s}^{*}})^{2}}\rightarrow\frac{\pi}{m_{D_{s}^{*}}\Gamma_{D_{s}^{*}}}\delta(p_{D_s\pi}^2-m_{D_{s}^{*}}^{2}).
\end{equation}
Therefore, the five-fold decay rate distribution can be written in terms of four-fold differential rate as given below
\begin{equation}
	\frac{d^{5}\Gamma}{dp_{D_s\pi}^2dq^{2}d\cos\theta_{D_{s}^{*}}d\cos\theta_{\ell}d\phi} \rightarrow \frac{d^{4}\Gamma}{dq^{2}d\cos\theta_{D_{s}^{*}}d\cos\theta_{\ell}d\phi}\delta(p_{D_s\pi}^2-m_{D_{s}^{*}}^{2}).
\end{equation}
Thus, when calculating the relevant observables, an integration over $p_{D_s\pi}^2$ would collapse it over the invariant mass squared directly onto the $D_s^*$ mass shell ($p_{D_s\pi}^2 = m_{D_{s}^{*}}^2$). As a result, the $p_{D_s\pi}^2$-dependence would get factored out and leave the differential decay width on the remaining four kinematic variables, i.e., $q^{2}$, $\theta_{D_{s}^{*}},~\theta_{\ell}$ and $\phi$. This is the reason why the invariant mass dependence has been omitted in our four-body analysis.

In eq.\eqref{eqn:differential decay width angular} the functions $f_{i}(\theta_{D_{s}^{*}},\theta_{\ell},\phi)$s encode all the necessary angular information. Their forms are shown in table \ref{table:anglularf}. The detailed mathematical expressions of the angular coefficients $I_i(q^2)$, which can be expressed in terms of various transversity amplitudes, are as given below:	

	\begin{table}[t!]
	\renewcommand{\arraystretch}{1.5}
	\begin{tabular}{|cc|cc|}
		\hline
		Angular Observables&$f_{i}(\theta_{D_{s}^{*}},\theta_{\ell},\phi)$&Angular Observables&$f_{i}(\theta_{{D_{s}}^{*}},\theta_{\ell},\phi)$\\
		\hline
		$I_{1s}(q^{2})$&$\sin^{2}\theta_{D_{s}^{*}}$&$I_{5}(q^{2})$&$\sin 2\theta_{D_{s}^{*}}\sin\theta_{\ell}\cos \phi$\\
		$I_{1c}(q^{2})$&$\cos^{2}\theta_{D_{s}^{*}}$&$I_{6s}(q^{2})$&$\sin^{2}\theta_{D_{s}^{*}}\cos\theta_{\ell}$\\
		$I_{2s}(q^{2})$&$\sin^{2}\theta_{D_{s}^{*}}\cos2\theta_{\ell}$&$I_{7}(q^{2})$&$\sin 2\theta_{D_{s}^{*}}\sin\theta_{\ell}\sin\phi$\\
		$I_{2c}(q^{2})$&$\cos^{2}\theta_{D_{s}^{*}}\cos2\theta_{\ell}$&$I_{8}(q^{2})$&$\sin 2\theta_{D_{s}^{*}}\sin 2\theta_{\ell}\sin\phi$\\
		$I_{3}(q^{2})$&$\sin^{2}\theta_{D_{s}^{*}}\sin^{2}\theta_{\ell}\cos 2\phi$&$I_{9}(q^{2})$&$\sin^{2}\theta_{D_{s}^{*}}\sin^{2}\theta_{\ell}\sin 2\phi$\\
		$I_{4}(q^{2})$&$\sin 2\theta_{D_{s}^{*}}\sin 2\theta_{\ell}\cos\phi$&&\\
		\hline
	\end{tabular}
	\caption{Expressions for $f_{i}(\theta_{D_{s}^{*}},\theta_{\ell},\phi)$. The corresponding angular coefficients are also mentioned the detailed mathematical expressions of which are given in the text.}
	\label{table:anglularf}
\end{table}
		
\begin{equation}
	\begin{split}
		I_{1s}=&\left(\frac{3}{4}-\hat{m_{\ell}}^{2}\right)\left(|\mathcal{A}_{\parallel}^{L}|^{2}+|\mathcal{A}_{\perp}^{L}|^{2}+|\mathcal{A}_{\parallel}^{R}|^{2}+|\mathcal{A}_{\perp}^{R}|^{2}\right)+4\hat{m_{\ell}}^{2}\text{Re}\left[\mathcal{A}_{\perp}^{L}\mathcal{A}_{\perp}^{R*}+\mathcal{A}_{\parallel}^{L}\mathcal{A}_{\parallel}^{R*}\right],\\
		I_{1c}=&|\mathcal{A}_{0}^{L}|^{2}+|\mathcal{A}_{0}^{R}|^{2}+4\hat{m_{\ell}}^{2}\left(|\mathcal{A}_{t}|^{2}+2\text{Re}[\mathcal{A}_{0}^{L}\mathcal{A}_{0}^{R*}]\right),\\
		I_{2s}=&\beta_{\ell}^{2}\frac{|\mathcal{A}_{\parallel}^{L}|^{2}+|\mathcal{A}_{\parallel}^{R}|^{2}+|\mathcal{A}_{\perp}^{L}|^{2}+|\mathcal{A}_{\perp}^{R}|^{2}}{4},\\
		I_{2c}=&-\beta_{\ell}^{2}\left(|\mathcal{A}_{0}^{L}|^{2}+|\mathcal{A}_{0}^{R}|^{2}\right),\\
		I_3=&\beta_{\ell}^{2}\frac{|\mathcal{A}_{\perp}^{L}|^{2}+|\mathcal{A}_{\perp}^{R}|^{2}-|\mathcal{A}_{\parallel}^{L}|^{2}-|\mathcal{A}_{\parallel}^{R}|^{2}}{2},\\
		I_4=&\beta_{\ell}^{2}\frac{\text{Re}\left[\mathcal{A}_{0}^{L}\mathcal{A}_{\parallel}^{L*}+\mathcal{A}_{0}^{R}\mathcal{A}_{\parallel}^{R*}\right]}{\sqrt{2}},\\
		I_5=&\sqrt{2}\beta_{\ell}\text{Re}\left[\mathcal{A}_{0}^{L}\mathcal{A}_{\perp}^{L*}-\mathcal{A}_{0}^{R}\mathcal{A}_{\perp}^{R*}\right],\\
		I_{6s}=&2\beta_{\ell}\text{Re}\left[\mathcal{A}_{\parallel}^{L}\mathcal{A}_{\perp}^{L*}-\mathcal{A}_{\parallel}^{R}\mathcal{A}_{\perp}^{R*}\right],\\
		I_7=&\sqrt{2}\beta_{\ell}\text{Im}\left[\mathcal{A}_{0}^{L}\mathcal{A}_{\parallel}^{L*}-\mathcal{A}_{0}^{R}\mathcal{A}_{\parallel}^{R*}\right],\\
		I_8=&\beta_{\ell}^{2}\frac{\text{Im}\left[\mathcal{A}_{0}^{L}\mathcal{A}_{\perp}^{L*}+\mathcal{A}_{0}^{R}\mathcal{A}_{\perp}^{R*}\right]}{\sqrt{2}},\\
		I_9=&\beta_{\ell}^{2}\text{Im}\left[\mathcal{A}_{\parallel}^{L*}\mathcal{A}_{\perp}^{L}-\mathcal{A}_{\parallel}^{R*}\mathcal{A}_{\perp}^{R}\right].\\
	\end{split}
	\label{eqn:angular observables}
\end{equation}
All these transversity amplitudes can further be expressed in terms of Wilson Coefficients $C_{7,9}^{eff}(\mu)$ and the form factors $A_{0,1,2}(q^{2})$, $V(q^{2})$ and $T_{1,2,3}(q^{2})$, respectively. We have presented the detailed mathematical expressions of all these transversity amplitudes and other related information in appendix \ref{sec:apptranversity}.  
			
The observables $I_{i}$s are hard to extract directly in experiments unless the $B_{c}$ meson is tagged. Experimental measurements are often done with no distinction between $B_{c}^{+}$ or $B_{c}^{-}$. Therefore, it is more preferable to define a set of CP averaged observables, which are more easily measurable. For this, we define differential decay width of the CP conjugate decay mode $B_{c}^{+}\rightarrow D_{s}^{*+}(\rightarrow D_{s}^{+}\pi^{0}) \ell^{+}\ell^{-}$ as
			\begin{equation}
				\frac{d^{4}\bar{\Gamma}}{dq^{2}d\cos\theta_{D_{s}^{*}}d\cos\theta_{l}d\phi}=\frac{9}{32\pi}\sum_{i}\bar{I_{i}}(q^{2})f_{i}(\theta_{D_{s}^{*}},\theta_{\ell},\phi),
			\end{equation}
			where $f_{i}(\theta_{D_{s}^{*}},\theta_{\ell},\phi)$ has the same functional form as in table \ref{table:anglularf}. The angular observables $\bar{I}_{i}$ can be obtained from $I_{i}$ by the replacements
			\begin{equation}
				\begin{split}
					&I_{1s,1c,2s,2c,3,4,7}\rightarrow \bar{I}_{1s,1c,2s,2c,3,4,7},\\
					&I_{5,6s,8,9}\rightarrow -\bar{I}_{5,6s,8,9}.
				\end{split}
			\end{equation}
			This happens because for the CP conjugate mode, roles of $\ell^{-}$ and $\ell^{+}$ get interchanged along with momentum flips, and handedness of decay planes which flip under parity, leading to
			\begin{equation}
				\theta_{\ell}\rightarrow \pi-\theta_{\ell},\qquad \phi\rightarrow -\phi.
			\end{equation}
			With these we can define a set of CP averaged observables $S_{i}$ and CP violating observables $A_{i}$ defined by
			\begin{equation}
				S_{i}=\frac{I_{i}+\bar{I}_{i}}{d(\Gamma+\bar{\Gamma})/dq^{2}}, \qquad A_{i}=\frac{I_{i}-\bar{I}_{i}}{d(\Gamma+\bar{\Gamma})/dq^{2}},
			\end{equation} 
			where the observables have been normalized by the CP averaged differential decay width in order to reduce the form factor uncertainties. In addition, there are a number of physical observables that can be derived from the above observables.
			\begin{itemize}
				\item Integrating the differential decay width mentioned before over the angles $\theta_{\ell}\in[0,\pi]$, $\theta_{D_{s}^{*}}\in[0,\pi]$ and $\phi\in[0,2\pi]$ the CP averaged differential decay width for $B_{c}^{-}\rightarrow D_{s}^{*-}(\rightarrow D_{s}^{-}\pi^{0}) \ell^{+}\ell^{-}$ can be expressed as
				\begin{equation}
					\begin{split}
						\frac{d\Gamma_{CPavg}}{dq^{2}}=\frac{1}{2}\left(\frac{d\Gamma}{dq^{2}}+\frac{d\bar{\Gamma}}{dq^{2}}\right)
						=\frac{1}{4}\left(3I_{1c}+6I_{1s}-I_{2c}-2I_{2s}\right).
					\end{split}
					\label{eqn:CP averaged branching fraction}
				\end{equation}
				\item CP averaged lepton forward backward asymmetry defined as
				\begin{equation}
					A_{FB}=\frac{3}{4}S_{6s}.
					\label{eqn:forward backward asymmetry}
				\end{equation}
				\item The longitudinal and transverse polarization fractions of the $D_{s}^{*}$ meson can be expressed as
				\begin{equation}
					\begin{split}
						F_{L}=&\frac{1}{4}(3S_{1c}-S_{2c}),\\
						F_{T}=&\frac{1}{2}(3S_{1s}-S_{2s}),
					\end{split}
					\label{eqn:lepton polarization}
				\end{equation}
				respectively.
				\item Clean angular observables $P_{1,2,3}$ and $P^{'}_{4,5,6,8}$ expressed as \cite{Matias:2012xw,Descotes-Genon:2013vna}
				\begin{eqnarray}
					\begin{split}
						P_{1}&=\frac{S_{3}}{2S_{2s}},\qquad &P_{4}^{'}&=\frac{S_{4}}{\sqrt{S_{1c}S_{2s}}},\\
						P_{2}&=\frac{\beta_{l}S_{6s}}{8S_{2s}},\qquad &P_{5}^{'}&=\frac{\beta_{l}S_{5}}{2\sqrt{S_{1c}S_{2s}}},\\
						P_{3}&=-\frac{S_{9}}{4S_{2s}},\qquad &P_{6}^{'}&=-\frac{\beta_{l}S_{7}}{2\sqrt{S_{1c}S_{2s}}},\\
						&	  \qquad	&P_{8}^{'}&=-\frac{S_{8}}{\sqrt{S_{1c}S_{2s}}}.
					\end{split}
					\label{eqn:Clean observables}
				\end{eqnarray}
			\end{itemize}
		\subsection{Form factors}
		\label{subsection:form factors}
		
		In the previous subsection and in the appendix, we have mentioned that the transversity amplitudes are dependent on the QCD form factors.   
		These form factors are functions that parametrize the hadronic matrix elements governing the $B_{c}$ meson decays, and encode the QCD dynamics of the transition. Depending on the final state meson, whether it is a pseudo-scalar or a vector meson, the transition matrix elements can be parametrized as
		\begin{equation}
			\begin{split}
				\langle D_{s}(P_{2})|\bar{q}(0)\gamma_{\mu}b(0)|B_{c,s}(P_{1})\rangle=&\left[(P_{1}+P_{2})_{\mu}-\frac{M^{2}-m^{2}}{q^{2}}q_{\mu}\right]F_{+}(q^{2})\\&+\left[\frac{M^{2}-m^{2}}{q^{2}}q_{\mu}\right]F_{0}(q^{2}),
			\end{split}
			\label{eqn:matrix element BD}
		\end{equation}
		for a $B_{c}$ or $B_{s}$ meson decaying into $D_{s}$ meson, governed by a vector current. $M$ and $m$ represent the masses of initial and final state mesons, respectively. For calculations in pQCD, however, it is much more convenient to express the form factors in terms of two auxiliary form factors $f_{1}(q^{2})$ and $f_{2}(q^{2})$ defined as
		\begin{equation}
			\left\langle D_{s}(P_{2})|\bar{q}(0)\gamma_{\mu}b(0)|B_{c,s}(P_{1})\right\rangle=f_{1}(q^{2})P_{1\mu}+f_{2}(q^{2})P_{2\mu},
		\end{equation}
		and are related to $F_{+}(q^{2})$ and $F_{0}(q^{2})$ as
		\begin{equation}
			\begin{split}
				F_{+}(q^{2})=&\frac{1}{2}[f_{1}(q^{2})+f_{2}(q^{2})],\\
				F_{0}(q^{2})=&\frac{1}{2}f_{1}(q^{2})\left[1+\frac{q^{2}}{M^{2}-m^{2}}\right]+\frac{1}{2}f_{2}(q^{2})\left[1-\frac{q^{2}}{M^{2}-m^{2}}\right].
			\end{split}
		\end{equation}
		Next, for a $B_{c}$ or $B_{s}$ meson decaying into $D_{s}^{*}$ meson, the matrix element can be parametrized as
		\begin{equation}
			\langle D_{s}^{*}(P_{2})|\bar{q}(0)\gamma_{\mu}b(0)|B_{c,s}(P_{1})\rangle=\epsilon_{\mu\nu\alpha\beta}\epsilon^{\nu*}P_{1}^{\alpha}P_{2}^{\beta}\frac{2\cdot V{q^{2}}}{M+m},
		\end{equation}
		and
		\begin{equation}
			\begin{split}
				\langle D_{s}^{*}(P_{2})|\bar{q}(0)\gamma_{\mu}\gamma_{5}b(0)|B_{c,s}(P_{1})\rangle=&i\left[\epsilon^{*}_{\mu}-\frac{\epsilon^{*}\cdot q}{q^{2}}q_{\mu}\right](M+m)A_{1}(q^{2})\\
				&-i\left[(P_{1}-P_{2})_{\mu}-\frac{M^{2}-m^{2}}{q^{2}}q_{\mu}\right](\epsilon^{*}\cdot q)\frac{A_{2}(q^{2})}{M+m}\\
				&+i\left[\frac{2m(\epsilon^{*}\cdot q)}{q^{2}}q_{\mu}\right]A_{0}(q^{2}),
			\end{split}
		\end{equation}
		where the transition is governed by vector and axial-vector currents. In addition to the above, we also have the transition matrix elements governed by a tensor current as
		\begin{equation}
			\left\langle D_{s}(P_{2})|\bar{q}(0)\sigma_{\mu\nu}b(0)|B_{c,s}(P_{1})\right\rangle=i[P_{2\mu}q_{\nu}-q_{\mu}P_{2\nu}]\frac{2F_{T}(q^{2})}{M+m},
		\end{equation} 
		and
		\begin{equation}
			\begin{split}
				\langle D_{s}^{*}(P_{2})|\bar{q}(0)\sigma_{\mu\nu}q^{\nu}(1+\gamma_{5})&b(0)|B_{c,s}(P_{1})\rangle=i\epsilon_{\mu\nu\alpha\beta}\epsilon^{*\nu}P_{1}^{\alpha}P_{2}^{\beta}2T_{1}(q^{2})\\
				&+\left[\epsilon_{\mu}^{*}(M^{2}-m^{2})-(\epsilon^{*}\cdot q)(P_{1}+P_{2})_{\mu}\right]T_{2}(q^{2})\\
				&+(\epsilon^{*}\cdot q)\left[q_{\mu}-\frac{q^{2}}{M^{2}-m^{2}}(P_{1}+P_{2})_{\mu}\right]T_{3}(q^{2}).
			\end{split}
		\end{equation} 
		where $q^{\mu}=(P_{1}-P_{2})^{\mu}$ is the momentum transferred to the lepton part. $P_{1}$ and $P_{2}$ are the momenta carried by the initial and final state mesons and are expressed as
		\begin{equation}
			P_{1}=\frac{M}{\sqrt{2}}(1,1,0_{\perp}),\qquad P_{2}=\frac{M}{\sqrt{2}}(r\eta^{+},r\eta^{-},0_{\perp}),
			\label{eqn: kinematics momentum}
		\end{equation}
		respectively, in the light cone coordinate system, with $r=m/M$ and $\eta^{\pm}=\eta\pm\sqrt{\eta^{2}-1}$. The term $\eta$ is expressed as
		\begin{equation}
			\eta=\frac{1+r^{2}}{2r}-\frac{q^{2}}{2rM^{2}}.
		\end{equation}  
		Momenta of the spectator quarks in the initial and final state mesons can be expressed as
		\begin{equation}
			k_{1}=\left(0,x_{1}\frac{M}{\sqrt{2}},k_{1\perp}\right),\qquad k_{2}=\left(x_{2}\frac{M}{\sqrt{2}}r\eta^{+},x_{2}\frac{M}{\sqrt{2}}r\eta^{-},k_{2\perp}\right),
		\end{equation}
		with $x_{1}$ and $x_{2}$ being the fraction of the total momentum carried by respective quarks, and M and m being the masses of initial and final state mesons, respectively. A point to be noted is that these form factors are not independent, but are connected by some constraints arising specifically at $q^{2}=0$ in order to cancel the poles that appear at maximum recoil. These constraints go as
		\begin{equation}
			\begin{split}
			&F_{+}(0)=F_{+}(0),\\
			&2rA_{0}(0)=(1+r)A_{1}(0)-(1-r)A_{2}(0),\\ &T_{1}(0)=T_{2}(0).
			\end{split}
			\label{eqn:form factor QCD constraints}
		\end{equation} 
	
 \paragraph{\bf Form factors calculated in pQCD framework:}	
		The form factors defined above are the quantities that we intend to find out first. In pQCD framework \cite{Li:1994iu}, the form factors are expressed as a convolution of distribution amplitudes of the participaing mesons, which encode the  non-perturbative contributions and is process independent, a hard kernel, which enocodes perturbative contributions and is process dependent, and an exponential term known as the Sudakov factor that reinforces the applicablity of pQCD by suppressing the long distance contributions. They are expressed as
			\begin{equation}
				F_{i}\propto \phi_{B_{s,c}(x,b)}\otimes H(x,t)\otimes\phi_{M}(x,b)\otimes\exp\left[-S(P,b)\right],
			\end{equation}
		where $\phi$, $H(x,t)$ and $S(P,b)$ represent the distribution amplitudes of the participating mesons, hard kernel of the process and Sudakov factor, respectively.
		\begin{itemize}
			\item The hard kernel describes the probability amplitude for a hard scattering event where the active quark from the $B_{s}$ or $B_{c}$ meson transfers momentum to either the spectator or the active quark of the final state meson, via a hard gluon exchange. It represents the short-distance interactions and is calculated perturbatively. We have presented the explicit expressions for the hard kernels used in this work in appendix \ref{section:appendix hard functions}. 
			\item As has been discussed in \cite{Li:1994iu} for $B\rightarrow \pi$ form factor and in \cite{Kurimoto:2002sb} for $B\rightarrow D^{(*)}$ form factors, they generate two kinds of double logarithmic enhancements, which must be resummed for the perturbative framework to remain valid.
			\begin{itemize}
				\item Threshold logarithms of the form $\ln^{2}(1-x)$ or $\ln^{2}x$ arise when the longitudinal momentum fraction of the parton approaches the end-point regions, $x\to 0$ or $x\to 1$ \cite{Li:2001ay}. These logarithms signal the breakdown of fixed-order perturbation theory and therefore require all-order resummation. This is achieved through threshold (jet) resummation, which systematically suppresses contributions from the end-point regions and regulates the associated divergences.
				
		In practical applications, the resummed result is commonly implemented in momentum space using a phenomenological threshold factor,
				\begin{equation}
					\label{eqn:jet function}
					S_{t}(x)
					=
					\frac{2^{\,1+2c}\,\Gamma\!\left(\tfrac{3}{2}+c\right)}
					{\sqrt{\pi}\,\Gamma(1+c)}
					\,[x(1-x)]^{c},
				\end{equation}
				where the parameter $c$ controls the strength of suppression near $x=0$ and $x=1$. Although eq.~\eqref{eqn:jet function} is a simple and convenient form for numerical studies, the choice of the parameter $c$ is not arbitrary. Instead, it is fixed by comparing the exact Sudakov factor $S_{t}(N)$ (defined below) obtained in Mellin space, where threshold resummation is performed in a systematic manner, with the Mellin transform of the phenomenological momentum-space form $S_{t}(x)$. Through this matching procedure, an effective value of $c$ is determined such that the momentum-space parametrisation faithfully reproduces the dominant threshold logarithms contained in $S_{t}(N)$. In this sense, the parameter $c$ encapsulates the underlying resummation dynamics and ensures that the phenomenological jet function provides a reliable approximation to the full threshold-resummed result, while maintaining the desired suppression of end-point contributions. 
				
			The Sudakov factor $S_{t}(N)$ in moment space $N$ upto next-to-leading logarithm accuracy can be expressed as \cite{Li:2001ay,Kurimoto:2001zj}
				\begin{equation}
					S_{t}(N)=\exp\left[\frac{1}{2}\int_{0}^{1-\frac{1}{N}}\frac{dz}{1-z}\int_{1-z}^{(1-z)^{2}}\frac{d\lambda}{\lambda}\gamma_{K}\left(\alpha_{s}\left(\frac{\lambda M^{2}}{2}\right)\right)\right],
					\label{eqn:threshold resummation funciton}
				\end{equation}
				with 
				\begin{equation}
					\gamma_{K}=\frac{\alpha_{s}}{\pi}C_{F}+\left(\frac{\alpha_{s}}{\pi}\right)^{2}C_{F}\left[C_{A}\left(\frac{67}{36}-\frac{\pi^{2}}{12}\right)-\frac{5}{18}n_{f}\right],
				\end{equation}
				representing the anomalous dimension containing the two-loop contributions, where $C_{F}=4/3$, $C_{A}=3$ and $n_{f}$ represents the number of quark flavors. The running coupling constant $\alpha_{s}$ scales as $\alpha_{s}(\frac{\lambda M^{2}}{2})$, where $M$ represents the mass of either $B_{s}$ or $B_{c}$ meson.
				
				To obtain its Mellin-space representation from the above $S_t(x)$ defined in eq.~\eqref{eqn:jet function}, one need to compute the following integral
					\begin{equation}
						S_{t}^{\text{simplified}}(N) = \int_0^1 dx \, x^{N-1} S_t(x).
					\end{equation}
					After solving the above integration, we get
					\begin{equation}
						\frac{S_{t}^{\text{simplified}}(N)}{N}=\frac{2^{1+2c}\Gamma(3/2+c)}{\sqrt{\pi}~\Gamma(1+c)}B(1+c,c+N).
						\label{eqn:threshold resummation fit}
					\end{equation}
The choice of $N$ is not arbitrary and has to lie in a suitable region. The region is selected keeping in mind that:
\begin{itemize}
    \item If $N$ is chosen too small ($N<3$), the threshold logarithm $\ln N$ is small. Consequently, the next-to-leading logarithm perturbative series converges naturally, thus making threshold resummation mathematically unnecessary.
    \item If $N$ is chosen too large ($N>7$), the running coupling constant $\alpha_{s}$ scaling as $\lambda M^{2}/2N^{2}$ in the moment space diverges to the Landau pole. This causes the perturbative expansion to break down.
\end{itemize}
We calculate $S_{t}(N)/N$ using eq.\eqref{eqn:threshold resummation funciton} for $J$ discrete values of $N$ in the range $3<N<7$. We then perform a least-squares fit by minimizing the sum of squared residuals defined as
\begin{equation}
    SSE=\sum_{i=1}^{J}\left[\frac{S_{t}(N)}{N}-\frac{S_{t}^{\text{simplified}}(N)}{N}\right]^{2},
\end{equation}
which gives us the following values of $c$ for $B_{s}\to D_{s}^{(*)}$ and $B_{c}\to D_{s}^{(*)}$ form factors:
\begin{itemize}
    \item For $B_{s}\to D_{s}^{(*)}$ transitions, we get c=0.313.
    \item For $B_{c}\to D_{s}^{(*)}$ transitions, we get c=0.236.
\end{itemize}
The values of $c$ are different for the two channels. This is because the running coupling constant in Sudakov function of eq.\eqref{eqn:threshold resummation funciton} depends on mass of the initial state meson. Hence, values of Sudakov function in eq.\eqref{eqn:threshold resummation funciton} will be different for the two channels.
				
				\item Transverse logarithms of the form $\ln^{2}(Pb)$, $b$ being the impact parameter, and is Fourier conjugate to the transverse momentum $k_{T}$. When $b\to \infty$ or $k_{T}\to 0$, the double logarithm becomes large, signifying the uncontrollable growth of soft gluon contributions, leading to the form factors becoming divergent or unstable, making the perturbative theory unreliable. To regulate this behavior, these transverse logarithms are resummed to all orders, resulting in the Sudakov factor, which suppresses the contributions from large transverse separations and ensures the reliability of the pQCD calculation, and has the form as
				\begin{equation}
					\begin{split}
						s(Q,b)=\int_{1/b}^{Q}\frac{d\mu}{\mu}\left[\ln\left(\frac{Q}{\mu}\right) A(\alpha_{s}(\mu))+B(\alpha_{s}(\mu))\right],
					\end{split}
				\end{equation}
				for decay modes of heavy light meson, with $A(\alpha_{s}(\mu))$ and $B(\alpha_{s}(\mu))$ being the anomalous dimensions to two loops and one loop respectively, with their explicit expressions being taken from \cite{Li:1994iu}.
				
				The $B_{c}$ meson, being a heavy-heavy bound system, involves multiple scales, making resummation of such systems much more complicated compared to that for $B_{(s)}$ meson decays.  However, taking the limit $m_{b}\rightarrow \infty$, but keeping $m_{c}$ finite, the $B_{c}$ meson can be treated as a heavy-light system and analysis of the decays can be carried out in conventional pQCD approach for B meson decays \cite{Kurimoto:2002sb}. This approximation, first introduced in \cite{PhysRevD.97.113001}, modifies the Sudakov factors, thus leading to a modified pQCD formalism. The Sudakov factor thus derived has the form
				
				\begin{equation}
					\begin{split}
						s_{c}(Q,b)=&s(Q,b)-s(m_{c},b),\\
						=&\int_{m_{c}}^{Q}\frac{d\mu}{\mu}\left[\int_{1/b}^{\mu}\frac{d\bar{\mu}}{\bar{\mu}}A(\alpha_{s}(\bar{\mu}))+B(\alpha_{s}(\bar{\mu}))\right].
					\end{split}
				\end{equation}
				
				However it is to be noted that in case of $B_{s}$ mesons, this modified pQCD framework is not necessary. This is because, unlike $B_{c}$ mesons, the $B_{s}$ meson is formed by the $b$ and $s$ quarks, the former being a heavy quark and the latter a light quark. Thus, the analysis of $B_{s}$ mesons can be carried on using the conventional pQCD approach \cite{Kurimoto:2002sb}, and the introduction of a finite charm quark mass scale is not required in this case. We present explicit expressions of the form factors, along with appropriate references in appendix \ref{section:appendix pQCD form factor expressions}.
			\end{itemize}
			
		\end{itemize}
		\subsection{Light Cone distribution amplitudes}
		\label{subsection:LCDAs}
	
		In the last subsection, we discussed that the form factors in pQCD have distribution amplitude of the participating mesons in the convolution. In this subsection, we briefly discuss the form of the distribution amplitudes used in this work. Light cone distribution amplitudes (LCDAs) encode how the momentum of a fast-moving hadron is distributed among the constituent quarks along the light cone direction. These serve as crucial non-perturbative inputs to the form factor expressions in pQCD, and their shape offers a degree of flexibility to obtain predictions of the form factors, by constraining them through model-independent techniques. For this work, we will be discussing the LCDAs of the $B_{s}$, $B_{c}$ and $D_{s}^{(*)}$ mesons. Reason for considering the $B_{s}$ meson LCDA will be explained in the following section.
		
		 \begin{itemize}
			\item For $B_{s}$ meson, the wavefunction has the form as
			\begin{equation}
				\Phi_{B_{s}}(p,x)=\frac{i}{\sqrt{2N_{c}}}\left(\slashed{p}_{B_{s}}+M\right)\gamma_{5}\phi_{B_{s}}(x,b),
			\end{equation}
			where $\phi_{B_{s}}(x,b)$ represents the $B_{s}$ meson LCDA, which assumes an approximate Gaussian form as \cite{Kurimoto:2002sb,Xiao:2011tx} 
			\begin{equation}
				\phi_{B_{s}}(x,b)=\frac{f_{B_{s}}}{2\sqrt{2N_{c}}}N_{B_{s}}x^{2}(1-x)^{2}\exp\left[-\frac{x^{2}M^{2}}{2\omega_{B_{s}}^{2}}-\frac{1}{2}\omega_{B_{s}}^{2}b^{2}\right],
			\end{equation}
			with $M=m_{B_{s}}$, $\omega_{B_{s}}$ the shape parameter defining the shape of the distribution amplitude, and $N_{B_{s}}$ the normalization constant fixed by the relation
			\begin{equation}
				\int_{0}^{1}\phi_{B_{s}}(x,b=0)dx=\frac{f_{B_{s}}}{2\sqrt{2N_{c}}},
			\end{equation}
			where $f_{B_{s}}$ is the decay constant of $B_{s}$ meson.
			\item For $B_{c}$ meson, the wavefunction has the form as
			\begin{equation}
				\Phi_{B_{c}}(p,x)=\frac{i}{\sqrt{2N_{c}}}\left(\slashed{p}_{B_{c}}+M\right)\gamma_{5}\phi_{B_{c}}(x,b),
			\end{equation}
			where $M=m_{B_{c}}$ and $\phi_{B_{c}}(x,b)$ represents the $B_{c}$ meson LCDA, which we consider to have an approximate Gaussian form as \cite{PhysRevD.97.113001}
			\begin{equation}
				\begin{split}
					\phi_{B_{c}}(x,b)=\frac{f_{B_{c}}}{2\sqrt{2N_{c}}}N_{B_{c}} x (1-x) exp\left[ -\frac{(1-x)m_{c}^{2}+xm_{b}^{2}}{8\omega_{B_{c}}^{2}x(1-x)}\right]\exp[-2~\omega_{B_{c}}^{2}x(1-x)b^{2}],
				\end{split}
			\end{equation}
			The normalization constant $N_{B_{c}}$ is fixed by the relation
			\begin{equation}
				\int_{0}^{1}\phi_{B_{c}}(x,b=0)dx=\frac{f_{B_{c}}}{2\sqrt{2N_{c}}},
			\end{equation}
			and the parameter $b$ being the impact parameter, which is \textcolor{blue}{in fact} Fourier conjugate to the transverse momentum $k_{T}$, $\omega_{B_{c}}$ being the shape parameter of the $B_{c}$ meson distribution amplitude and $f_{B_{c}}$ the decay constant of $B_{c}$ meson.
			\item For $D_{s}$ and $D_{s}^{*}$ mesons, the wavefunctions have the same form as in \cite{Hu:2019bdf}
			\begin{equation}
				\footnotesize
				\begin{split}
					\Phi_{D_{s}}(p,x)=&\frac{i}{\sqrt{2N_{c}}}\gamma_{5}\left(\slashed{p}_{D_{s}}+m\right)\phi_{D_{s}}(x,b),\\
					\Phi_{D_{s}^{*}}(p,x)=&-\frac{i}{\sqrt{2N_{c}}}\left[\slashed{\epsilon}_{L}\left(\slashed{p}_{D_{s}^{*}}+m\right)\phi_{D_{s}^{*}}^{L}(x,b)+\slashed{\epsilon}_{T}\left(\slashed{p}_{D_{s}^{*}}+m\right)\phi_{D_{s}^{*}}^{T}(x,b)\right],
				\end{split}
			\end{equation}
			where $m=m_{D_{s}^{(*)}}$, $\phi_{D_{s}}(x,b)$, $\phi_{D_{s}^{*}}^{L}(x,b)$ and $\phi_{D_{s}^{*}}^{T}(x,b)$ the LCDAs of respective mesons. In this work we consider these to have a simple polynomial form as \cite{Li:2008ts}
			\begin{equation}
				\label{eqn:wf1}
				\phi_{D_{s}^{(*)}}(x,b)=\frac{f_{D_{s}^{(*)}}}{2\sqrt{2N_{c}}}6x(1-x)\left[1+C_{D_{s}^{(*)}}(1-2x)\right]\cdot \exp\left[-\frac{\omega_{D_{s}^{(*)}}^{2}b^{2}}{2}\right],
			\end{equation}
			where $C_{D_{s}^{(*)}}$ and $\omega_{D_{s}^{(*)}}$ are parameters that control the shape of the corresponding distribution amplitudes. $\phi_{D_{s}^{(*)}}$ satisfies the normalization condition
			\begin{equation}
				\int_{0}^{1}\phi_{D_{s}^{(*)}}(x,0)dx=\frac{f_{D_{s}^{(*)}}}{2\sqrt{2N_{c}}},
			\end{equation}
			with $f_{D_{s}^{(*)}}$ being the decay constant of the respective meson.
		\end{itemize}
        
        In most existing studies on $B_s$ and $B_c$ decays in pQCD framework \cite{PhysRevD.90.094018,PhysRevD.97.113001,Hu:2019bdf}, the authors simply treat the shape parameters ($\omega_{B_{s}}$, $\omega_{B_{c}}$, etc.) as fixed numbers. They also restrict their use of Lattice QCD data to just the maximum momentum transfer ($q^2_{\text{max}}$). Our work takes a different approach. Instead of fixing these values beforehand, we treat them as free parameters and find their best values by fitting them to the data. By analyzing the $B_c \to D_{s}$ and $B_{s} \to D_{s}^{(*)}$ channels together, we get a consistent set of parameters for everyone. This combined method also lets us see how the parameters are connected to each other (the correlation matrix), something we cannot infer if we analyze the channels separately. Understanding these connections is key to getting reliable error estimates for our final predictions.	
		\begin{table}[t!]
		\renewcommand{\arraystretch}{1.5}
		\centering
		\begin{tabular}{|c|lll|}
			\hline 
			\textbf{Mass (\text{GeV})} & $m_{B_{c}}=6.274$ & $m_{B_{s}}=5.367$ &  \\ 
			
			& $m_{D_{s}}=1.969$ & $m_{D_{s}^{*}}=2.112$ &  \\ 
			
			& $m_{e}=0.511\times 10^{-3}$&$m_{\mu}=0.105$&$m_{\tau}=1.776$\\
			\hline 
			\textbf{Decay}&$f_{B_{c}}=0.427(6)$\cite{McNeile:2012qf}&$f_{B_{s}}=0.229(5)$\cite{Lubicz:2017asp}&\\
			\textbf{constants (\text{GeV})}&$f_{D_{s}}=0.2480(25)$\cite{Lubicz:2017asp}&$f_{D_{s}^{*}}=0.2688(65)$\cite{Lubicz:2017asp}&\\
			\hline
			\textbf{CKM}&$|V_{cb}|=41.1(1.2)\times 10^{-3}$&&\\
			\textbf{matrix elements}&$|V_{us}|=0.22431(85)$&$|V_{ub}|=3.82(20)\times 10^{-3}$&\\
			\cite{ParticleDataGroup:2024cfk}&$|V_{ts}|=41.5(9)\times 10^{-3}$&$|V_{tb}|=1.010(27)$&\\
			\hline
			\textbf{Lifetime (ps)}&$\tau_{B_{c}}=0.510(9)$\cite{ParticleDataGroup:2024cfk}&&\\
			\hline
		\end{tabular}
		\caption{Values of input parameters used in this work.}
		\label{table:input parameters}
	\end{table}

	\section{Extraction of LCDA shape parameters and form factors at $q^{2}=0$}
	\label{section:Extraction of LCDA parameters}
	
	With all the theoretical prerequisites discussed, we now move on to extracting the shape parameters of LCDAs of the participating mesons. For the $B_{c}\rightarrow D_{s}$ mode, HPQCD \cite{Cooper:2021bkt} has obtained information of the form factors over full $q^{2}$ region using Bourrley-Caprini-Lellouch (BCL) \cite{Bourrely:2008za} parametrization, through which we can conveniently obtain the relevant form factors at $q^{2}=0$. Additionally, to better constrain the $D_{s}^{(*)}$ LCDA shape parameters, we consider the $B_{s}\rightarrow D_{s}^{(*)}$ modes, information of which has been supplied by HPQCD \cite{McLean:2019qcx,Harrison:2021tol}. In addition to the lattice inputs, we also use inputs of the respective form factors presented in the LCSR approach \cite{Bordone:2019guc}. Note that our objective is to perform a comprehensive global analysis by incorporating all well-established and reliable theoretical inputs available in the literature. In particular, lattice QCD provides robust and model-independent determinations of the relevant form factors in the high-$q^2$ region, where its predictions are most reliable. These constraints are complemented by light-cone sum rule (LCSR) calculations at $q^2=0$, which are known to provide reliable information in the low-$q^2$ region. By combining these two non-perturbative approaches, we exploit their complementary regions of validity across the kinematic range. This strategy allows the meson distribution-amplitude parameters to be constrained simultaneously by precise lattice QCD inputs at large $q^2$ and LCSR predictions at small $q^2$, thereby making use of the full range of currently available theoretical information.

	The numerical values of the form factor data for the first chi-square minimization are shown in table \ref{table:lattice inputs for BDStar form factors}. Corresponding correlation matrix between these inputs are presented in table \ref{table:Extraction of LCDA shape parameters input correlations}. As for the other inputs, such as CKM matrix elements, mass of intial and final state mesons and decay constants we take their values preented in table \ref{table:input parameters}.
	
	Finally, we construct a chi-square function with the shape parameters as free parameters, and then minimize it. The chi-square function has the form \cite{Cowan:1998}
	\begin{equation}
		\label{eqn:chi square function}
		\chi^{2}=\sum_{i,j}(\mathcal{O}_{i}^{\text{th}}-\mathcal{O}_{i}^{\text{data}})V_{ij}^{-1}(\mathcal{O}_{j}^{\text{th}}-\mathcal{O}_{j}^{\text{data}})^{T}+\chi^{2}_{\text{nuis}},
	\end{equation}
	where $\mathcal{O}_{i}^{\text{th}}$ represents the pQCD expressions for form factors at $q^{2}=0$, $\mathcal{O}_{i}^{\text{data}}$ represents the inputs on the corresponding form factors at $q^{2}=0$ and $V_{ij}$ represents the covariance matrix between the inputs.  These covariance matrices are derived as
		\begin{equation}
			V_{ij}=C_{ij}\sigma_{i}\sigma_{j},
		\end{equation}
		with $\sigma_{i}$ and $C_{ij}$ being the $1\sigma$ uncertainty of each input and the correlation matrix between them, respectively. As for the analysis in this section, uncertainties in the inputs are presented in table \ref{table:lattice inputs for BDStar form factors} and the corresponding correlation matrix in table \ref{table:Extraction of LCDA shape parameters input correlations}. In later sections of the paper, wherever $\chi^2$-minimization method has been adopted to extract the relevant parameters, the inputs considered and corresponding correlation matrices have been quoted in the text.
	
	The quantity $\chi^2_{\text{nuis}}$ denotes the chi-square contribution associated with the nuisance parameters and is defined as
		\begin{equation}
			\chi^2_{\text{nuis}} =
			\sum_{k,l=1}^{n}
			\left( \mathcal{P}_k - \mathcal{P}_k^{\text{th}} \right)
			V^{-1}_{kl}
			\left( \mathcal{P}_l - \mathcal{P}_l^{\text{th}} \right),
		\end{equation}
		where $\mathcal{P}_k$ represents the floating value of the $k$-th nuisance parameter, $\mathcal{P}_k^{\text{th}}$ denotes its corresponding central (theoretical or experimental) value, and $V_{kl}$ is the covariance matrix encoding the uncertainties and correlations among the nuisance parameters. The summation runs over all parameters chosen to be treated as nuisance parameters.
		In the absence of correlations among the nuisance parameters, the covariance matrix becomes diagonal,
		\(
		V_{kl} = \delta_{kl}\,\sigma_k^2,
		\)
		where $\sigma_k$ is the standard deviation associated with $\mathcal{P}_k$. In this case, the inverse covariance matrix is simply given by $V^{-1}_{kl} = \delta_{kl}/\sigma_k^{2}$, and the above expression reduces to a sum of squared pulls weighted by the inverse variances. In this analysis, we have included the correction terms $\delta_i$ for each of the form factors and the masses $m_{b}$, $m_{c}$ as nuisance parameters. These variables are listed in the right column of table \ref{table:extracted value of Bc and D shape parameters}.

 In a standard $\chi^{2}$-function, a free parameter is completely unconstrained by any kind of external priors. Its central value and uncertainty are determined entirely through $\chi^{2}$ minimization based on the dataset being used as inputs. On the other hand, a nuisance parameter is one whose preliminary estimate of central value and uncertainty is available to us via some prior independent extraction method. These parameters themselves are not the primary parameters of interest, and are only added to act as a penalty function to the $\chi^{2}$-function constructed for free parameters. The minimization algorithm allows the nuisance parameter to fluctuate within its known prior uncertainty, thus ensuring that this error propagates into the final extracted estimate of the free parameters. 
	
	\begin{table}[t]
		\centering
		\renewcommand{\arraystretch}{1.1}
		\begin{tabular}{|c|c|c|c|}
			\hline
			\textbf{Decay}&\textbf{Form}&\multicolumn{2}{c|}{\textbf{Values at }$\boldsymbol{q^{2}=0}$}\\
			\cline{3-4}
			\textbf{Channel}&\textbf{Factors}&\textbf{Lattice}&\textbf{LCSR}\\
			&&\textbf{ (HPQCD)}&\\
			\hline
			$B_{s}\rightarrow D_{s}$&$F_{+}(0)$&0.665(12)&0.708(153)\\
			\hline
			&$A_{0}(0)$&0.611(55)&0.731(187)\\
			$B_{s}\rightarrow D_{s}^{*}$&$A_{1}(0)$&0.591(40)&0.633(143)\\
			&$V(0)$&0.977(153)&0.735(176)\\
			\hline
			$B_{c}\rightarrow D_{s}$&$F_{+}(0)$&0.217(18)&-\\
			&$F_{T}(0)$&0.299(54)&-\\
			\hline
		\end{tabular}
		\caption{Form factor inputs for $B_{s}\rightarrow D_{s}^{(*)}$ and $B_{c}\rightarrow D_{s}$ semileptonic channels at $q^{2}=0$.}
		\label{table:lattice inputs for BDStar form factors}
	\end{table}
	
	Coming back to our analysis, we take the charm and bottom quark masses, $m_{c}$ and $m_{b}$, as the arithmetic averages of their values in the pole, $ \overline{\text{MS}}$, and kinetic schemes. To ensure a scheme-independent and inclusive treatment of mass uncertainties, we assign relative errors of 25\% for $m_{c}$ and 10\% for $m_{b}$, chosen to encompass the full range of variation across these schemes. The quark masses are presented in table \ref{table:scheme dependent masses}.
	
	\begin{table}[t!]
		\centering
		\begin{tabular}{|c|cc|}
			\hline
			\textbf{Scheme}&$\boldsymbol{m_{b}}$ (\text{GeV})&$\boldsymbol{m_{c}}$ (\text{GeV})\\
			\hline
			Pole mass&4.78&1.67\\
			$\overline{MS}$&4.18&1.273\\
			Kinetic&4.56&1.091\\
			\hline
			Average&4.506(451)&1.345(336)\\
			\hline
		\end{tabular}
		\caption{Values of $m_{b}$ and $m_{c}$ in three different schemes and their average value \cite{ParticleDataGroup:2024cfk}.}
		\label{table:scheme dependent masses}
	\end{table}
	
	With these inputs, we can now construct the relevant chi-square function and then minimize it to extract the required shape parameters of the meson LCDAs, i.e. $\omega_{B_{c}},~C_{D_{s}^{(*)}}$ and $\omega_{D_{s}^{(*)}}$ along with bottom and charm quark masses as nuisance parameters. It should be noted that the shape parameters associated with the $B_s$ and $B_c$ meson distribution amplitudes enter the meson wave functions through exponential terms, resulting in a highly non-linear dependence of the pQCD form factors on these parameters. Although, in principle, one could construct a database of precomputed form factors over the entire parameter space, such an approach becomes computationally impractical in our analysis due to the high dimensionality of the fit, which involves 16 independent parameters. Consequently, instead of relying on precomputed grids and interpolation, the form factors are evaluated dynamically during the $\chi^2$ minimization, ensuring that the theoretical predictions are calculated directly for each parameter point explored by the fitting algorithm. The fit is performed by comparing a data vector, containing all available lattice QCD and LCSR inputs together with their covariance matrix, with a theoretical prediction vector whose elements correspond to the relevant pQCD form factors. These theoretical predictions depend not only on the six LCDA shape parameters but also on a number of nuisance parameters entering the meson distribution amplitudes and other theoretical inputs. Each element of the prediction vector is obtained through the numerical evaluation of multidimensional pQCD integrals. 
	
	The determination of pQCD the form factors at $q^2=0$ therefore requires specifying physically motivated ranges for all free parameters. The chosen parameter intervals are not arbitrary; rather, they are guided by existing theoretical studies and available phenomenological constraints in the literature \cite{Hu:2019bdf,PhysRevD.97.113001,PhysRevD.90.094018,Wang:2012lrc}. For the six LCDA shape parameters, the initial ranges are selected to encompass all currently available estimates. In particular, we take $\omega_{B_c}\in[0.8,1.2]\,\mathrm{GeV}$ and $\omega_{B_s}\in[0.3,0.7]\,\mathrm{GeV}$, while for the $D_s^{(*)}$ meson distribution amplitudes we adopt $C_{D_s^{(*)}}\in[0.3,0.7]$ and $\omega_{D_s^{(*)}}\in[0.05,0.3]$. Similarly, the initial ranges of the nuisance parameters, including the heavy-quark masses $m_b$ and $m_c$, are chosen to fully cover the corresponding $1\sigma$ intervals listed in table~\ref{table:scheme dependent masses}. The same strategy is applied to the correction parameters $\delta_{f_i}$, whose ranges are determined from their respective uncertainties. The motivation for introducing these correction terms and their role in the fit are discussed in detail later in this section.
	
	Since the evaluation of a single form factor typically requires about 5-6 minutes, a serial computation of all form factors entering the fit would be prohibitively expensive. To address this issue, we implemented a parallelized framework in \textit{Mathematica} using \texttt{ParallelSubmit} and \texttt{WaitAll}, allowing independent form-factor calculations to be performed simultaneously and reducing the effective computation time per parameter point. The global minimization was then carried out with \texttt{NMinimize}\footnote{The global minimization was performed using \texttt{NMinimize} in \textit{Mathematica}, which offers several optimization algorithms, including \texttt{DifferentialEvolution}, \texttt{RandomSearch}, \texttt{NelderMead}, and \texttt{SimulatedAnnealing}. We tested multiple methods and found that all of them yielded consistent best-fit values, indicating the robustness of the fit. However, owing to the highly non-linear dependence of the pQCD form factors on the fit parameters and the presence of small numerical fluctuations arising from multidimensional integrations, \texttt{RandomSearch} exhibited the most stable and reproducible convergence behaviour. Since this method performs a broad exploration of the parameter space through multiple random starting points before applying local optimization, it is less susceptible to local structures in the $\chi^2$ landscape and provides a more reliable identification of the global minimum. For this reason, although other algorithms produced compatible results, the final results reported in this work are obtained using the \texttt{RandomSearch} method.}, starting from broad parameter ranges and iteratively refining the search region to ensure convergence to the global minimum. In addition, a simple normalization of the $\chi^2$ function was introduced to improve numerical stability and convergence efficiency without affecting the location of the minimum or the extracted parameter values. Consequently, our analysis relies on a fully dynamic, parallelized, and numerically optimized fitting framework rather than a precomputed database.

	Our current analysis already incorporates radiative corrections up to $\mathcal{O}(\alpha_s)$ and $\mathcal{O}(\alpha_s^2)$, following the calculations in ref.~\cite{Keum:2000wi} for $B$ mesons and in \cite{Liu:2020upy} for $B_c$ meson wave functions. The calculations are relevant for the $B_{s} \to P(V)$ and $B_c \to P(V)$ form factors within the pQCD framework. We emphasize that the pQCD form factors employed in this analysis incorporate only the leading-order (LO) contributions in the hard kernel, and the meson wave functions are defined within the leading-twist approximation. Furthermore, unlike the expressions presented in \cite{PhysRevD.90.094018}, which were derived in the limits $m_{b}\to m_{B_{c}}$ (or $r_{b}\to 1$) and $m_{s}\to 0$ (or $r_{s}\to 0$), we have derived our expressions by incorporating finite $r_{b}$ and $r_{s}$ corrections. We have presented the updated expressions in appendix \ref{section:appendix pQCD form factor expressions}.
    
LCDAs describe how the momentum of a hadron is shared among its constituent partons when projected onto the light cone. In our analysis, we employ a QCD-inspired distribution amplitude, meaning that the longitudinal part of the light-cone wave function is constrained by principles of Quantum Chromodynamics (QCD), rather than being purely phenomenological as in the simple Gaussian ansatz.

A Gaussian ansatz is mathematically convenient but does not accurately represent QCD dynamics; in particular, it suppresses the endpoint regions $x \to 0$ and $x \to 1$ too strongly. QCD-inspired DAs avoid this issue because they incorporate information from QCD sum rules, lattice QCD, and the known perturbative-QCD asymptotic behavior. Consequently, they provide realistic endpoint behavior and a more accurate momentum-fraction dependence.

In our framework, the full light-cone wave function is written as the product of two components:
\begin{enumerate}
	\item the longitudinal distribution amplitude, taken from a QCD-inspired model, and
	\item a transverse-momentum profile, for which we use a scale-independent Gaussian function.
\end{enumerate}
This approach preserves the simplicity of a Gaussian transverse profile while improving the physical accuracy of the longitudinal structure through QCD-based constraints.

In the light-cone formalism, the twist of an operator is defined as the difference between its mass dimension and its spin. The leading contribution arises from twist-2 operators, which correspond to the simplest quark--antiquark configuration and dominate the behavior at large momentum transfer. In this work, we restrict our attention to the leading-twist light-cone distribution amplitudes (LCDAs).

Higher-twist terms (twist-3, twist-4, etc.) originate from several sources, such as
\begin{itemize}
	\item intrinsic transverse momentum of quarks,
	\item quark-gluon interactions,
	\item and more involved multi-parton correlations.
\end{itemize}
Although these contributions are formally suppressed by powers of $\Lambda_{\text{QCD}}/m_Q$, they may still play a significant role, especially at low and intermediate values of $q^2$, where nonperturbative effects become more pronounced. Twist-3 amplitudes typically involve pseudoscalar and tensor components, while twist-4 terms incorporate explicit quark--gluon operator structures. Including these higher-twist effects introduces additional nonperturbative parameters but can enhance the precision and reliability of form-factor predictions.

However, in the present analysis we do not have sufficient input to reliably constrain all higher-twist contributions. To account for the possible impact of these missing effects, we introduce an additional theoretical uncertainty in the LCDA shape parameters and in the resulting form factors. Previous studies of $B \to \pi$ and $B \to \rho$ form factors within the pQCD framework (see refs.~\cite{Wang:2012ab,Cheng:2014fwa,Mahajan:2004dx}) show that next-to-leading-order (NLO) corrections to the hard kernel, together with higher-twist effects, can induce shifts of about 20-30\% relative to leading-order (LO) predictions. Motivated by these findings, we conservatively assign a 30\% uncertainty to our LO form factor predictions to cover the dominant theoretical uncertainties associated with neglected higher-order radiative corrections and higher-twist LCDA contributions.

This uncertainty is implemented through a multiplicative nuisance parameter, $\delta_{f_i}$, which is constrained to lie within this range. Treating $\delta_{f_i}$ as a nuisance parameter during the minimization procedure ensures that its effect is consistently propagated into the final uncertainties of the extracted shape parameters.

When constructing the $\chi^{2}$ function, we do not include the uncertainties of the meson decay constants in $\chi^{2}_{\text{nuis}}$, since their errors are typically only at the level of 2-3\% (see table~\ref{table:input parameters}). Compared to the much larger theoretical uncertainties associated with the form factors, the impact of varying these decay constants on the fit results is negligible.

	\begin{table}[t]
	\centering
	\renewcommand{\arraystretch}{1.3}
	\begin{tabular}{|cl|cl|}
		\hline
		\multicolumn{2}{|c|}{\textbf{Free Parameters}}  &  \multicolumn{2}{c|}{\textbf{Nuisance Parameters}} \\
		\hline
		\textbf{Parameters}& \textbf{Fit Results} & \textbf{Parameters}& \textbf{Fit Results} \\
		\hline
		$\omega_{B_{c}}$ &1.019(28)~\text{GeV} & $m_{b}$ & 4.508(354)~\text{GeV} \\
		$\omega_{B_{s}}$ &0.496(36)~\text{GeV} & $m_{c}$ & 1.332(79)~\text{GeV} \\
		$C_{D_{s}}$ & 0.495(15) & $\delta_{f_{1}}^{B_{s}\to D_{s}}$ & -0.072(367) \\
		$C_{D_{s}^{*}}$ & 0.505(39) & $\delta_{f_{2}}^{B_{s}\to D_{s}}$ & -0.004(172) \\
		$\omega_{D_{s}}$& 0.104(12) & $\delta_{A_{0}}^{B_{s}\to D_{s}^{*}}$ & -0.104(153) \\
		$\omega_{D_{s}^{*}}$&0.098(15)&$\delta_{A_{1}}^{B_{s}\to D_{s}^{*}}$ & -0.022(383)\\
		&&$\delta_{V}^{B_{s}\to D_{s}^{*}}$ & 0.041(44)\\
		&&$\delta_{f_{1}}^{B_{c}\to D_{s}}$ & -0.011(163)\\
		&&$\delta_{f_{2}}^{B_{c}\to D_{s}}$ & 0.014(130)\\
		&&$\delta_{F_{T}}^{B_{c}\to D_{s}}$ & 0.009(123)\\
		\hline
		\textbf{DOF}&\multicolumn{3}{|c|}{4}\\
        \hline
		$\boldsymbol{\chi^{2}_{\text{min}}/DOF}$&\multicolumn{3}{|c|}{0.605}\\
		\hline
		\textbf{p-Value}&\multicolumn{3}{|c|}{65.93\%}\\
		\hline
	\end{tabular}
	\caption{Extracted values of LCDA parameters obtained by fitting pQCD form factors of $B_{c}\rightarrow D_{s}$ and $B_{s}\rightarrow D_{s}^{(*)}$ transitions with corresponding lattice and LCSR inputs at $q^{2}=0$.}
	\label{table:extracted value of Bc and D shape parameters}
\end{table}

We present our estimates of the thus extracted parameters in table \ref{table:extracted value of Bc and D shape parameters} along with the corresponding correlation matrix in table \ref{table: correlation LCDA BcD}. Checking table \ref{table:extracted value of Bc and D shape parameters} we see that $\chi^{2}_{min}/\text{DOF}< 1$, signifying that our fit is statistically good. We also see that our estimate of $\omega_{B_{c}}$ is in good agreement with our previously extracted value in \cite{Dey:2025xdx,Dey:2025xjg}, thereby justifying its estimate through three independent analyses. Our estimates of shape parameters of $B_{s}$ and $D_{s}^{(*)}$ mesons are also in good agreement with existing model-dependent estimates.

Note that in ref.~\cite{Biswas:2023bqz}, the parameter $\omega_{J/\psi}$ was extracted using experimental data on the leptonic decay width $\Gamma(J/\psi \to e^+e^-)$ together with lattice-QCD inputs for the decay constant $f_{J/\psi}$. Similarly, in the same analysis, the parameter $\omega_{\eta_c}$ was determined from the two-photon decay width $\Gamma(\eta_c \to \gamma\gamma)$ combined with lattice input on $f_{\eta_c}$. In that work, both $\omega_{J/\psi}$ and $\omega_{\eta_c}$ were treated as \emph{free fit parameters} and were determined through a $\chi^2$ minimization procedure, rather than being introduced as nuisance parameters.

In contrast, in our earlier work \cite{Dey:2025xdx}, these parameters were already available from ref.~\cite{Biswas:2023bqz} with well-controlled uncertainties, dominantly constrained by quarkonium decay data and lattice inputs. Since the uncertainties associated with the $B_{c} \to J/\psi$ transition form factors are comparatively larger, these observables do not provide competitive constraints on $\omega_{J/\psi}$ or $\omega_{\eta_c}$. As a result, repeating the extraction using the same decay-width inputs would be redundant and would not lead to any meaningful improvement in their determination.

For this reason, in \cite{Dey:2025xdx} we directly adopted the previously extracted values of $\omega_{J/\psi}$ and $\omega_{\eta_c}$ as external inputs and treated them as \emph{nuisance parameters} in the subsequent $\chi^2$ minimization. This procedure avoids unnecessary repetition of the analysis while ensuring consistency across the two studies, and it correctly reflects the fact that these parameters are predominantly constrained by quarkonium decay data rather than by the $B_{c}\to J/\psi$ transition form factors. In the present work, we perform a simultaneous extraction of all parameters associated with the $D_s^{(*)}$ meson light-cone wave functions. To the best of our knowledge, there exists no prior analysis in the literature that independently constrains the unknown shape parameters of the $D_s^{(*)}$ meson wave functions. Consequently, the corresponding $\chi^{2}$ function has been constructed by treating the shape parameters of both the initial- and final-state mesons as \emph{free parameters}.

 In particular, the parameters of the $D_s$ meson, namely $C_{D_s}$ and $\omega_{D_s}$, are constrained by lattice-QCD inputs for the $B_c \to D_s$ and $B_s \to D_s$ transition form factors. Likewise, the shape parameters of the $D_s^{*}$ meson, $C_{D_s^{*}}$ and $\omega_{D_s^{*}}$, are constrained using lattice inputs for the $B_s \to D_s^{*}$ form factors. Since these lattice determinations currently provide the most reliable and precise information on the relevant heavy-to-heavy transition dynamics, they play a crucial role in fixing the $D_s^{(*)}$ wave-function parameters within our global fit. This behaviour is expected given the high precision and extensive kinematic coverage of the available lattice results. By contrast, the influence of the LCSR inputs at $q^2=0$ is comparatively weaker, as these inputs carry larger uncertainties and represent constraints at a single kinematic point. Consequently, their impact on the extracted shape parameters is naturally subdominant relative to that of the lattice determinations.

Nevertheless, we believe that the inclusion of the LCSR information remains important for several reasons. First, a global analysis should incorporate all reliable and independent theoretical inputs available in the literature rather than relying exclusively on a single framework. Second, the LCSR predictions provide complementary information in the low-$q^2$ region, where lattice calculations are less directly accessible, and are obtained only through suitable extrapolation techniques. Third, excluding the LCSR inputs would prevent us from quantitatively assessing their consistency with the lattice-constrained fit and from evaluating their actual impact on the extracted distribution-amplitude parameters. Therefore, although the final results are primarily driven by the more precise lattice QCD inputs, the inclusion of the LCSR constraints is essential for completeness, for testing the mutual consistency of different theoretical approaches, and for providing a fully comprehensive determination of the meson distribution-amplitude shape parameters. This strategy ensures a consistent and data-driven determination of all previously unconstrained $D_s^{(*)}$ meson wave-function parameters, while avoiding model-dependent assumptions or ad hoc inputs.

Apart from providing complementary low-$q^2$ information, the inclusion of the LCSR inputs is also statistically important. Our fit involves six independent shape parameters, and using only the available lattice QCD inputs would leave us with six constraints for six parameters, corresponding to zero degrees of freedom. In that case, the analysis would effectively reduce to solving a system of equations rather than performing a genuine statistical fit, making the determination of uncertainties and correlations through the Hessian matrix unreliable. The LCSR inputs supply additional independent constraints, rendering the fit overconstrained and allowing for a robust extraction of the shape parameters together with meaningful uncertainty estimates.

    
To verify the consistency of our extracted parameters, in table \ref{table:form factor predictions BsDStar}, we evaluate the $B_{s}\to D_{s}^{(*)}$ and $B_{c}\to D_{s}$ form factors using the parameters extracted in table \ref{table:extracted value of Bc and D shape parameters} as inputs into the expression of appendix \ref{section:appendix pQCD form factor expressions} and compare them with available lattice and LCSR predictions. From the table, we observe that the form factors calculated using our extracted parameters are well consistent with the inputs. Our predicted central values for both the $B_s$ and $B_c$ form factors fall well within the $1\sigma$ error margins of the lattice QCD and LCSR inputs, thus confirming the reliability of the extracted LCDA shape parameters. 
    \begin{table}[t]
		\renewcommand{\arraystretch}{1.5}
		\centering
		\begin{tabular}{|c|ccc|}
			\hline
			\textbf{Form Factors}& \textbf{This work}&\textbf{Lattice QCD}\cite{Cooper:2021bkt,McLean:2019qcx,Harrison:2021tol}&\textbf{LCSR}\cite{Bordone:2019guc} \\
			\hline
			$F_{+}^{B_{s}\rightarrow D_{s}}(0)$& 0.644(14) & 0.665(12) & 0.709(153)\\
			
			$A_{0}^{B_{s}\rightarrow D_{s}^{*}}(0)$& 0.640(83) & 0.611(80) & 0.731(188)\\
			
			$A_{1}^{B_{s}\rightarrow D_{s}^{*}}(0)$& 0.554(35) & 0.591(40)& 0.633(143) \\
			
			$V^{B_{s}\rightarrow D_{s}^{*}}(0)$& 0.903(132) & 0.977(153) & 0.736(176)\\
			
			$F_{+}^{B_{c}\rightarrow D_{s}}(0)$& 0.203(16) & 0.217(18) & -\\

            $F_{T}^{B_{c}\rightarrow D_{s}}(0)$& 0.293(57) & 0.299(54) & -\\
			\hline
		\end{tabular}
		\caption{Prediction of form factors of $B_{s}\rightarrow D_{s}^{(*)}$ and $B_{c}\rightarrow D_{s}$ transitions at $q^{2}=0$ along with comparison with lattice HPQCD and LCSR results used as inputs in this analysis.}
		\label{table:form factor predictions BsDStar}
	\end{table}
 
A statistically meaningful assessment of fit quality should be based on the agreement between predictions and inputs within their quoted uncertainties, rather than on the absolute numerical values of the best-fit parameters alone. In a properly constructed $\chi^2$ framework, a fit is considered satisfactory when the fitted predictions lie within the one--sigma uncertainty bands of the corresponding input quantities, as this indicates consistency at the level expected from Gaussian statistics.
 
It is important to emphasise that best-fit values by themselves do not carry intrinsic statistical significance: they depend on the particular realisation of the data, correlations among parameters, and the chosen parametrisation. Two fits with different best-fit points may be statistically equivalent if both are well within the allowed uncertainty region. Consequently, comparing best-fit values in isolation-without reference to uncertainties, correlations, or the resulting $\chi^2$ probability-can be misleading and is not a reliable measure of goodness of fit.
 
In contrast, the $\chi^2$ probability (or equivalently, the pull of each observable) provides a statistically well-defined criterion. If the fitted observables are within one standard deviation of their input values, the resulting $\chi^2/\mathrm{d.o.f.}$ is naturally of order unity, corresponding to a good fit probability. Deviations at this level are not only acceptable but expected due to statistical fluctuations. Therefore, one-sigma agreement should be interpreted as evidence that the model adequately describes the data within its stated uncertainties.
 
 For this reason, our assessment of fit quality is based on the consistency of the predictions with the input constraints within their full covariance structure, rather than on the numerical proximity of individual best-fit parameters. This approach correctly accounts for correlations, uncertainties, and degeneracies, and provides a statistically robust measure of the goodness of fit.
 
Further, for $B_{c}\to D_{s}^{*}$ channel we present our predictions, along with comparisons with previous pQCD and other model-dependent predictions in table \ref{table:form factor predictions BcDStar}, and the corresponding correlation matrix in table \ref{table:correlation form factors BcDStar}. In these predictions, we have explicitly applied a $30\%$ correction factor to account for uncertainties arising from loop-level corrections and next-to-leading-twist LCDAs. Furthermore, the errors associated with the decay constants have also been propagated independently to the final results. In these tables we have not quoted prediction of form factors $A_{2}(0)$ and $T_{2}(0)$, since these form factors are dependent on $A_{0}(0)$, $A_{1}(0)$ and $T_{1}(0)$ via kinematic constraints at $q^{2}=0$ discussed in eq.\eqref{eqn:form factor QCD constraints}.
	
	\begin{table}[t]
		\renewcommand{\arraystretch}{1.5}
		\centering
		\begin{tabular}{|c|cccc|}
			\hline
			\textbf{Form Factors}& \textbf{This work}&\textbf{Previous pQCD}\cite{PhysRevD.90.094018}&\textbf{CLFQM}\cite{Wang:2008xt}&  \\
			\hline
			$A_{0}^{B_{c}\rightarrow D_{s}^{*}}(0)$& 0.219(65) & 0.21(4) &$0.17^{+0.01+0.01}_{-0.01-0.01}$ &\\
			
			$A_{1}^{B_{c}\rightarrow D_{s}^{*}}(0)$& 0.236(77) & 0.23(4) &$0.14^{+0.01+0.02}_{-0.01-0.01}$ &\\
			
			$V^{B_{c}\rightarrow D_{s}^{*}}(0)$& 0.319(112) & 0.33(6)& $0.23^{+0.02+0.03}_{-0.02-0.02}$ &\\
			
			$T_{1}^{B_{c}\rightarrow D_{s}^{*}}(0)$& 0.269(95) & 0.28(6) &- &\\
			
			$T_{3}^{B_{c}\rightarrow D_{s}^{*}}(0)$& 0.234(82) & 0.27(6) &- &\\
			\hline
		\end{tabular}
		\caption{Prediction of form factors of $B_{c}\rightarrow D_{s}^{(*)}$ transition at $q^{2}=0$ along with comparison with previous predictions.}
		\label{table:form factor predictions BcDStar}
	\end{table}
	
	Revisiting the form factor expressions in appendix \ref{section:appendix pQCD form factor expressions}, we can see that the integrations over $b_{1}$ and $b_{2}$ have been done up to a cut-off $b_{c}$. In this work, we have set it at around 90\% of $1/\Lambda_{QCD}$. This has been done to keep our calculations well within the perturbative QCD region and to avoid them from including any non-perturbative contributions. Regarding the error estimates for each form factor in table \ref{table:form factor predictions BcDStar}, we observe that they are much larger than those of previous pQCD predictions. This is mainly due to the 30\% uncertainty that we have propagated as systematic error. In addition, we have the parameters $\omega_{B_{c}}$, $C_{D_{s}^{(*)}}$, $\omega_{D_{s}^{(*)}}$, $m_{b}$ and $m_{c}$ that also contribute to the total error. On the contrary, the predictions in ref. \cite{PhysRevD.90.094018} have used model-dependent values of all these parameters, with a 25\% and 10\% uncertainties for $C_{D_{s}^{(*)}}$ and $\omega_{B_{c}}$. A point worth mentioning at this point is that as a result of incorporating these $r_b$, $r_c$, and $r_s$ terms and other discussed corrections, the numerical values of the calculated form factors at $q^2=0$ shifted by a minimum of 2\% for $T_{1}$ and a maximum of 26\% for $V$, compared to the predictions done without $r_b$, $r_c$, and $r_s$. Furthermore, the inclusion of uncertainties from finite $m_{b}$ and $m_{c}$ mass contributions impacted the theoretical error estimates, modifying the relative uncertainties by a minimum of 3.5\% for $A_{0}$ and a maximum of 14.0\% for $V$.
	
\section{Obtaining form factor information over full physical $q^{2}$ range}
	\label{section:Extrapolation to full physical range}

	In the previous section, we had calculated the form factors at $q^{2}=0$. However, when it comes to prediction of physical observables, this is not enough, since, as we have seen in section \ref{section:Theoretical background}, we need form factor information over full physical $q^{2}$ range for doing so. pQCD as a framework, is itself not enough since it is more reliable in the smaller $q^{2}$ region, i.e., near $q^{2}=0$. To overcome this limitation, we derive certain symmetry relations valid at high $q^{2}$ region to connect the $B_{c}\rightarrow D_{s}$ form factors, whose information we have, with $B_{c}\rightarrow D_{s}^{*}$ form factors, whose information we want to obtain. Following this, we adopt certain parametrization method to obtain information of $B_{c}\rightarrow D_{s}^{*}$ form factors over the full $q^{2}$ region. We divide this section into three subsections. In subsection \ref{subsection: connecting the form factors} we derive the symmetry relations between the relevant form factors. In subsection \ref{subsection:extracting the universal functions} we extract necessary parameters, and in subsection \ref{subsection:Bc to Ds* form factors}, we perform the appropriate parametrization and obtain form factor information over the full physical $q^{2}$ region.
	
		\subsection{Connecting $B_{c}\rightarrow D_{s}^{*}$ with $B_{c}\rightarrow D_{s}$ form factors}
		\label{subsection: connecting the form factors}
		
		In this subsection we briefly discuss about the relations connecting the $B_{c}\rightarrow D_{s}$ and $B_{c}\rightarrow D_{s}^{*}$ form factors. We use the Heavy-Quark-Effective-Theory (HQET) trace formalism, retaining both leading and subleading contributions in the heavy-quark limit. This allows us to express all the form factors by just two universal form factors.
		
		Following the trace formalism developed in \cite{Jenkins:1992nb,Colangelo:2021dnv}, the weak  matrix element in HQET for transitions with a generic Dirac bilinear $\Gamma$ can be expressed as
		\begin{equation}
			\langle D_{s}^{(*)}(v,k)|\bar{q}\Gamma Q|B_{c}(v)\rangle=-\sqrt{M m}~\text{Tr}[\bar{H}^{(\bar{c})}\Sigma(v,a_{0}k)\Gamma H^{(c\bar{b})}],
		\end{equation} 
		with the initial state meson and the final state mesons carry four momenta $P_{1}=Mv$ and $P_{2}=mv+k$  respectively, with $k$ being a small residual momentum. The $B_{c}^{+}$ and $B_{c}^{*+}$ doublet comprising of two heavy quarks $\bar{b}$ and $c$ is represented by the effective field
		\begin{equation}
			H^{c\bar{b}}=\frac{1+\slashed{v}}{2}[B_{c}^{*\mu}\gamma_{\mu}-B_{c}\gamma_{5}]\frac{1-\slashed{v}}{2},
		\end{equation}
		and the $D_{s}$ and $D_{s}^{*}$ doublet with a single heavy quark $c$ is represented by the effective fields
		\begin{equation}
			H^{c}=[D_{s}^{*\mu}\gamma_{\mu}-D_{s}\gamma_{5}]\frac{1-\slashed{v}}{2}.
		\end{equation}
		
		The term $\Sigma(v,a_{0}k)$ can be generalized in terms of two dimensionless functions as
		\begin{equation}
			\Sigma(v,a_{0}k)=\Sigma_{1}+\slashed{k}a_{0}\Sigma_{2},
		\end{equation}
		with $\Sigma_{1}$ and $\Sigma_{2}$ being the universal form factors that encode the non-perturbative QCD dynamics of the light degrees of freedom. The function $\Sigma_{1}$ arises at leading order contribution to the HQET expansion of the matrix element and reflects the structure dictated by Heavy Quark Spin Symmetry (HQSS), which becomes exact in the infinite mass limit. The second function $\Sigma_{2}$, however, captures the leading corrections to this symmetry. These corrections are associated with residual momentum $k$ of $D_{s}^{(*)}$ meson, and introduce the $\mathcal{O}(1/m_{Q})$ symmetry breaking effects. $a_{0}$ has the dimension of length and is typically the Bohr radius of the $B_{c}$ meson. The $a_{0}$ factor suppresses $\Sigma_{2}$ compared to $\Sigma_{1}$, thus making the symmetry breaking effects small. The weak matrix element for $B_{c}\rightarrow D_{s}$ transition is then obtained as
		\begin{equation}
			\langle D_{s}(v,k)|\bar{q}\gamma_{\mu}Q|B_{c}(v)\rangle=2\sqrt{Mm}[\Sigma_{1}(w)v_{\mu}+a_{0}\Sigma_{2}(w)k_{\mu}],
			\label{eqn:matrix element HQET}
		\end{equation}
		where the recoil parameter $w$ is related is $q^{2}$ as
		\begin{equation}
			w=\frac{M^{2}+m^{2}-q^{2}}{2Mm}.
		\end{equation}
		Following the same method, the $B_{c}\rightarrow D_{s}$ matrix element induced by the tensor curent can be expressed as
		\begin{equation}
			\langle D_{s}(v)|\bar{q}\sigma_{\mu\nu}Q|B_{c}(v)\rangle=-2i\sqrt{Mm}a_{0}\Sigma_{2}(w)(v_{\mu}k_{\nu}-v_{\nu}k_{\mu}),
			\label{eqn:matrix element HQET FT}
		\end{equation}
		and the $B_{c}\rightarrow D_{s}^{*}$ matrix elements induced by vector, axial-vector and tensor currents can be expressed as
		\begin{equation}
			\begin{split}
				\langle D_{s}^{*}(v,k,\epsilon)|\bar{q}\gamma_{\mu}Q|B_{c}(v)\rangle=2\sqrt{Mm}~\epsilon_{\mu\nu\alpha\beta}\epsilon^{*\nu}v^{\alpha}k^{\beta}a_{0}\Sigma_{2}(w),
			\end{split}
			\label{eqn:matrix element HQET 1}
		\end{equation}
		\begin{equation}
			\begin{split}
				\langle D_{s}^{*}(v,\epsilon)|\bar{q}\gamma_{\mu}\gamma_{5}Q|B_{c}(v)\rangle=2i\sqrt{Mm}&\bigg[\epsilon^{*}_{\mu}\left(\Sigma_{1}(w)+v\cdot k a_{0}\Sigma_{2}(w)\right)\\&-\left(v_{\mu}-\frac{k_{\mu}}{m}\right)\epsilon^{*}\cdot k a_{0}\Sigma_{2}(w)\bigg],
			\end{split}
			\label{eqn:matrix element HQET 2}
		\end{equation}
		\begin{equation}
			\normalsize
			\begin{split}
				\langle D_{s}^{*}(v,\epsilon)|\bar{q}\sigma_{\mu\nu}q^{\nu}Q|B_{c}(v)\rangle=2i\sqrt{Mm}~\epsilon_{\mu\nu\alpha\beta}\epsilon^{*\beta}\left[v^{\alpha}\Sigma_{1}(w)+k^{\alpha}a_{0}\Sigma_{2}(w)\right]\left[(M-m)v^{\nu}-k^{\nu}\right],
			\end{split}
			\label{eqn:matrix element HQET 3}
		\end{equation}
		and
		\begin{equation}
			\begin{split}
				\langle D_{s}^{*}(v,\epsilon)|\bar{q}\sigma_{\mu\nu}q^{\nu}\gamma_{5}Q|B_{c}(v)\rangle=&2\sqrt{Mm}\bigg[ \epsilon^{*}_{\mu}(v_{\nu}\Sigma_{1}(w)+k_{\nu}a_{0}\Sigma_{2}(w))\\&-\epsilon^{*}_{\nu}(v_{\mu}\Sigma_{1}(w)+k_{\mu}a_{0}\Sigma_{2}(w))\bigg]\left[(M-m)v^{\nu}-k^{\nu}\right],
			\end{split}
			\label{eqn:matrix element HQET 4}
		\end{equation}
		respectively. Matching these HQET matrix elements with the ones previously discussed in subsection \ref{subsection:form factors} we can express all the full QCD form factors in terms of $\Sigma_{1}(w)$ and $\Sigma_{2}(w)$. For $B_{c}\rightarrow D_{s}$ form factors, we get the relations as
		\begin{equation}
			\begin{split}
				F_{+}(w)&=\sqrt{\frac{m}{M}}\left(\Sigma_{1}(w)+(M-m)a_{0}\Sigma_{2}(w)\right),\\
				F_{0}(w)&=\frac{2\sqrt{Mm}}{M^{2}-m^{2}}\left[(M-mw)\Sigma_{1}(w)+m(M+m)(w-1)a_{0}\Sigma_{2}(w)\right],\\
				F_{T}(w)&=\sqrt{\frac{m}{M}}(M+m)a_{0}\Sigma_{2}(w),
			\end{split}
			\label{eqn:Full QCD to HQET Bc to Ds form factors}
		\end{equation}
		and for the $B_{c}\rightarrow D_{s}^{*}$ form factors, we get the relations as
		\begin{equation}
			\begin{split}
				V(w)&=\sqrt{\frac{m}{M}}(M+m)a_{0}\Sigma_{2}(w),\\
				A_{0}(w)&=\frac{2M\Sigma_{1}(w)+((M-2m)(2mw-M)+(2m^{2}+M^{2}-2Mmw))a_{0}\Sigma_{2}(w)}{2\sqrt{Mm}},\\[0.5em]
				A_{1}(w)&=\frac{2\sqrt{Mm}}{M+m}\left(\Sigma_{1}(w)+m(w-1)a_{0}\Sigma_{2}(w)\right),\\[0.5em]
				A_{2}(w)&=\frac{\sqrt{Mm}(M-2m)(M+m)a_{0}\Sigma_{2}(w)}{M^{2}},
			\end{split}
			\label{eqn:Full QCD to HQET Bc to D* form factors 1}
		\end{equation}
		for the vector and axial-vector form factors, and
		\begin{equation}
			\begin{split}
				T_{1}(w)&=\sqrt{\frac{m}{M}}\left[\Sigma_{1}(w)+(M-m)a_{0}\Sigma_{2}(w)\right],\\[0.5em]
				T_{2}(w)&=\frac{2\sqrt{Mm}}{M^{2}-m^{2}}\left[\Sigma_{1}(w)\left(M-mw\right)+\left(m(M+m)(w-1)\right)a_{0}\Sigma_{2}(w)\right],\\[0.5em]
				T_{3}(w)&=\sqrt{\frac{m}{M}}\left[-\Sigma_{1}(w)+(M+m)a_{0}\Sigma_{2}(w)\right],
			\end{split}
			\label{eqn:Full QCD to HQET Bc to D* form factors 2}
		\end{equation}
		for the tensor form factors. In all the calculations, since we have not neglected the symmetry breaking corrections, we have not neglected $v\cdot k$, and also have considered the contributions coming from $k_{\mu}/m$. At this point we would like to re-iterate that these relations were presented in our earlier work \cite{Dey:2025xjg}. However, we include a brief discussion here as well to make the present work self-contained and the notations easier to follow.

		\subsection{Extracting the universal functions $\Sigma_{1}$ and $\Sigma_{2}$}
		\label{subsection:extracting the universal functions}

		With expressions for the form factors in terms of soft functions $\Sigma_{1}$ and $\Sigma_{2}$ derived, we move onto extracting the shape of these functions in this subsection. These universal functions are well defined around $q_{max}^{2}$, or near $w=1$, where the initial and final meson states have the same velocities, thus making it possible to expand these functions into a Taylor series around $w=1$, enabling us to express them in a parametric form defined as
			\begin{equation}
			\begin{split}
				\Sigma_{1}(w)&=\Sigma_{1}(1)+\Sigma_{1}^{\prime}(w-1)+\frac{1}{2}\Sigma_{1}^{\prime\prime}(w-1)^{2},\\[1em]
				a_{0}\Sigma_{2}(w)&=a_{0}\Sigma_{2}(1)+a_{0}\Sigma_{2}^{\prime}(w-1)+\frac{1}{2}a_{0}\Sigma_{2}^{\prime\prime}(w-1)^{2},\\
			\end{split}
			\label{eqn:sigma functions}
		\end{equation}
		
		where $\Sigma_{1}(1)$, $\Sigma_{1}^{\prime}$, $\Sigma_{1}^{\prime\prime}$, $a_{0}\Sigma_{2}(1)$, $a_{0}\Sigma_{2}^{\prime}$ and $a_{0}\Sigma^{\prime\prime}_{2}$ are the coefficients that control the shape of these functions. 
        
        To extract these coefficients, we construct a chi-square function with $B_{c}\to D_{s}$ lattice form factors at $w=1.0,~1.15\text{ and }1.3$ as inputs (refer to table \ref{table:input BcDs}) and their corresponding expressions presented in eq.\eqref{eqn:Full QCD to HQET Bc to Ds form factors}, and then minimize it. The thus extracted coefficients, along with the corresponding correlation matrix is presented in table \ref{table:Soft function parameters}.
			\begin{table}[htb!]
	\renewcommand{\arraystretch}{1.5}
	\centering
    \resizebox{\textwidth}{!}{
	\begin{tabular}{|c|c|cccccc|}
		\hline
		\textbf{Parameters}&\textbf{Our}&&&&\textbf{Correlation}&&\\
		\cline{3-8}
		&\textbf{estimates}&$\Sigma_{1}(1)$&$\Sigma_{1}^{\prime}$&$\Sigma_{1}^{\prime\prime}$&$a_{0}\Sigma_{2}(1)~(\text{GeV}^{-1})$&$a_{0}\Sigma_{2}^{\prime}~(\text{GeV}^{-1})$&$a_{0}\Sigma_{2}^{\prime\prime}~(\text{GeV}^{-1})$\\
		\hline
		$\Sigma_{1}(1)$&0.860(12)&1.0&-0.551&0.357&0.104&-0.114&0.106\\
		$\Sigma_{1}^{\prime}$&-3.111(89)&&1.0&-0.883&-0.372&0.298&-0.285\\
		$\Sigma_{1}^{\prime\prime}$&8.645(420)&&&1.0&0.414&-0.450&0.460\\
		$a_{0}\Sigma_{2}(1)~(\text{GeV}^{-1})$&0.444(12)&&&&1.0&-0.861&-0.795\\
		$a_{0}\Sigma_{2}^{\prime}~(\text{GeV}^{-1})$&-1.577(83)&&&&&1.0&-0.985\\
		$a_{0}\Sigma_{2}^{\prime\prime}~(\text{GeV}^{-1})$&4.244(278)&&&&&&1.0\\
		\hline
\multicolumn{8}{|c|}{\textbf{DOF: 3} \hspace{2cm} $\boldsymbol{\chi^{2}/DOF:}$ \textbf{1.916} \hspace{2cm} \textbf{p-Value: 12.46\%}} \\
\hline
	\end{tabular}
    }
	\caption{Extracted estimate of the coefficients of $\Sigma_{1}$ and $a_{0}\Sigma_{2}$.}
	\label{table:Soft function parameters}
\end{table}

\begin{figure}[htb!]
	\centering
	\subfloat[\centering]{\includegraphics[width=7.0cm]{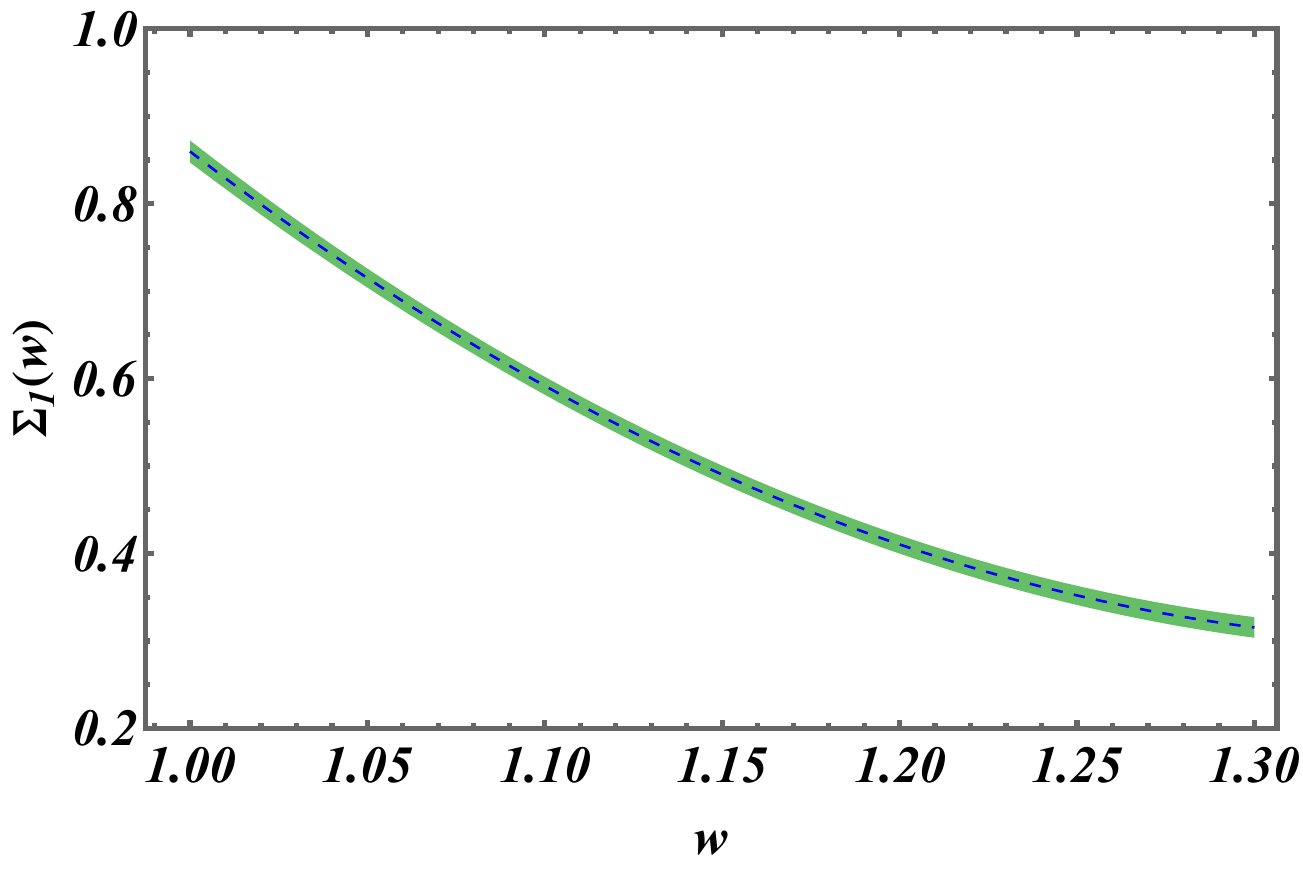}}
	\qquad
	\subfloat[\centering]{\includegraphics[width=7.0cm]{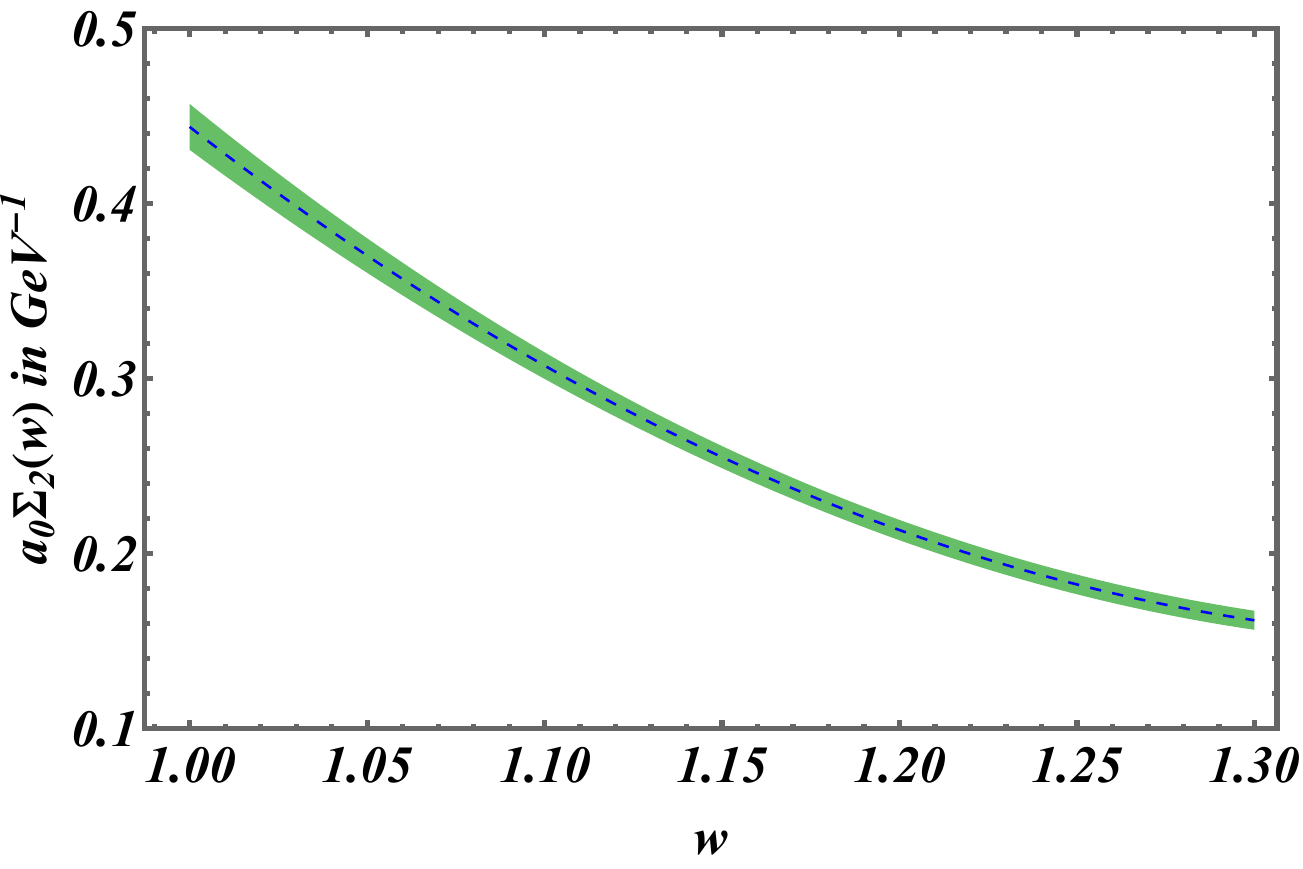}}
	\caption{Plots showing the $w$ distribution of $\Sigma_{1}(w)$ and $a_{0}\Sigma_{2}(w)$.}
	\label{fig:universal functions HQSS}
\end{figure}

From table \ref{table:Soft function parameters} we observe that estimates of $a_{0}\Sigma_{2}$ coefficients are about 50\% suppressed compared to those for $\Sigma_{1}$, signifying that the symmetry-breaking contributions to the matrix elements, although not zero, are significantly suppressed compared to the leading-order contribution. Using the coefficients in table \ref{table:Soft function parameters} as inputs in eq.\eqref{eqn:sigma functions}, we can now obtain the $w$ distribution of both $\Sigma_{1}$ and $a_{0}\Sigma_{2}$. We present the plots of these functions in figure \ref{fig:universal functions HQSS}. The error estimate is calculated by propagating the error estimates of HPQCD form factor inputs. Since this method is more reliable near $w=1$, we confine our plots in the region $w\in (1,1.3)$.
		
\begin{figure}[htb!]
			\centering
			\subfloat[\centering]{\includegraphics[width=7.0cm]{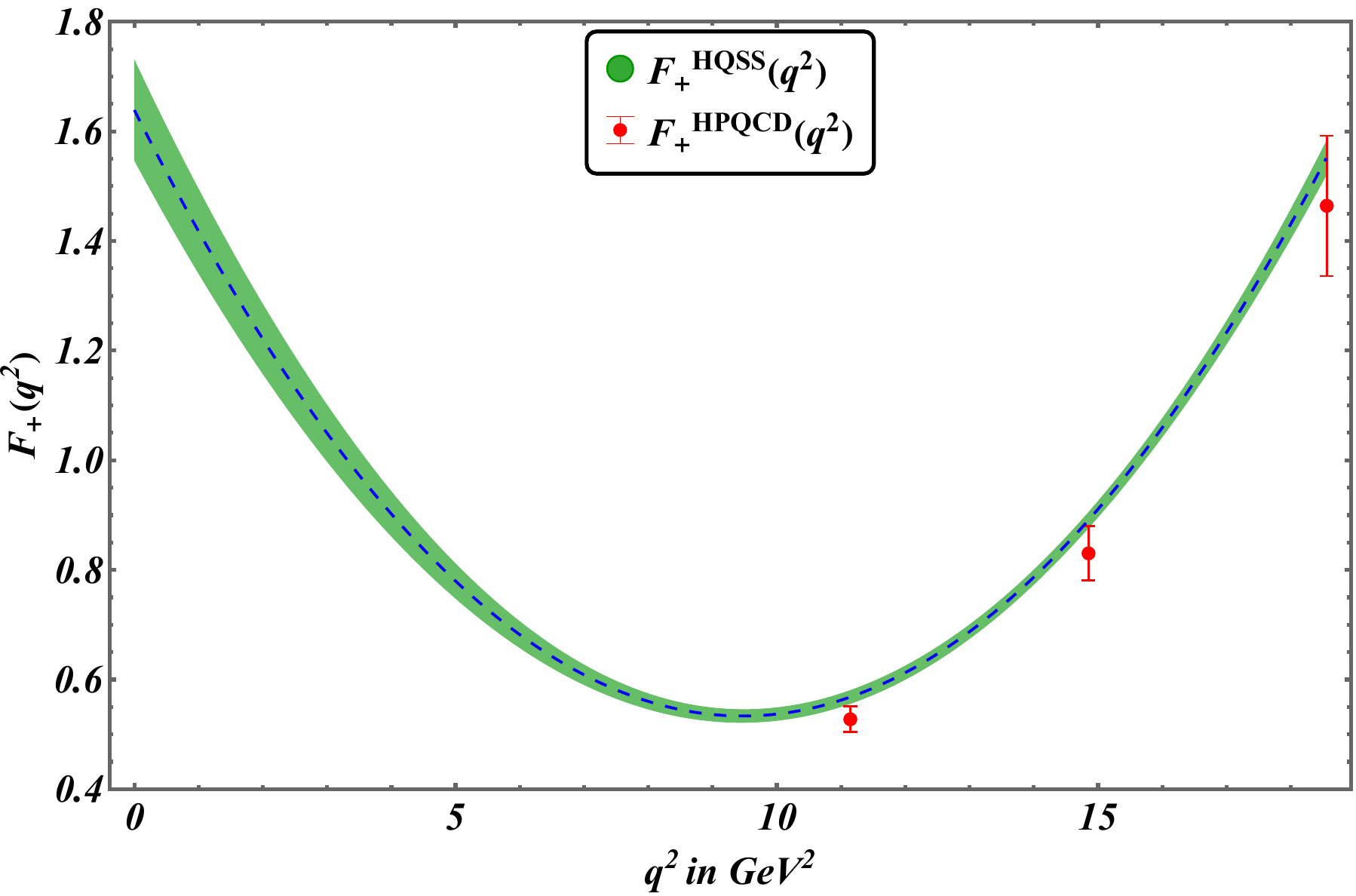}}
			\qquad
			\subfloat[\centering]{\includegraphics[width=7.0cm]{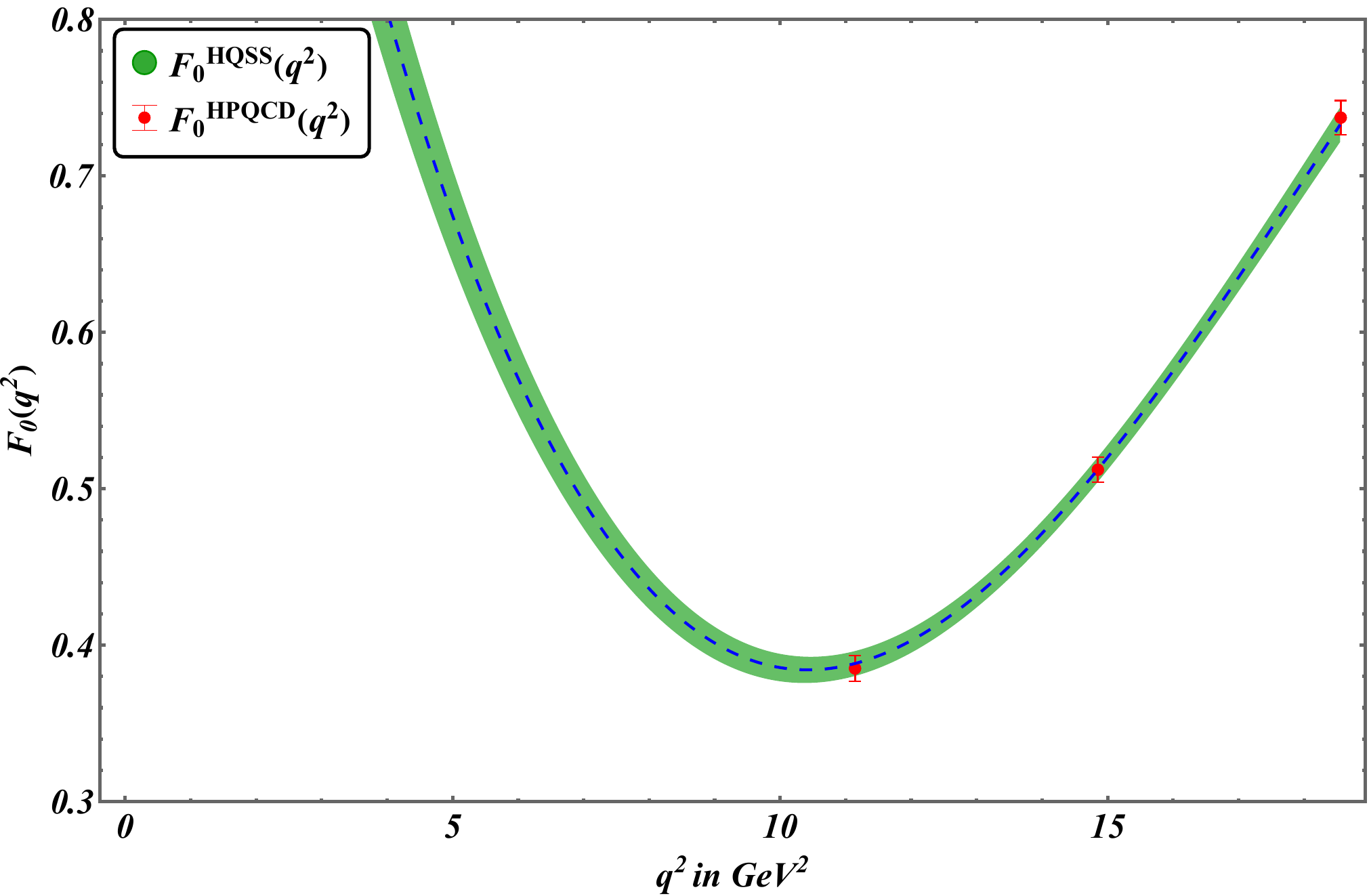}}
			\qquad
			\subfloat[\centering]{\includegraphics[width=7.0cm]{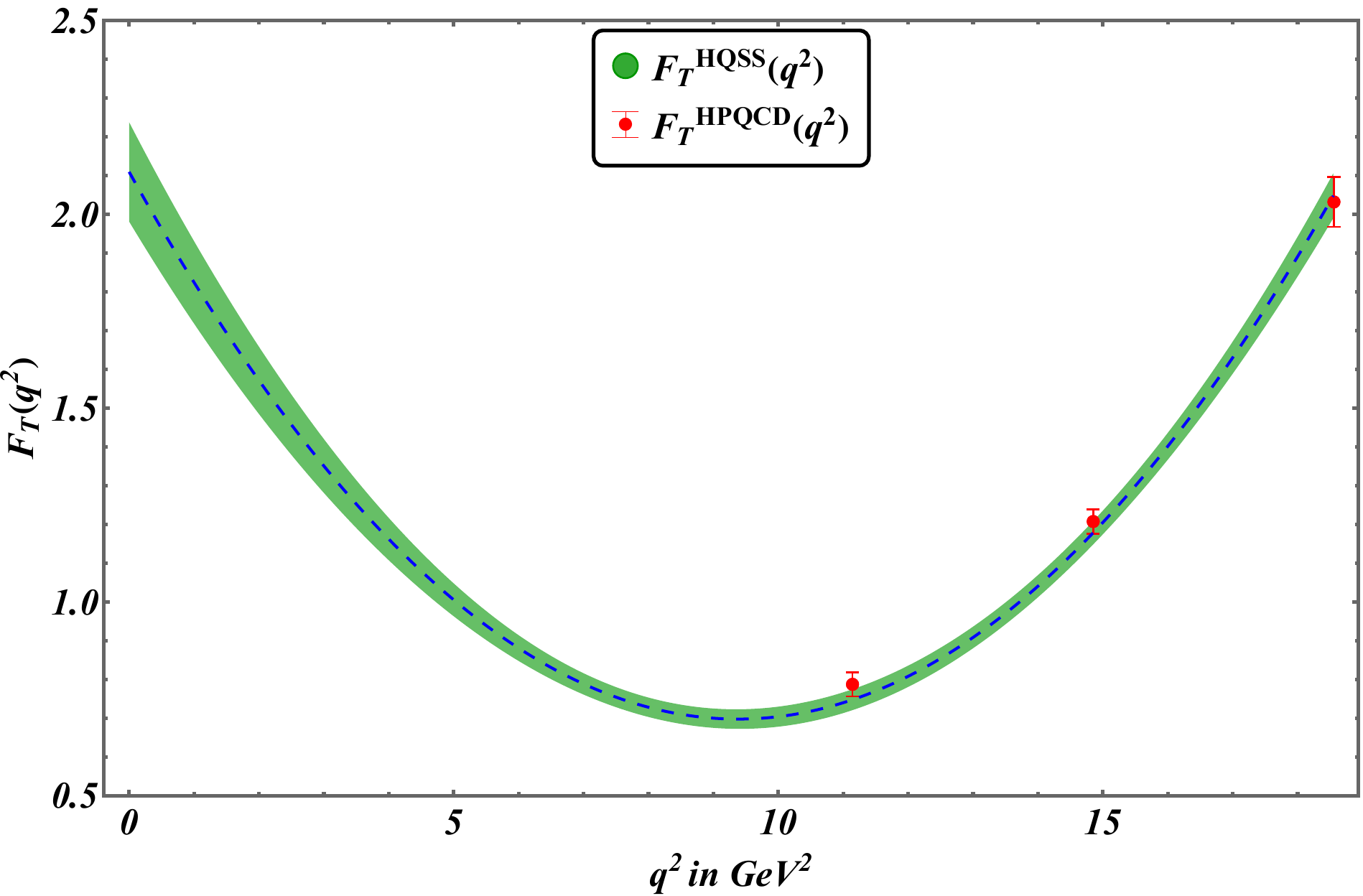}}
			\caption{Plots showing the $q^{2}$ distribution of $B_{c}\rightarrow D_{s}$ form factors $F_{+}$, $F_{0}$ and $F_{T}$. The red markers denote the synthetic data points generated using the BCL parameters supplied by HPQCD.}
			\label{fig:Form factors F+F0}
		\end{figure}

		Additionally, for reference to the readers, in figure \ref{fig:Form factors F+F0} we also present $q^{2}$ distribution of the $B_{c}\rightarrow D_{s}$ form factors $F_{+}$, $F_{0}$ and $F_{T}$, using the extracted universal function parameters as inputs. The uncertainty in each form factor is controlled primarily by that of $F_{0}$ and $F_{T}$, which ranges between 1.5 and 3.0\%. This results in the uncertainty in $F_{+}$ being tightly constrained relative to the input. Furthermore, we observe that the shapes are well-behaved and monotonically decreasing up to $q^{2}\approx 11~\text{GeV}^2$, which corresponds to $w\approx 1.3$. Below this point, the shape of the form factors rises abruptly, thus hinting at the non-reliability of the approach at the low $q^{2}$ region. Hence we restrict the validity of the above results upto $w=1.3$. 
		
		\subsection{Obtaining $q^{2}$ distribution of rest of the $B_{c}\rightarrow D_{s}^{*}$ form factors}
		\label{subsection:Bc to Ds* form factors}
		Once we have information on the soft functions near $w=1$, we can use them in eqs.\eqref{eqn:Full QCD to HQET Bc to D* form factors 1} and \eqref{eqn:Full QCD to HQET Bc to D* form factors 2} to obtain information of all the $B_{c}\rightarrow D_{s}^{*}$ form factors at high $q^{2}$ region. But as we just saw, this approach does not give reliable results for low and mid $q^{2}$ region. To obtain the shape in this region, we adopt a suitable parametrization method. In this work, we employ the Boyd-Grinstein-Lebed (BGL) parametrization \cite{Boyd:1995cf,Boyd:1995sq,Boyd:1997kz} due to its model-independent foundation. The parametrization is done by mapping the physical $q^{2}$ region onto a disk $|z|\le 1$ via conformal transformation $q^{2}\to z$, where $z$ is the new kinematic variable. Each form factor can then be expressed as a convergent power series in the variable $z$, ensuring a well-behaved expansion. This parametrization method also respects analyticity and crossing symmetry, and eliminates unphysical singularities associated with intermediate resonances through an appropriate pole factor.  The form factors in this parametrization, defined through a z-expansion, take the form		
		\begin{equation}
			f_{i}(q^{2})=\frac{1}{P_{i}(q^{2})\phi_{i}(q^{2})}\sum_{n=0}^{N}a_{n}^{i}z(q^{2})^{n},
		\end{equation}
		where $a_{n}^{i}$ are the expansion coefficients that encode the shape of each of the form factors over the kinematic region. These coefficients are intrinsically constrained by the unitarity condition
		\begin{equation}
			\sum_{n=0}^{N}|a_{n}^{i}|^{2}\le 1.
		\end{equation}
		The conformal variable $z(q^{2},t_{0})$ is defined as
		\begin{equation}
			z(q^{2},t_{0})=\frac{\sqrt{t_{+}-q^{2}}-\sqrt{t_{+}-t_{0}}}{\sqrt{t_{+}-q^{2}}+\sqrt{t_{+}-t_{0}}},
		\end{equation}
		with $t_{+}=(m_{B_{0}}+m_{K/K^{*}})^{2}$ and $t_{0}$, an arbitrary reference point is chosen to be \cite{Biswas:2022lhu}
		 \begin{equation}
			t_{0}=t_{opt}=t_{+}\left(1-\sqrt{1-\frac{t_{-}}{t_{+}}}\right),
		\end{equation}
		to center the z-expansion around mid $q^{2}$ region, so as to facilitate a faster convergence.\footnote{Note that in this work we adopt a different choice of $t_{0}$ compared to HPQCD \cite{Cooper:2021bkt}. This choice of $t_{0}$ maps the physical semileptonic region into a symmetric interval in $z$, i.e., $z(q^{2}=0)=z(q^{2}=t_{-})$, so that the physical $q^{2}$ region corresponds to $-z_{max}\leq z\leq z_{max}$. As a result, the maximum $|z|$ is minimized, which means the truncated z-series converges much faster compared to either $t_{0}=0$ or $t_{0}=t_{-}$.}
				
		At this point, an issue needs to be addressed with the current form-factor convention. The issue mainly arises during the calculation of $\phi_{i}(q^{2})$, the outer function, whose form is chosen to be such that the unitarity constraints are satisfied. Checking the formalism described in section 3.3 of \cite{Bharucha:2010im}, the authors write the unitarity inequality as
		\begin{equation}
			\frac{1}{\pi}\int_{t_{+}}^{\infty}\frac{dt}{t-t_{0}}\sqrt{\frac{t_{+}-t_{0}}{t-t_{+}}}|\phi_{i}(t)A_{i}(t)|^{2}\le 1,
		\end{equation}
		where $A_{i}(q^{2})$ represent the helicity-based form factors, and $i$ runs over each helicity basis. Now these form factors have been calculated to be \cite{Bharucha:2010im}
		\begin{equation}
			\begin{split}
				|A_{T}^{V-A}|^{2}=\sum_{i=0}^{2}|\mathcal{B}_{V,i}|^{2}, \qquad |A_{L}^{V-A}|^{2}=3|\mathcal{B}_{V,t}|^{2}, \qquad |A_{T}^{T+AT}|^{2}=q^{2}\sum_{i=0}^{2}|\mathcal{B}_{T,i}|^{2},
			\end{split}
		\end{equation}
		where for vector and axial-vector currents, the form factors are given by \cite{Bharucha:2010im}
		\begin{equation}
			\begin{split}
				\mathcal{B}_{V,0}(q^{2})&=\frac{(M+m)(M^{2}-m^{2}-q^{2})A_{1}(q^{2})-\lambda(q^{2})A_{2}(q^{2})}{2m\sqrt{\lambda(q^{2})}(M+m)},\\
				\mathcal{B}_{V,1}(q^{2})&=\frac{\sqrt{2q^{2}}}{M+m}V(q^{2}),\\
					\mathcal{B}_{V,2}(q^{2})&=\frac{\sqrt{2q^{2}}(M+m)}{\sqrt{\lambda(q^{2})}}A_{1}(q^{2}),\\
					\mathcal{B}_{V,t}(q^{2})&=A_{0}(q^{2}),
			\end{split}
		\end{equation}
		and for tensor currents, the form factors are given by
		\begin{equation}
			\begin{split}
				\mathcal{B}_{T,0}(q^{2})&=\sqrt{q^{2}}\frac{(M^{2}-m^{2})(M^{2}+3m^{2}-q^{2})T_{2}(q^{2})-\lambda(q^{2})T_{3}(q^{2})}{2m\sqrt{\lambda(q^{2})}(M^{2}-m^{2})},\\
				\mathcal{B}_{T,1}(q^{2})&=\sqrt{2}T_{1}(q^{2}),\\
				\mathcal{B}_{T,2}(q^{2})&=\frac{\sqrt{2}(M^{2}-m^{2})}{\sqrt{\lambda(q^{2})}}T_{2}(q^{2}).
			\end{split}
		\end{equation}
		Now, if we continue our analysis in the current form factor convention, $|A_{T}^{V-A}|^{2}$ would give terms like $\mathrm{Re}(A_{1}^{*}A_{2})$ and $\mathrm{Re}(A_{1}A_{2}^{*})$ along with the usual $|A_{1}|^{2}$ and $|A_{2}|^{2}$ terms. Similarly $|A_{T}^{T+AT}|^{2}$ would give terms like $\mathrm{Re}(T_{2}^{*}T_{3})$ and $\mathrm{Re}(T_{2}T_{3}^{*})$ along with the usual $|T_{2}|^{2}$ and $|T_{3}|^{2}$ terms. These cross terms lead to a non-diagonal structure in the unitarity relations. Because of this mixing, it would become impossible to define unique outer functions $\phi_i(q^{2})$ for each form factor individually, as the unitarity bound no longer factorizes into independent form factors. To resolve this issue the authors in \cite{Bharucha:2010im,Gubernari:2023puw} suggested shifting the form factors to the helicity basis, where the form factors $A_{2}$ and $T_{3}$ get replaced by $A_{12}$ and $T_{23}$, defined as
		\begin{equation}
			\begin{split}
				A_{12}&=\frac{(M+m)^{2}(M^{2}-m^{2}-q^{2})A_{1}-\lambda(q^{2})A_{2}}{16Mm^{2}(M+m)},\\
				T_{23}&=\frac{(M^{2}-m^{2})(M^{2}+3m^{2}-q^{2})T_{2}-\lambda(q^{2})T_{3}}{8Mm^{2}(M-m)}.
			\end{split}
			\label{eqn:form factors in helicity basis}
		\end{equation}
		This choice of form factor basis eliminates all the cross terms and thus diagonalizes the unitarity relations. Also, we can now separately extract the outer functions for each form factor. Henceforth, we will be performing all our calculations using helicity basis form factors. The form of the outer function has been obtained by \cite{Gubernari:2023puw} as
		\begin{equation}
			\phi_{i}(q^{2})=\sqrt{\frac{\mathcal{N_{F}\eta^{\text{b}\to \text{s}}}}{32 \pi^{2}\chi_{i}}}\left(\frac{\lambda(q^{2})}{-z(q^{2},t_{-})}\right)^{\frac{m}{4}}\left(\frac{-z(q^{2},0)}{q^{2}}\right)^{\frac{n+p+1}{2}}\sqrt{\frac{4(1+z(q^{2},t_{0}))(t_{+}-t_{0})}{(z(q^{2},t_{0})-1)^{3}}}.
			\label{eqn:outer function}
		\end{equation}
		Construction of the outer function $\phi_{i}(q^{2})$ originates from the short-distance two-point correlator of the quark current and is therefore independent of the spectator quark. Consequently, we can use the same functional form and the relevant parameters universally for all transitions mediated by the same quark current. This allows us to adopt the same analytical form of $\phi_{i}(q^{2})$ and the corresponding $\chi_{i}$ functions derived for $b\to s$ transitions in \cite{Bharucha:2010im,Gubernari:2023puw}, while replacing relevant kinematic quantities. The parameters ${\mathcal{N_{F}},p,n,m}$ are form factor specific and are listed in table \ref{table:parameters of outer function}.
				\begin{table}[htb!]
			\centering
            \renewcommand{\arraystretch}{1.2}
			\begin{tabular}{|c|cccc|}
				\hline
				Form factor&$\mathcal{N_{F}}$&p&n&m\\
				\hline
				$A_{0}$&1&2&1&3\\
				$A_{1}$&$2t_{+}$&1&2&1\\
				$A_{12}$&$64M^{2}m^{2}$&2&2&1\\
				$V$&$2/t_{+}$&1&2&3\\
				$T_{1}$&2&1&3&3\\
				$T_{2}$&$2t_{+}t_{-}$&1&3&1\\
				$T_{23}$&$16M^{2}m^{2}/t_{+}$&0&3&1\\
				\hline
			\end{tabular}
			\caption{Parameters of the outer function for $B_{c}\rightarrow D_{s}^{*}$ form factors.}
			\label{table:parameters of outer function}
		\end{table}
		
		Additionally, the outer function depends on $\chi_{i}$ coefficient, which is a perturbatively calculable short-distance coefficient that comes from the derivative of two point current-current correlator:
		
		\begin{equation}
			\chi_{i}=\frac{1}{2}\frac{d^{n}}{dq^{2n}}\Pi_{i}(q^{2})|_{q^{2}=0},
		\end{equation}
		where $\Pi_{i}(q^{2})$ is the two-function correlation function. The values of $\chi_{i}$ used in this work are shown in table \ref{table:dispersion functions}, and have been taken from \cite{Bharucha:2010im}
		\begin{table}[htb!]
			\centering
            \renewcommand{\arraystretch}{1.3}
			\begin{tabular}{|c|ccccccc|}
				\hline
				Form Factor&  $A_{0}$ & $A_{1}$ & $A_{12}$ & $V$&$T_{1}$&$T_{2}$&$T_{23}$\\
				\hline
				$\chi_{i}\times 10^{2}$&1.57&$1.13/m_{b}^{2}$&$1.13/m_{b}^{2}$&$1.20/m_{b}^{2}$&$0.803/m_{b}^{2}$&$0.748/m_{b}^{2}$&$0.748/m_{b}^{2}$\\
				\hline
			\end{tabular}	
			\caption{List of the functions $\chi_{i}$ for each form factor considered in this work.}
			\label{table:dispersion functions}
			\end{table}
			
		 Finally, the function $P_{i}(q^{2})$ represents the Blaschke factor and prevents divergence in the BGL expansion by factoring out known resonance poles from the form factors, thus ensuring that the remaining piece is analytic and can be safely expanded in a rapidly converging power series in $z(q^{2},t_{0})$. It has the form
		 \begin{equation}
		 	P_{i}(q^{2})=\prod_{res}\frac{z(q^{2},t_{0})-z(M_{i,res}^{2},t_{0})}{1-z(q^{2},t_{0})z(M_{i,res}^{2},t_{0})},
		 \end{equation}
		 where the product runs over each resonance state. $M_{i,res}$ represents the $B_{s}^{*}$ resonance masses, values of which used in this work are shown in table \ref{table:resonance masses}, and has been taken from \cite{Bharucha:2015bzk}. Since we are only considering the lowest-lying resonance state for each form factor, the Blaschke factor reduces to a single pole form. 
				
		\begin{table}[htb!]
			\centering
            \renewcommand{\arraystretch}{1.1}
			\begin{tabular}{|c|ccccccc|}
				\hline
				Form Factor&  $A_{0}$ & $A_{1}$ & $A_{12}$ & $V$&$T_{1}$&$T_{2}$&$T_{23}$\\
				\hline
				$M_{i,\text{res}}$ in \text{GeV}&5.366&5.829&5.829&5.415&5.415&5.829&5.829\\
				\hline
			\end{tabular}	
			\captionof{table}{Masses of the low lying $B_{s}^{*}$ resonances.}
			\label{table:resonance masses}
		\end{table}
		
		With the z-expansion expressions of the form factors and the corresponding form factor inputs obtained using eqs.\eqref{eqn:Full QCD to HQET Bc to D* form factors 1} and \eqref{eqn:Full QCD to HQET Bc to D* form factors 2} and presented in tables \ref{table: inputs BcDStar A012V} and \ref{table: inputs BcDStar T123}, we can construct a chi-square function with the BGL coefficients as free parameters. But these inputs would constrain the form factor shape only at high $q^{2}$ region. To constrain the shapes at low $q^{2}$ region, we take pQCD form factor values at $q^{2}=0$, previously presented in table \ref{table:form factor predictions BcDStar}, as additional inputs into the chi-square function.\footnote{It is to be noted that the form factors calculated using pQCD and universal functions were in the previous basis, $A_{0,1,2}$, $V$ and $T_{1,2,3}$. Before incorporating these inputs into the chi-square function, we have transformed them into the helicity basis by computing the corresponding $A_{12}$ and $T_{23}$ form factors using eq.\eqref{eqn:form factors in helicity basis}} Once the total chi-square function is constructed, we then minimize it to extract the required BGL coefficients, estimates of which are given in table \ref{table:BGL Bc to Ds*} along with the corresponding correlation matrix in table \ref{table:correlation BGL parameters BcDsStar}.
		\begin{table}[htb!]
	\centering
    \renewcommand{\arraystretch}{1.5}
	\begin{tabular}{|c|ccc|}
		\hline
		\textbf{Form factors}&$\boldsymbol{a_{0}}$&$\boldsymbol{a_{1}}$&$\boldsymbol{a_{2}}$\\
		\hline
		$A_{0}$&0.0841(16)&-0.406(28)&0.688(350)\\
		$A_{1}$&0.0348(8)&-0.117(16)&0.704(163)\\
		$A_{12}$&-&-0.032(6)&0.279(65)\\
		$V$&0.0957(31)&-0.328(46)&0.288(282)\\
		$T_{1}$&0.0455(9)&-0.197(13)&0.356(147)\\
		$T_{2}$&-&-0.0638(45)&0.116(59)\\
		$T_{23}$&0.0452(15)&-0.350(22)&0.881(262)\\
		\hline
        \multicolumn{4}{|c|}{\textbf{DOF: 7} \qquad $\boldsymbol{\chi^{2}_{\text{min}}/}\textbf{DOF:}$ \textbf{5.074} } \\
        \hline
	\end{tabular}
	\caption{Extracted BGL coefficients for $B_{c}\rightarrow D_{s}^{*}$ form factors.}
	\label{table:BGL Bc to Ds*}
\end{table}
		\begin{figure}[t!]
	\centering
	\subfloat[\centering]{\includegraphics[width=7.0cm]{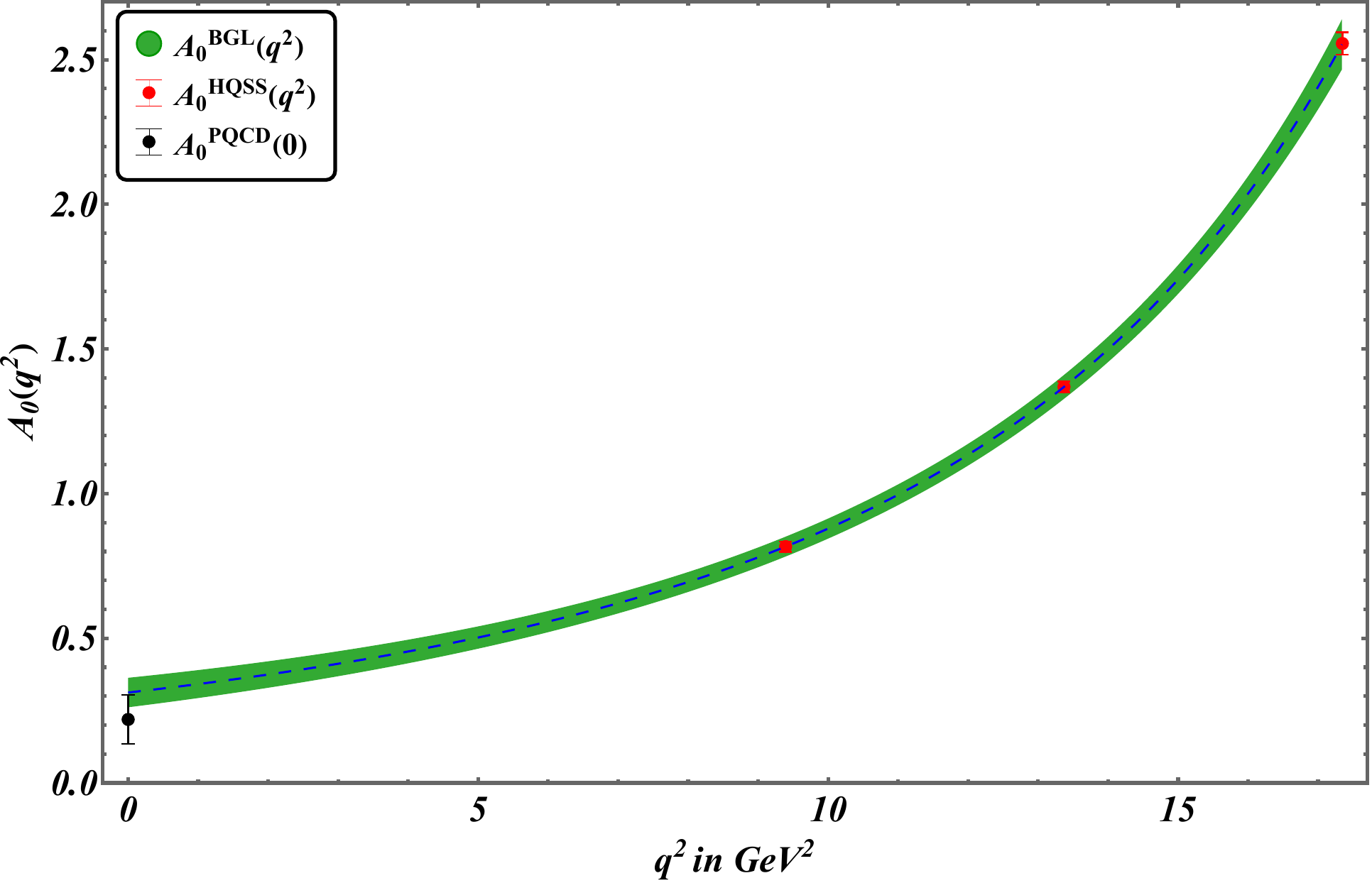}}
	\qquad
	\subfloat[\centering]{\includegraphics[width=7.0cm]{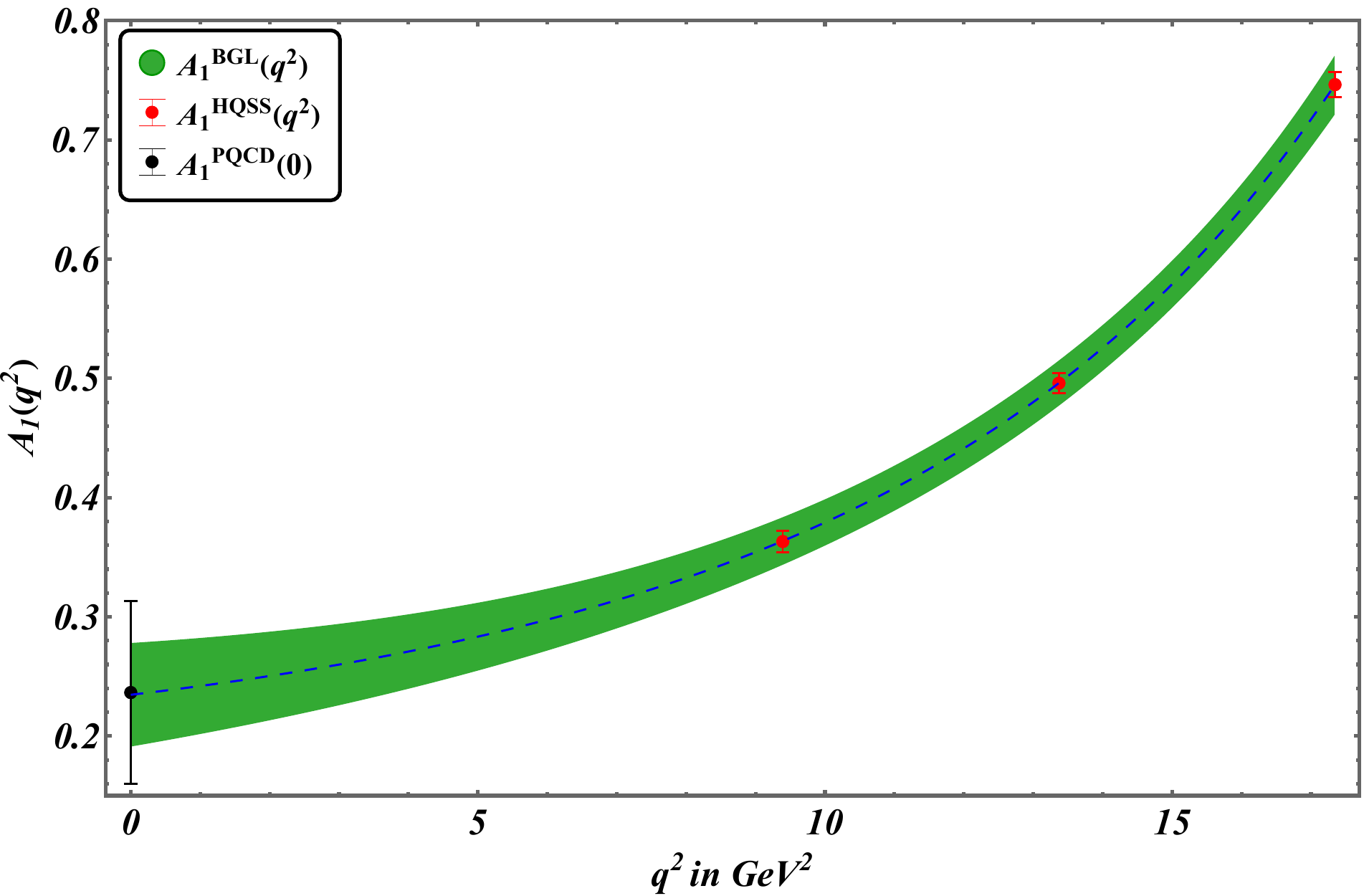}}
	\qquad
	\subfloat[\centering]{\includegraphics[width=7.0cm]{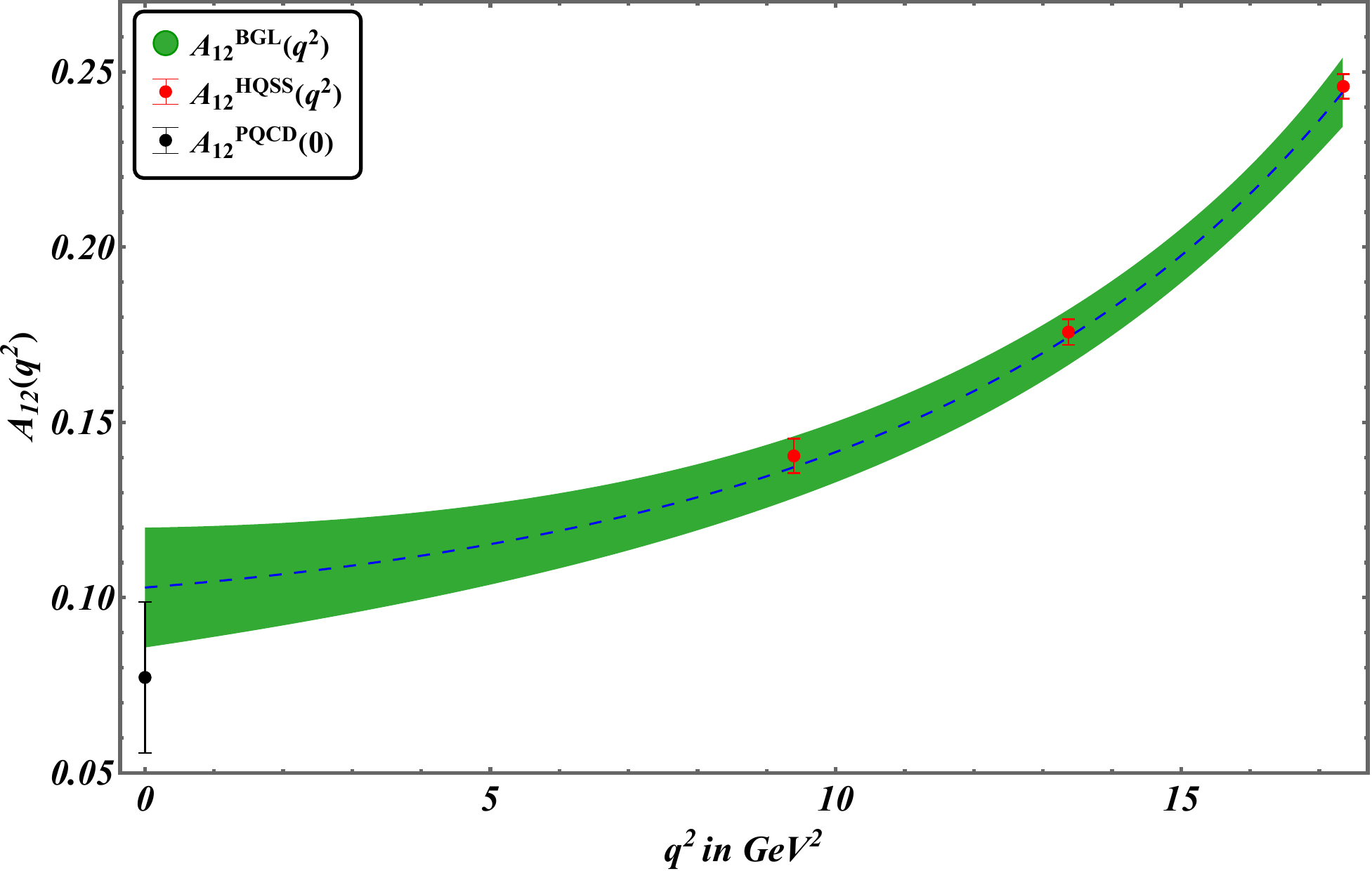}}
	\qquad
	\subfloat[\centering]{\includegraphics[width=7.0cm]{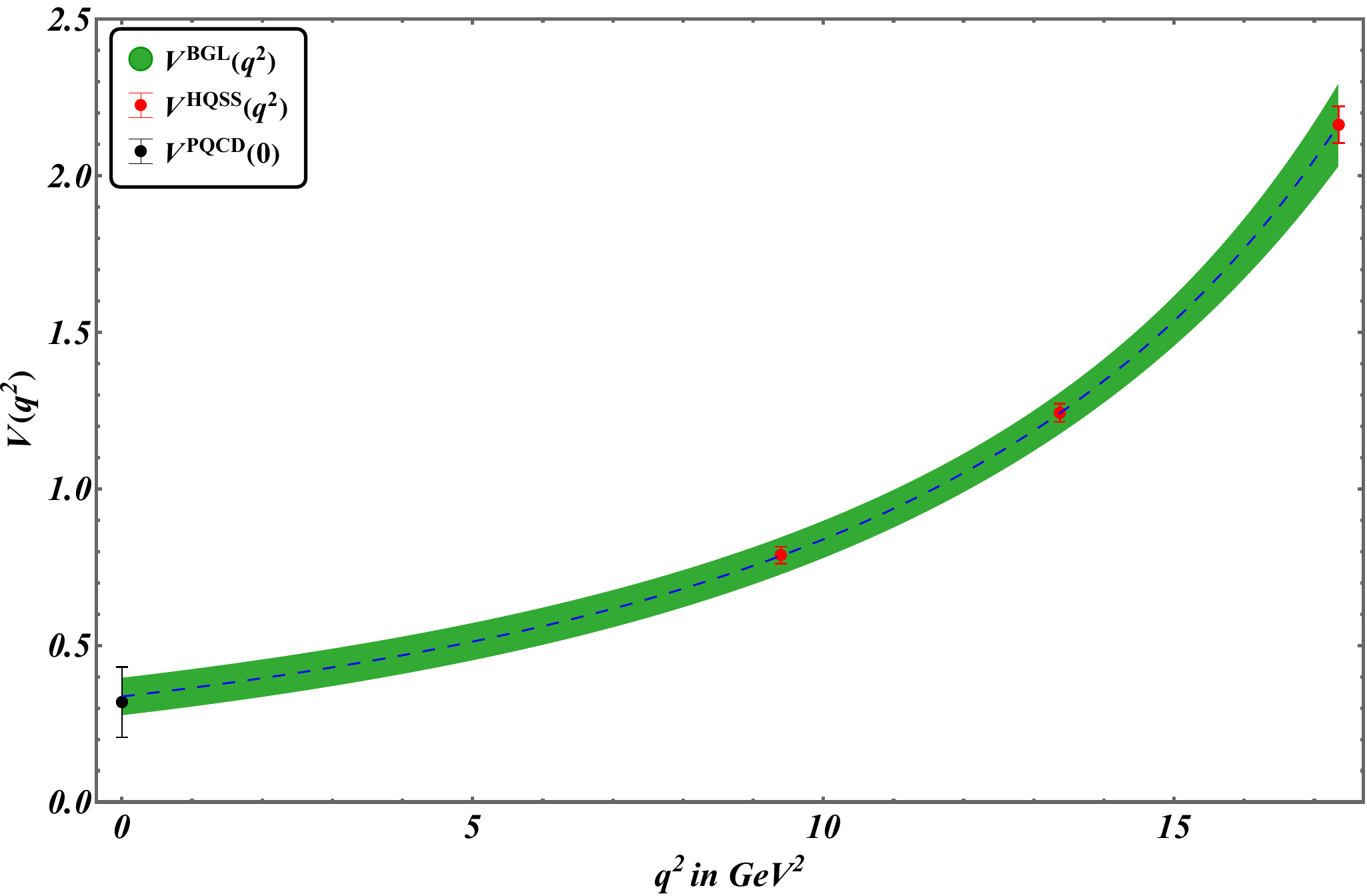}}
	\caption{$q^{2}$ distribution of $B_{c}\rightarrow D_{s}^{*}$ vector and axial-vector form factors. The red and black markers denote the form factor values obtained using the extracted soft functions and pQCD, respectively.}
	\label{fig:form factors Bc to Ds* 1}
\end{figure}

		A point to be noted from table \ref{table:BGL Bc to Ds*} is that we have not included the coefficients $a_{0}$ for $A_{2}$ and $T_{2}$. This omission is done to respect the QCD constraints, previously discussed in eq.\eqref{eqn:form factor QCD constraints}, and satisfied by the form factors at $q^{2}=0$, making $a_{0}$ for $A_{2}$ a function of BGL coefficients of $A_{0}$ and $A_{1}$, and $a_{0}$ for $T_{2}$ a function of BGL coefficients of $T_{1}$. Hence, treating $a_0$ for $A_2$ and $T_2$ as free parameters would violate these known QCD relations and induce redundancy during the chi-square minimization. 
        
        We also observe that the chi-square minimization yields a very poor goodness of fit estimate, i.e., $\chi^{2}_{\text{min}}/\text{DOF}>1$. This is primarily due to the very precise error estimates of the form factor inputs, ranging between 1.5-6.5\%. Further, the HQSS-derived form factor values at large $q^2$ dominantly control the overall slope of the BGL expansion, pulling the extrapolated result at $q^2=0$ slightly away from the pQCD input values. To account for this, Birge's rescaling, a method of rescaling the error estimate implemented by FLAG in \cite{FlavourLatticeAveragingGroupFLAG:2024oxs}, is adopted. The error estimates of the extracted coefficients are rescaled by a factor of $\chi^{2}_{\text{min}}/\text{DOF}=\sqrt{35.517/7}=2.252$. The correlation matrix, however, remains unaffected by this rescaling.
        
\begin{figure}[t!]
			\centering
			\subfloat[\centering]{\includegraphics[width=7.0cm]{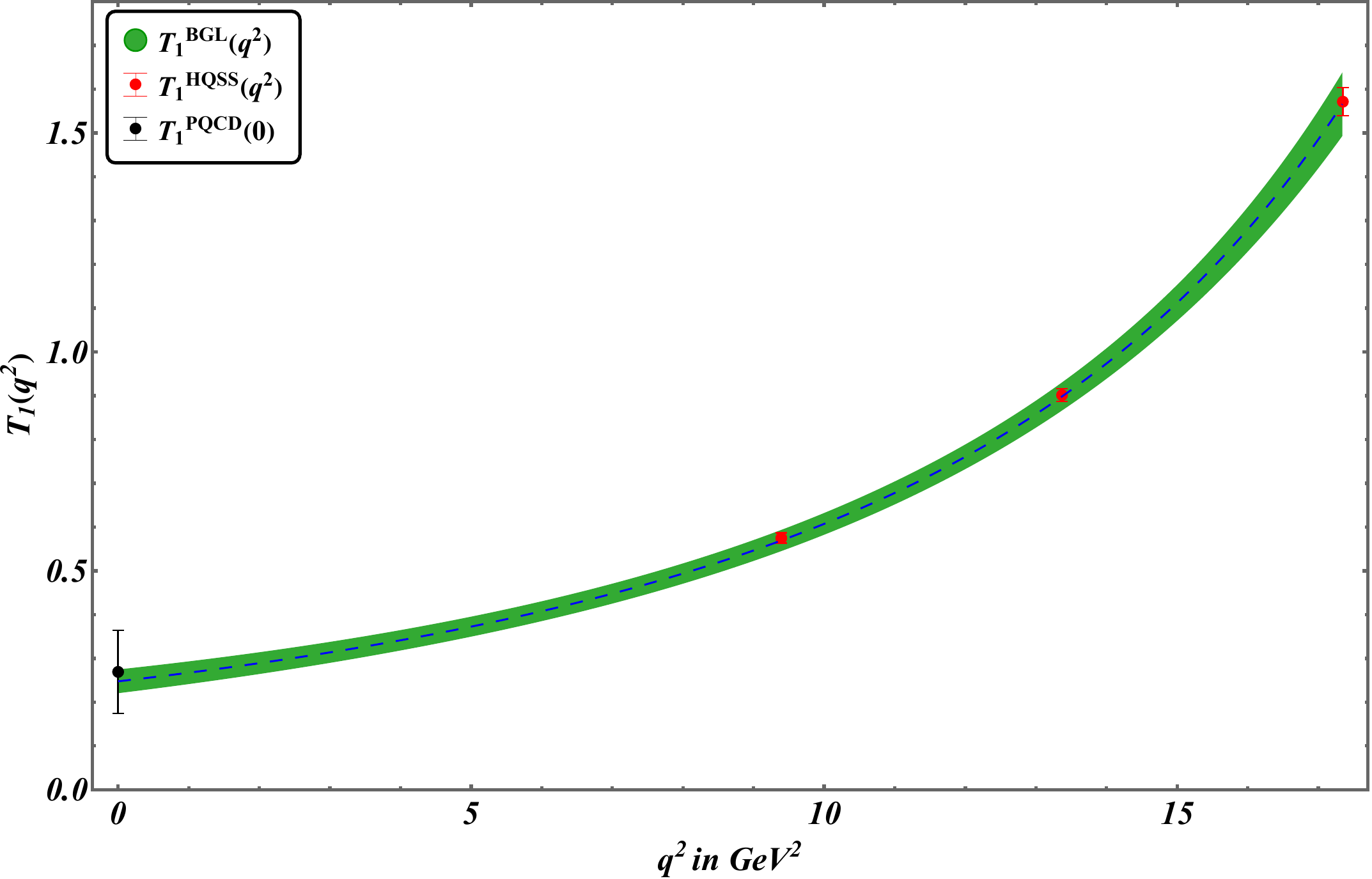}}
			\qquad
			\subfloat[\centering]{\includegraphics[width=7.0cm]{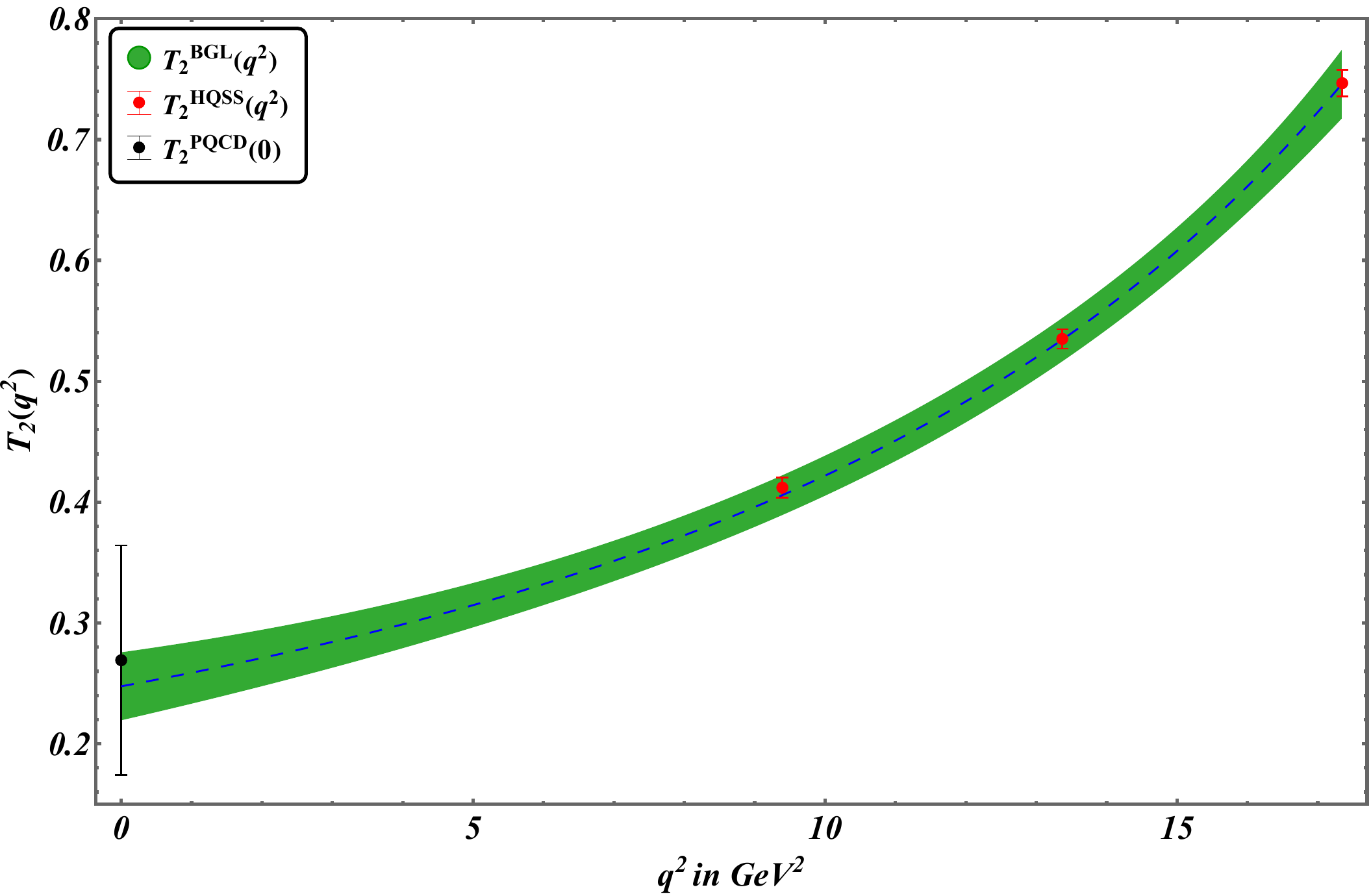}}
			\qquad
			\subfloat[\centering]{\includegraphics[width=7.0cm]{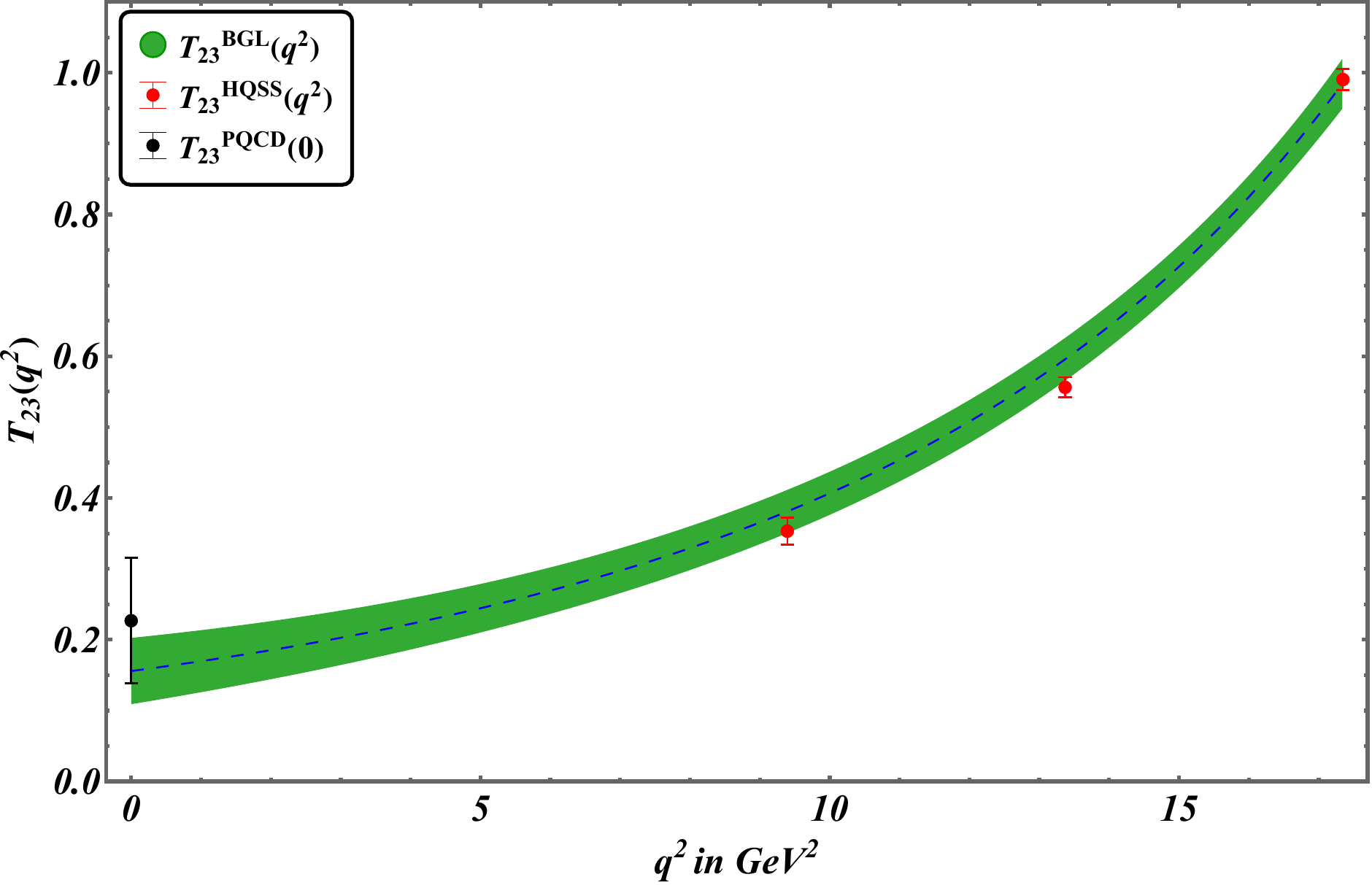}}
			\caption{$q^{2}$ distribution of $B_{c}\rightarrow D_{s}^{*}$ tensor form factors. The red and black markers denote the form factor values obtained using the extracted soft functions and pQCD, respectively.}
			\label{fig:form factors Bc to Ds* 2}
		\end{figure}

		Furthermore, we observe that error estimates of the parameters for each form factor increases as we go up in order of z-expansion series, thus signifying a reducing sensitivity of the form factor slopes with increasing order of the parameters. Further we also observe that the extracted parameters follow the hierarchy $a_{2}z(q^{2})^{2}<a_{1}z(q^{2})<a_{0}$, implying that our z-expansion is converging in nature. Additionally, since we have truncated the BGL series up to quadratic order, i.e., up to $z^{2}$, we introduce an additional error to take into account the error associated with missing higher-order terms. In particular, we estimate the possible effect of the next higher-order term in the expansion, i.e., $z^{3}$, neglecting the relatively small higher-order terms after that, as prescribed in \cite{Bourrely:2008za}. We then propagate this additional error as a systematic uncertainty into the form factors. The estimate of this systematic error is defined as follows:
			\begin{equation}
				\delta f_{i}(q^{2})=\frac{a_{3}^{max}|z(q^{2})^{3}|}{P(q^{2})},
				\label{eqn:truncation}
			\end{equation}
			where $a_{3}^{max}$ represents the maximum possible value of $|a_{3}|$ for each individual form factor and is estimated by using the unitarity constraints, where the coefficients $a_{0,1,2}$ are already extracted in table \ref{table:BGL Bc to Ds*}.
		With all the BGL coefficients extracted and truncation uncertainty included, we have all the information necessary to get the $q^{2}$ distribution of the form factors over the full semileptonic region. In figures \ref{fig:form factors Bc to Ds* 1} and \ref{fig:form factors Bc to Ds* 2} we showcase these $q^{2}$ distribution plots.
		
	\section{Prediction of physical observables}
	\label{section:Prediction of physical observables}
	
		With the $B_{c}\rightarrow D_{s}^{*}$ form factors defined and their information over the full physical $q^{2}$ region available to us, we are now in a position to predict some of the important physical observables. In subsection \ref{subsection:branchng ratios} we present our prediction of branching fractions of some of the relevant decay channels. In subsection \ref{subsection:angular observables} we perform a full angular analysis of FCNC process and present predictions of relevant observables. $\tau_{B_{c}}$ required to calculate the branching ratios is taken from table \ref{table:input parameters}.
	
		\subsection{Decay width and branching fractions}
		\label{subsection:branchng ratios}
		For the $B_{c}^{-}\rightarrow D_{s}^{*-}\ell^{+}\ell^{-}$ and $B_{c}^{-}\rightarrow D_{s}^{*-}\nu\bar{\nu}$ FCNC processes governed by $b\rightarrow s$ quark-level transition, the differential decay width is taken from subsection~\ref{subsubsection:decay width theory}. It is then integrated over the full physical $q^{2}$ region, i.e., $q^{2}\in (4m_{\ell}^{2},(M-m)^{2})$, and multiplied  by $\tau_{B_{c}}$ to obtain the branching fractions of the respective mode. We present our predictions for these branching fractions, along with a comparison with previous pQCD predictions, in table~\ref{table:predictions branching fractions rare BcDsStar}. These channels are more interesting since these are mediated exclusively by neutral current processes, and hence have a highly suppressed branching ratio, making them highly sensitive to possible physics beyond the SM. 
			\begin{table}[htb!]
			\renewcommand{\arraystretch}{1.3}
			\centering
			\begin{tabular}{|c|cc|}
				\hline
				\textbf{Decay Modes}&\textbf{This work}&\textbf{Previous pQCD}\cite{PhysRevD.90.094018}\\
				\hline
				$\mathcal{B}(B_{c}^{-}\rightarrow D_{s}^{*-}\ell^{+}\ell^{-})$&4.582(619)$\times 10^{-7}$&4.40$^{+2.11}_{-1.76}\times 10^{-7}$\\
				$\mathcal{B}(B_{c}^{+}\rightarrow D_{s}^{*+}\ell^{+}\ell^{-})$&4.576(618)$\times 10^{-7}$&-\\
				$\mathcal{B}(B_{c}^{-}\rightarrow D_{s}^{*-}\tau^{+}\tau^{-})$&0.637(106)$\times 10^{-7}$&0.52$^{+0.26}_{-0.22}\times 10^{-7}$\\
				$\mathcal{B}(B_{c}^{+}\rightarrow D_{s}^{*+}\tau^{+}\tau^{-})$&0.636(106)$\times 10^{-7}$&-\\
				$\mathcal{B}(B_{c}^{-}\rightarrow D_{s}^{*-}\nu\bar{\nu})$&3.506(260)$\times 10^{-6}$&4.04$^{+1.96}_{-1.62}\times 10^{-6}$\\
				\hline
			\end{tabular}
			\caption{Predictions for the branching fractions of the $B_{c}\rightarrow D_{s}^{*}$ FCNC transitions with $(l=e,\mu)$ along-with comparison with predictions in existing literature.}
			\label{table:predictions branching fractions rare BcDsStar}
		\end{table}
        
		In the results of table \ref{table:predictions branching fractions rare BcDsStar}, the primary contribution to the total uncertainty originates from the form factors, resulting in an overall uncertainty of about 13\% for the fgirst four channels and 7\% for the last channel considered. Our results show a significant improvement in the percentage error compared to previous pQCD predictions. Additionally, we can define an LFUV observable by taking the ratio of branching fractions of the $\tau$ and $\ell$ modes. The observable is defined as
		\begin{equation}
            R^{B_{c}^{-}\to D_{s}^{*-}}=\frac{\mathcal{B}(B_{c}^{-}\rightarrow D_{s}^{*-}\tau^{+}\tau^{-})}{\mathcal{B}(B_{c}^{-}\rightarrow D_{s}^{*-}\ell^{+}\ell^{-})}=0.138(5),
        \end{equation}
		where as expected, we observe the error percentage to be quite small. This can be tested once experimental measurements start to come up in future.

        Furthermore, we also observe a difference in the predictions of $\mathcal{B}(B_{c}^{-}\to D_{s}^{*-}\ell^{+}\ell^{-})$ channels and their respective CP-conjugate modes. This difference hints towards a non-zero value of CP asymmetry, thus making prediction of the CP asymmetry observable significant. The observable is defined as
\begin{equation}
	\mathcal{A}_{CP}(q^{2})=\frac{d\mathcal{B}(B_{c}^{-}\rightarrow D_{s}^{*-}\ell^{+}\ell^{-})/dq^{2}-d\mathcal{B}(B_{c}^{+}\rightarrow D_{s}^{*+}\ell^{+}\ell^{-})/dq^{2}}{d\mathcal{B}(B_{c}^{-}\rightarrow D_{s}^{*-}\ell^{+}\ell^{-})/dq^{2}+d\mathcal{B}(B_{c}^{+}\rightarrow D_{s}^{*+}\ell^{+}\ell^{-})/dq^{2}},
\end{equation}
and the $q^{2}$ averaged values of this observable, integrated over the full physical $q^{2}$ region, are presented in table \ref{table:CP asymmetry Bc to Ds*}.
		\begin{table}[htb!]
			\centering
            \renewcommand{\arraystretch}{1.5}
			\begin{tabular}{|c|c|}
				\hline
				\textbf{Decay Mode}&\textbf{Our prediction}\\
				\hline
				$\langle\mathcal{A}_{CP}\rangle(B_{c}\to D_{s}^{*}\ell^{+}\ell^{-})$&6.389(128)$\times 10^{-4}$\\
				$\langle\mathcal{A}_{CP}\rangle(B_{c}\to D_{s}^{*}\tau^{+}\tau^{-})$&3.930(82)$\times 10^{-4}$\\
				\hline
			\end{tabular}
            \caption{Predictions for CP asymmetry of the respective decay channels.}
	\label{table:CP asymmetry Bc to Ds*}
		\end{table}

        From table \ref{table:CP asymmetry Bc to Ds*} we can see that the CP asymmetry, although small, but remains non-zero. This non-vanishing value stems primarily from the phase terms carried by the CKM elements $V_{ub}$ within the effective coefficient $C_{9}^{\text{eff}}$. A comparison of the second row highlights the suppression in the $\tau$-lepton case arising due to heavy lepton mass effects and reduced phase space. 
        
		\subsection{Angular observables}
		\label{subsection:angular observables}
				\begin{figure}[t!]
			\centering
			\subfloat[\centering]{\includegraphics[width=4.5cm]{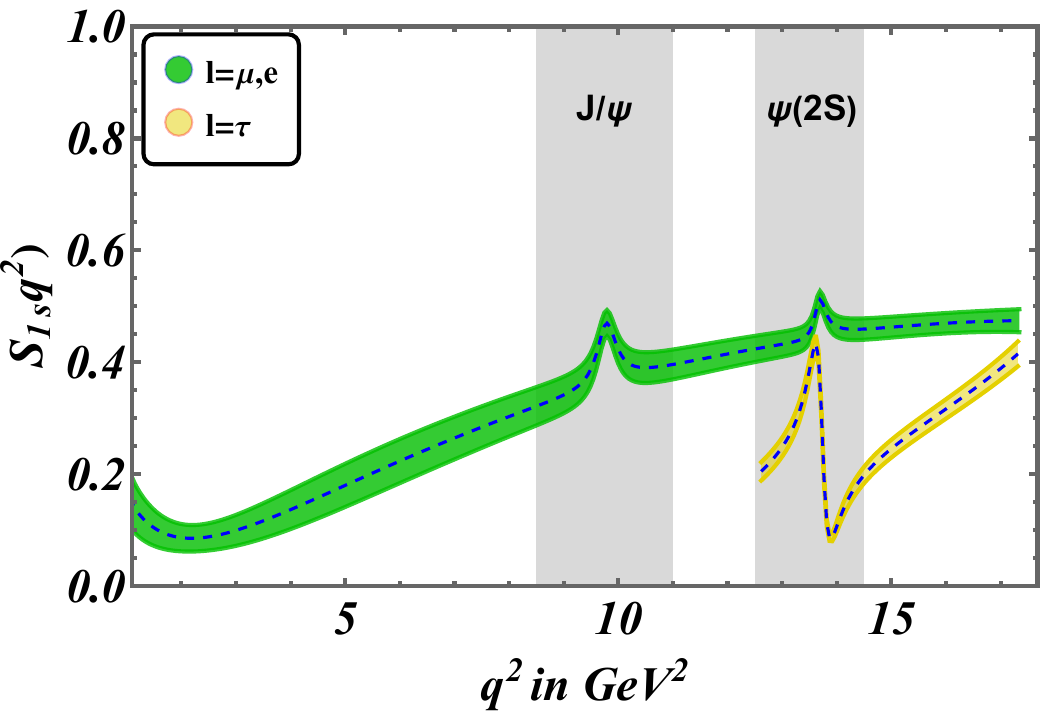}}
			\qquad
			\subfloat[\centering]{\includegraphics[width=4.5cm]{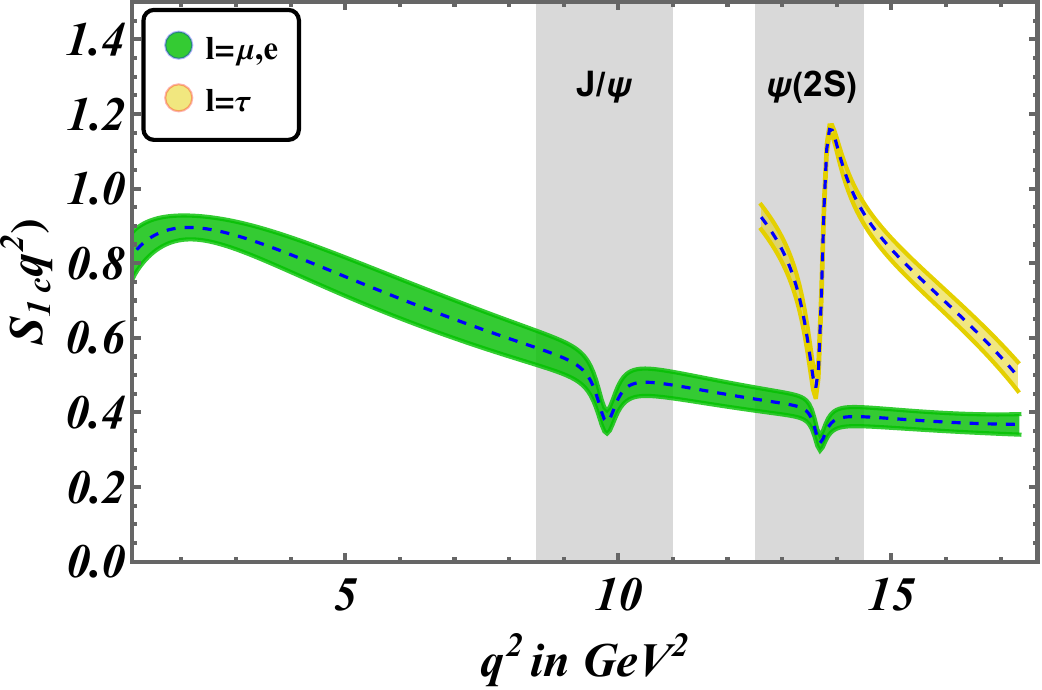}}
			\qquad
			\subfloat[\centering]{\includegraphics[width=4.5cm]{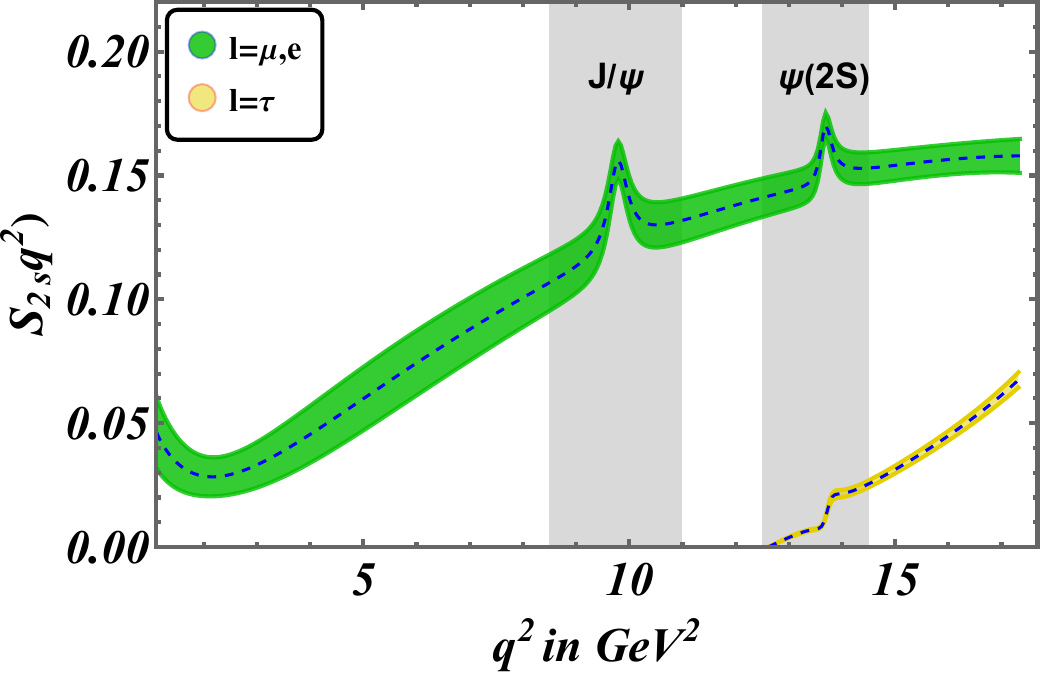}}
			\qquad
			\subfloat[\centering]{\includegraphics[width=4.5cm]{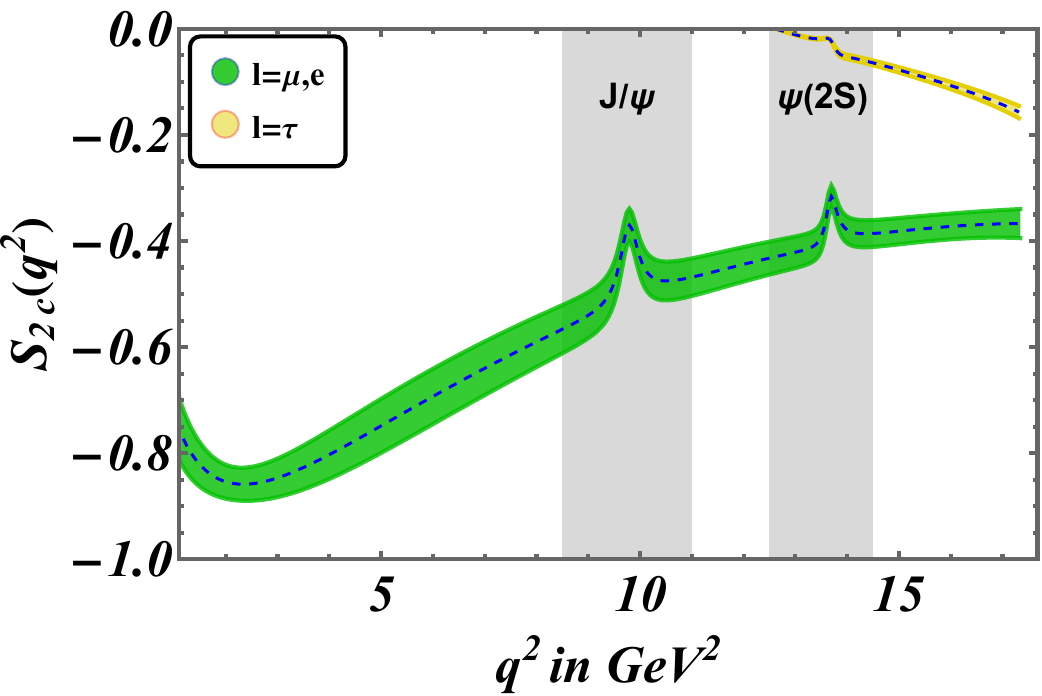}}
			\qquad
			\subfloat[\centering]{\includegraphics[width=4.5cm]{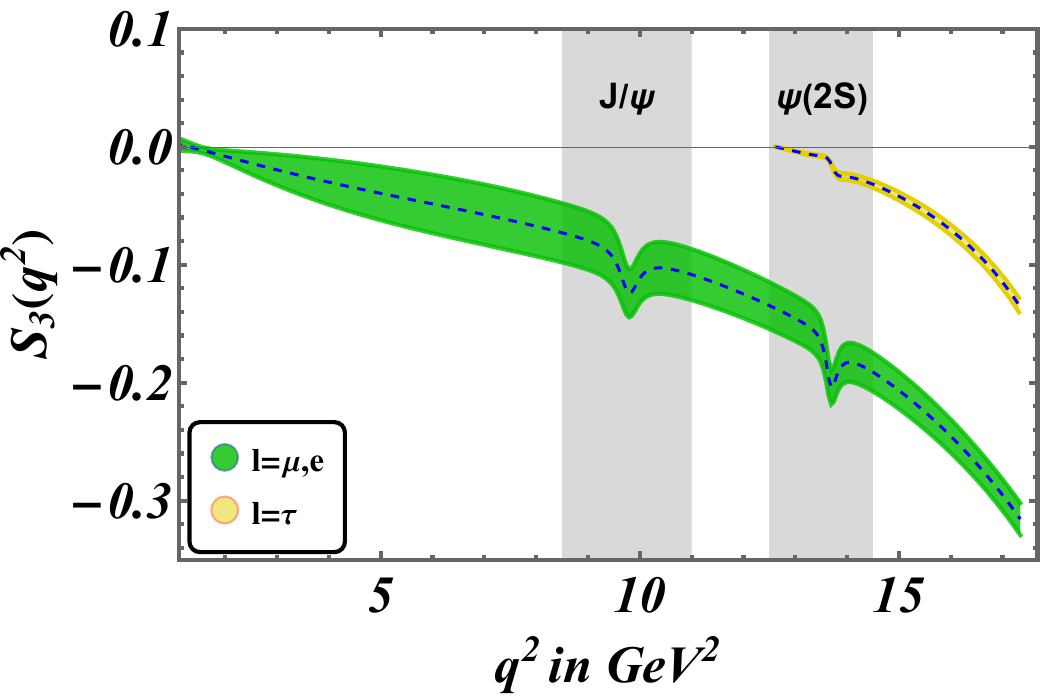}}
			\qquad
			\subfloat[\centering]{\includegraphics[width=4.5cm]{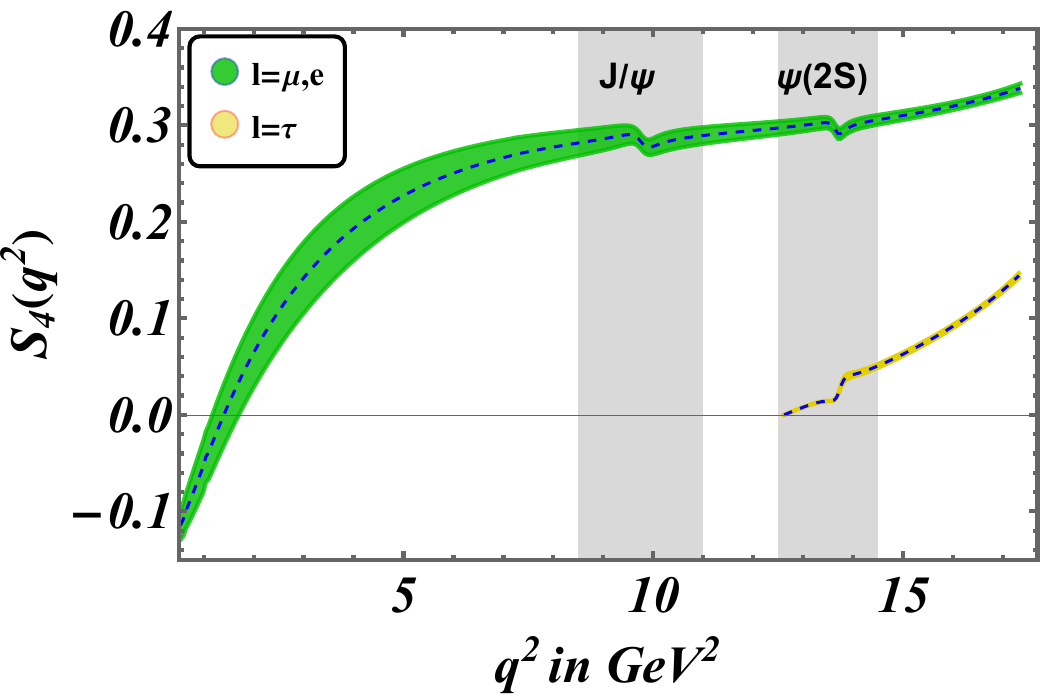}}
			\qquad
			\subfloat[\centering]{\includegraphics[width=4.5cm]{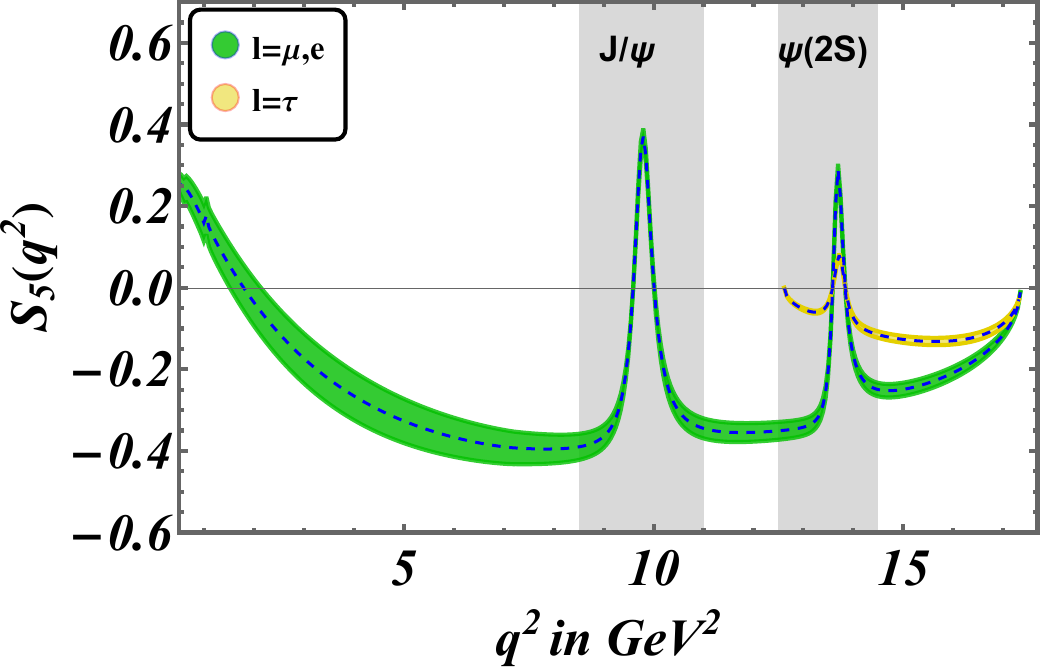}}
			\qquad
			\subfloat[\centering]{\includegraphics[width=4.5cm]{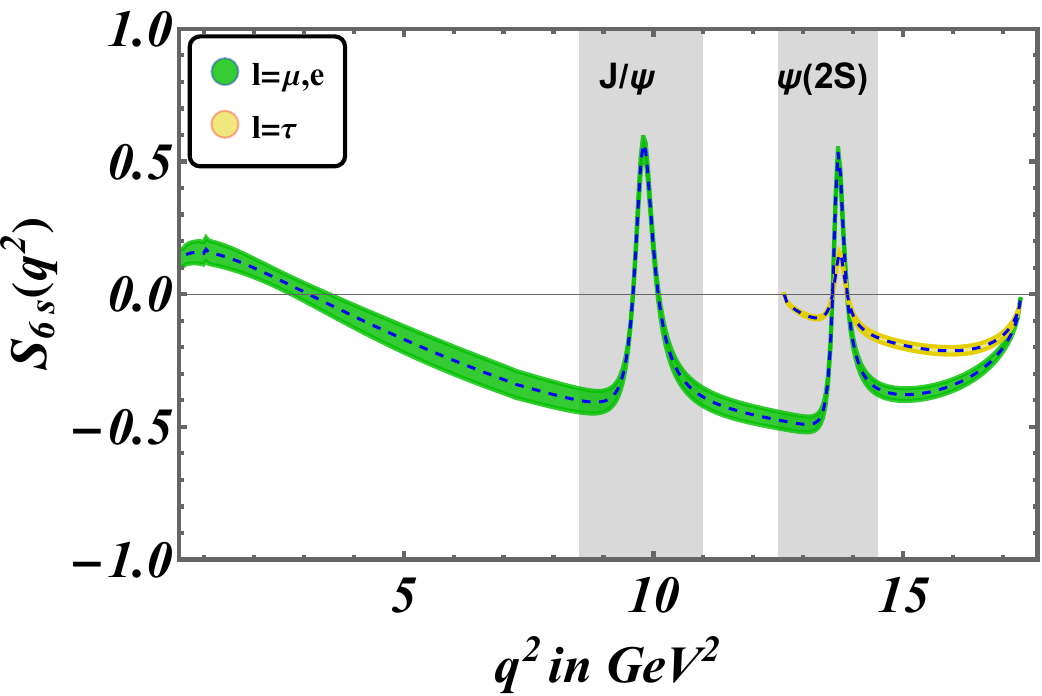}}
			\qquad
			\subfloat[\centering]{\includegraphics[width=4.5cm]{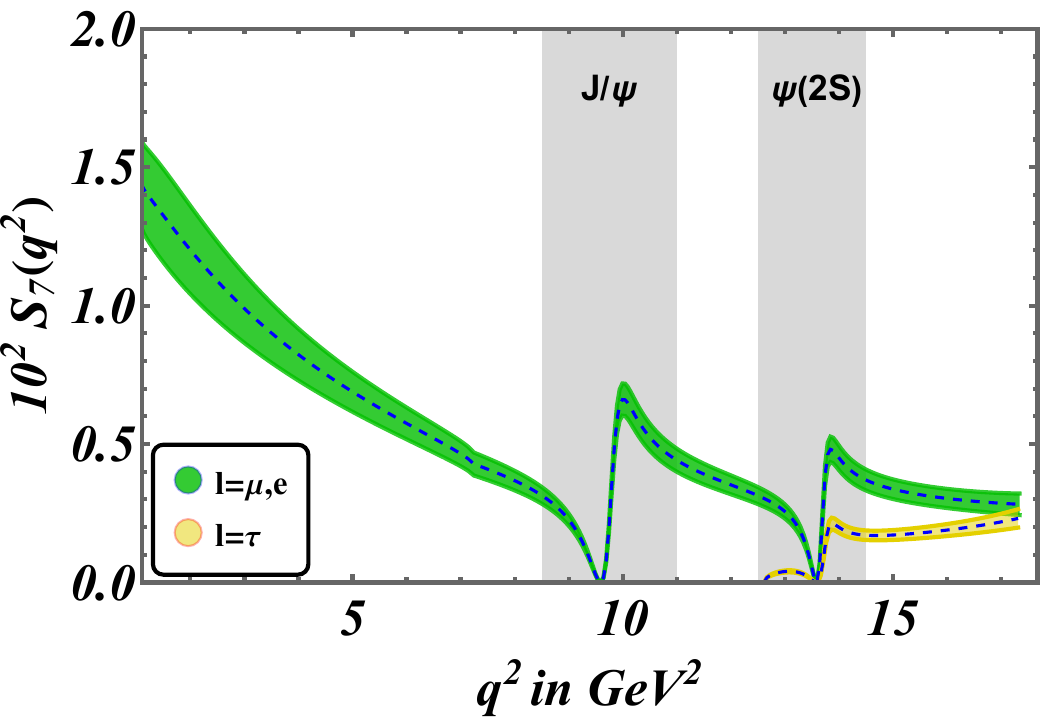}}
			\qquad
			\subfloat[\centering]{\includegraphics[width=4.5cm]{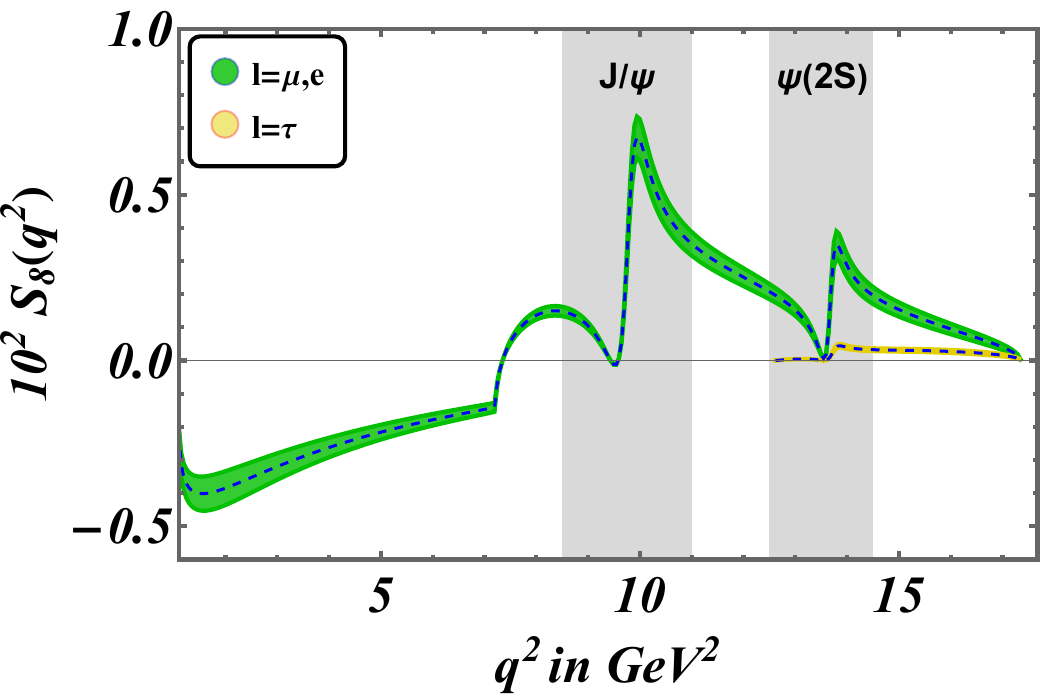}}
			\qquad
			\subfloat[\centering]{\includegraphics[width=4.5cm]{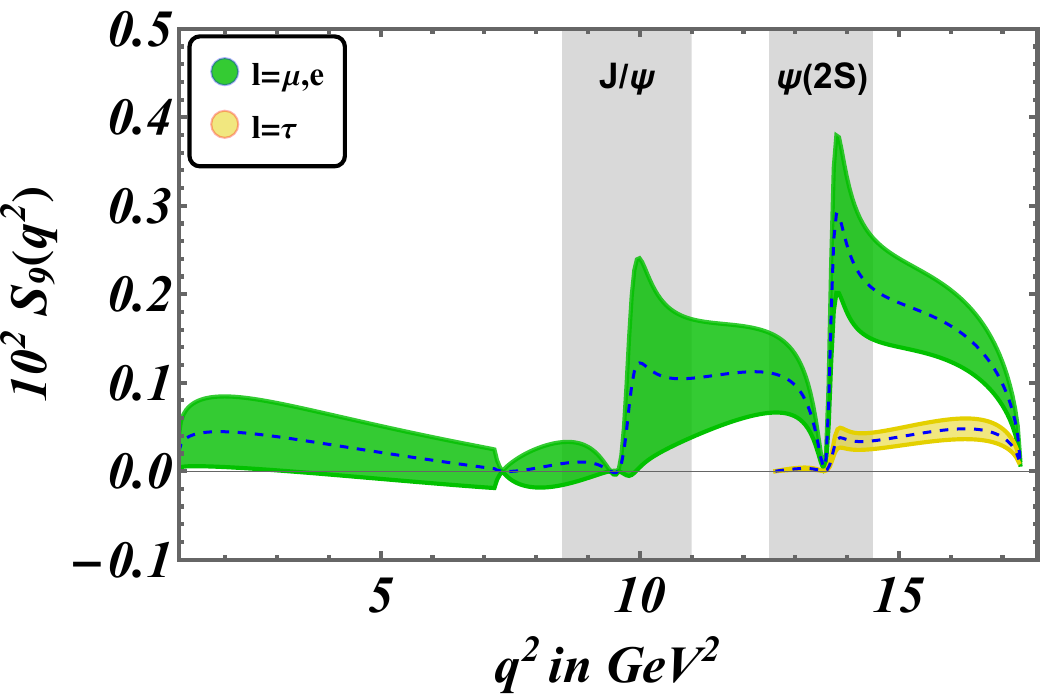}}
			\caption{The $q^{2}$ dependence of normalized CP averaged angular observables $S_{i}$. The green and yellow plots depict our results for the light lepton and $\tau$ lepton modes respectively and the gray bands denote the region of $J/\psi$ and $\psi(2S)$ resonances.}
			\label{fig:CP averaged angular observables}
		\end{figure}
		Following the prediction on branching fractions in the previous subsection, we next shift our focus onto studying the cascade decay mode $B_{c}^{-}\rightarrow D_{s}^{*-}(\rightarrow D_{s}^{-}\pi^{0}) \ell^{+}\ell^{-}$, and present predictions on a number of angular observables and some other observables derived from them, as was briefed in section \ref{subsubsection:Angular analysis theory}. 
		
		We start by presenting our results on the CP average angular coefficients $S_{i}$. In figure \ref{fig:CP averaged angular observables} we present the $q^{2}$ distribution of each of these angular coefficients. The green and yellow curves denote the distribution of observables corresponding to the light and heavy lepton modes respectively, and the grey bands denote the regions of $J/\psi$ amd $\psi(2S)$ resonances. From figure \ref{fig:CP averaged angular observables} we can see that the observables $S_{4}$, $S_{5}$, $S_{6s}$, $S_{8}$ and $S_{9}$ have a zero crossing within the physical $q^{2}$ region. However, $S_{4}$, $S_{8}$ and $S_{9}$ are themselves not clean and have very high sensitivity to hadronic inputs like form factors, making them highly sensitive to the choice of form factor parametrization and hence their zero crossing point is not that reliable. For $S_{8}$ and $S_{9}$, this point becomes relevant if we study possible NP scenarios. Therefore, we quote zero crossing $q^{2}$ points for the $S_{5}$ and $S_{6s}$, which are
		\begin{equation}
			\begin{split}
				 S_{5}:q_{0}^{2}=1.80(17)~\text{GeV}^{2}, \qquad S_{6s}:q_{0}^{2}=3.09(35)~\text{GeV}^{2}.
			\end{split}
		\end{equation}
		These points are interesting because they indicate a complete cancellation of the effects of left- and right-handed transversity amplitudes. Any higher-order QCD or non-factorizable corrections that modify these amplitudes will shift the zero point from the obtained values, thereby giving us an idea of the impact of corrections on these observables. Also, these points are sensitive to any possible new physics effects, since a small change in the Wilson coefficients will drastically change the zero crossing points. Out of all the observables, the zero crossing point of $S_{6s}$ is the most important one, since this directly gives the $q^{2}$ point where the forward-backward asymmetry also becomes zero, a point where the forward and backward rates of the decay process exactly balance each other. We also calculate $q^{2}$ averaged values of these observables separated into four $q^{2}$ bins and present the evaluated values in table \ref{table:angular observables in bins}. 
		
\begin{table}[htb!]
	\renewcommand{\arraystretch}{1.6}
	\centering
	\scriptsize
    \resizebox{\textwidth}{!}{%
	\begin{tabular}{|c|cc|cc|cc|cc|}
		\hline
		$\boldsymbol{q^{2}}$ \textbf{bins} $\boldsymbol{(\text{GeV}^{2})}$& \multicolumn{2}{c|}{$\boldsymbol{\langle S_{1s}\rangle}$} &\multicolumn{2}{c|}{$\boldsymbol{\langle S_{1c}\rangle}$ }&\multicolumn{2}{c|}{$\boldsymbol{\langle S_{2s}\rangle}$}&\multicolumn{2}{c|}{$\boldsymbol{\langle S_{2c}\rangle}$}\\
		\hline
		\textbf{Lepton Mode:}&$\boldsymbol{l=\mu,e}$&$\boldsymbol{l=\tau}$&$\boldsymbol{l=\mu,e}$&$\boldsymbol{l=\tau}$&$\boldsymbol{l=\mu,e}$&$\boldsymbol{l=\tau}$&$\boldsymbol{l=\mu,e}$&$\boldsymbol{l=\tau}$\\
		\hline
		$[1.1,6.0]$&0.147(34)&-&0.810(46)&-&0.049(11)&-&-0.784(45)&-\\
		$[6.0,8.0]$&0.273(39)&-&0.639(53)&-&0.091(13)&-&-0.629(52)&-\\
		$[11.0,12.5]$&0.410(25)&-&0.454(34)&-&0.136(8)&-&-0.450(34)&-\\
		$[15.0,17.0]$&0.468(19)&0.313(18)&0.376(25)&0.701(33)&0.156(6)&0.044(2)&-0.374(25)&-0.106(7)\\
		\hline
		\hline
		$\boldsymbol{q^{2}}$ \textbf{bins} $\boldsymbol{(\text{GeV}^{2})}$& \multicolumn{2}{c|}{$\boldsymbol{\langle S_{3}\rangle}$} &\multicolumn{2}{c|}{$\boldsymbol{\langle S_{4}\rangle}$ }&\multicolumn{2}{c|}{$\boldsymbol{\langle S_{5}\rangle}$}&\multicolumn{2}{c|}{$\boldsymbol{\langle  S_{6s}\rangle}$}\\
		\hline
		\textbf{Lepton Mode:}&$\boldsymbol{l=\mu,e}$&$\boldsymbol{l=\tau}$&$\boldsymbol{l=\mu,e}$&$\boldsymbol{l=\tau}$&$\boldsymbol{l=\mu,e}$&$\boldsymbol{l=\tau}$&$\boldsymbol{l=\mu,e}$&$\boldsymbol{l=\tau}$\\
		\hline
		$[1.1,6.0]$&-0.028(17)&-&0.158(33)&-&-0.201(56)&-&-0.079(39)&-\\
		$[6.0,8.0]$&-0.058(26)&-&0.265(18)&-&-0.388(40)&-&-0.328(52)&-\\
		$[11.0,12.5]$&-0.121(21)&-&0.293(8)&-&-0.352(23)&-&-0.439(33)&-\\
		$[15.0,17.0]$&-0.243(15)&-0.071(4)&0.319(4)&0.091(2)&-0.205(15)&-0.124(9)&-0.332(22)&-0.202(14)\\
		\hline
		\hline
		$\boldsymbol{q^{2}}$ \textbf{bins} $\boldsymbol{(\text{GeV}^{2})}$& \multicolumn{2}{c|}{$\boldsymbol{10^{2}\times\langle S_{7}\rangle}$} &\multicolumn{2}{c|}{$\boldsymbol{10^{2}\times\langle S_{8}\rangle}$}&\multicolumn{2}{c|}{$\boldsymbol{10^{2}\times\langle S_{9}\rangle}$}&&\\
		\hline
		\textbf{Lepton Mode:}&$\boldsymbol{l=\mu,e}$&$\boldsymbol{l=\tau}$&$\boldsymbol{l=\mu,e}$&$\boldsymbol{l=\tau}$&$\boldsymbol{l=\mu,e}$&$\boldsymbol{l=\tau}$&&\\
		\hline
		$[1.1,6.0]$&0.906(96)&-&-0.283(30)&-&0.033(36)&-&&\\
		$[6.0,8.0]$&0.454(45)&-&-0.058(6)&-&0.004(8)&-&&\\
		$[11.0,12.5]$&0.372(36)&-&0.273(29)&-&0.111(55)&-&&\\
		$[15.0,17.0]$&0.306(36)&0.189(23)&0.098(17)&0.026(5)&0.161(40)&0.045(11)&&\\
		\hline
	\end{tabular}
    }
	\caption{$q^{2}$ averaged estimates of the various CP averaged angular observables $S_{i}$ in separate $q^{2}$ bins.}
	\label{table:angular observables in bins}
\end{table}

The $q^{2}$ limits in each bin are chosen as follows:
\begin{itemize}
	\item For the first $q^{2}$ bin, the lower limit should ideally be $4 m_{\ell}^{2}$. But in our analysis we have set it at $q^{2}=1.1~\text{GeV}^{2}$. This is because at small $q^{2}$ value, the relevant decay amplitude is dominated mostly by the photon pole, and hence by just one Wilson Coefficient $C_{7}^{\text{eff}}$. Considering this region into the analysis will not add any new information in comparison to what is already available from analysis of $b\rightarrow s\gamma$ channel. Additionally including the small $q^{2}$ region might introduce contributions due to light resonances like $\rho,~\omega$ and $\phi$ \cite{Altmannshofer:2008dz}. Hence we avoid this region. The upper limit of $q^{2}=6.0~\text{GeV}^{2}$ is taken to avoid the $J/\psi$ resonance peak at $m_{J/\psi}^{2}=9.6~\text{GeV}^{2}$ by a safe margin, and also to stay in accordance to experimental conventions.
	\item For the second $q^{2}$ bin, we continue from the first bin, setting the lower limit at $q^{2}=6.0~\text{GeV}^{2}$. As for the upper limit, we set $q^{2}=8.0~\text{GeV}^{2}$ to stay below the $J/\psi$ resonance. A point to note is that, instead of having a single bin in the range $q^{2} \in (1.1, 8.0)~\text{GeV}^{2}$, we have divided the $q^{2}$ range into two bins. This is because the range $6.0-8.0~\text{GeV}^{2}$ lies close to the $J/\psi$ resonance and the non-local $c\bar{c}$ resonance effects are non-negligible in this region, while the range $1.1-6.0~\text{GeV}^{2}$ does not suffer from any such contributions. Hence mixing the two regions will contaminate the clean predictions in the low $q^{2}$ region.
	\item For the third $q^{2}$ bin, the $q^{2}$ range is chosen to lie in the region between the $J/\psi$ and $\psi(2S)$ resonances. For the lower limit, we set it at $q^{2}=11.0~\text{GeV}^{2}$, a value conveniently above $J/\psi$ resonance, while for the upper limit we set it at $q^{2}=12.5~\text{GeV}^{2}$, safely below the $\psi(2S)$ resonance peak.
	\item For the fourth $q^{2}$ bin, we choose the lower limit at $q^{2}=15.0~\text{GeV}^{2}$, a value greater than the $\psi(2S)$ resonance peak. The upper limit of $q^{2}=17.0~\text{GeV}^{2}$ is chosen to be in accordance with experimental convention.
	\end{itemize}
From figure \ref{fig:CP averaged angular observables} and table \ref{table:angular observables in bins}, we can see that among the observables $S_{1s}$ to $S_{5}$, $S_{3}$ has a smaller magnitude compared to the other observables, but is comparable to $S_{6s}$. This is primarily due to partial cancellation between the transversity amplitudes, as can be seen in eq.\eqref{eqn:angular observables}. The cancellation becomes weaker as we move from the first to the fourth bin, leading to an increase in the absolute value. The reason for this can be traced back to analytic expressions in eq.\eqref{eqn:transversity amplitudes}, where it can be clearly seen that they are dependent upon the imaginary part of the transversity amplitudes, and hence of the Wilson Coefficients $C_{7}^{\text{eff}}$ and $C_{9}^{\text{eff}}$, whose imaginary parts, upon checking numerically, is found out to be highly suppressed compared to the real part.
\begin{figure}[t!]
	\centering
	\subfloat[\centering]{\includegraphics[width=7.0cm]{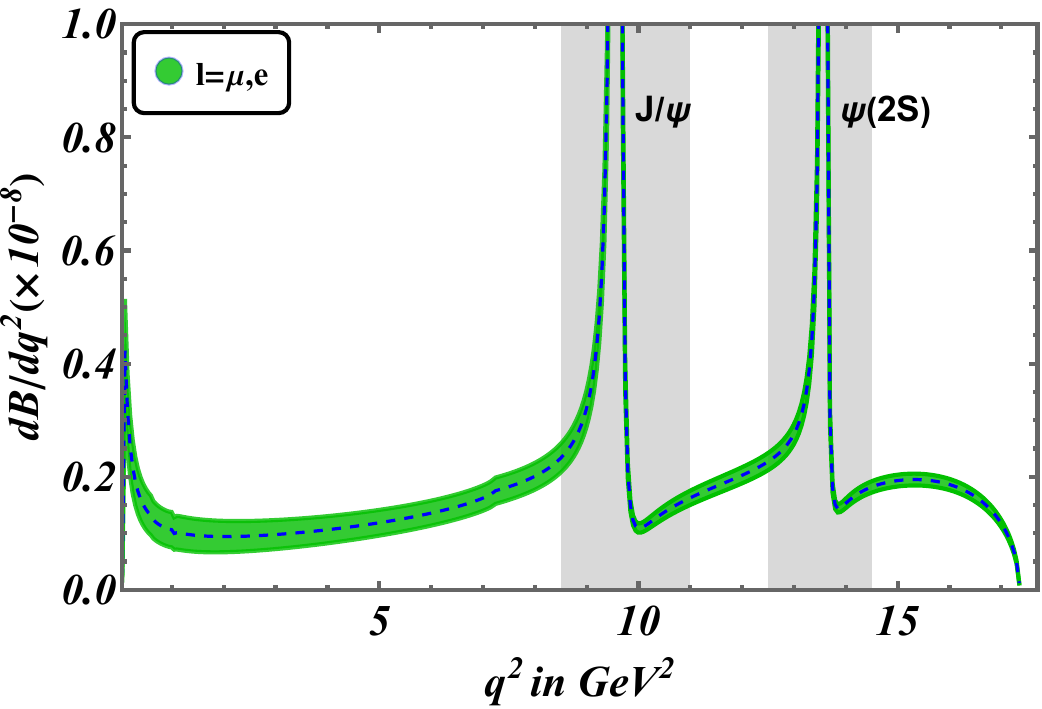}}
	\qquad
	\subfloat[\centering]{\includegraphics[width=7.0cm]{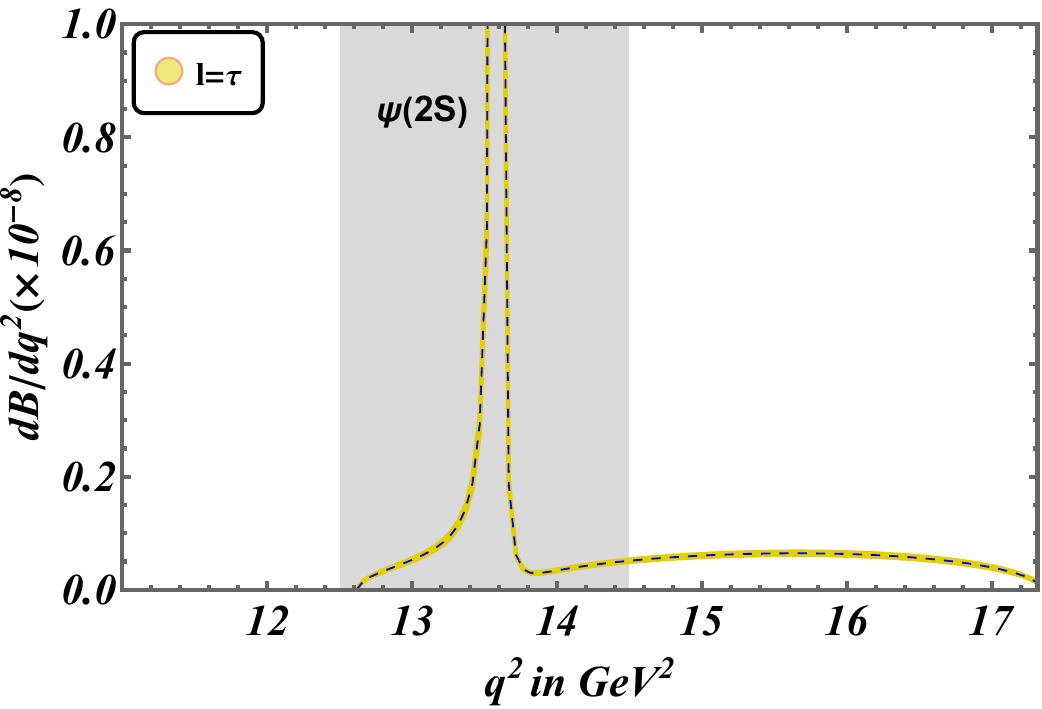}}
	\caption{The $q^{2}$ dependence of differential branching fractions $d\mathcal{B}/dq^{2}$ of $B_{c}^{-}\rightarrow D_{s}^{*-}(\rightarrow D_{s}^{-}\pi^{0}) \ell^{+}\ell^{-}$. The first plot depicts our result for the light lepton mode, and the second plot depicts our result for the $\tau$ lepton mode, and the gray bands denote the region of $J/\psi$ and $\psi(2S)$ resonances.}
	\label{fig:branching fractions in terms of angular observables}
\end{figure}

Next, we calculate the CP-averaged branching fraction, taking the expressions previously discussed in eq.\eqref{eqn:CP averaged branching fraction} and multiplying them by $\tau_{B_{c}}$. In figure \ref{fig:branching fractions in terms of angular observables} we present $q^{2}$ distribution of branching fractions for the light and heavy lepton modes. Following this, in table \ref{table:average decay width in terms of angular observables} we calculate $q^{2}$ averaged values of the branching fractions in the same four $q^{2}$ bins, i.e., $[1.1,6.0]$, $[6.0,8.0]$, $[11.0,12.5]$ and $[15.0,17.0]$ $\text{GeV}^{2}$.

\begin{table}[htb!]
	\renewcommand{\arraystretch}{1.4}
	\centering
	\begin{tabular}{|c|c|c|}
		\hline
		$\boldsymbol{q^{2}}$ \textbf{bins} $\boldsymbol{(\text{GeV}^{2})}$&$\boldsymbol{\mathcal{B}(l=e,\mu)}$&$\boldsymbol{\mathcal{B}(l=\tau)}$\\
		\hline
		$[1.1,6.0]$&0.530(108)&-\\
		$[6.0,8.0]$&0.331(40)&-\\
		$[11.0,12.5]$&0.290(19)&-\\
		$[15.0,17.0]$&0.354(19)&0.119(6)\\
		\hline
	\end{tabular}
	\caption{$q^{2}$ averaged estimates of branching fractions $(\times 10^{-8})$ of $B_{c}^{-}\rightarrow D_{s}^{*-}(\rightarrow D_{s}^{-}\pi^{0}) \ell^{+}\ell^{-}$ in separate $q^{2}$ bins.}
	\label{table:average decay width in terms of angular observables}
\end{table}

We have verified the results of table \ref{table:average decay width in terms of angular observables} by calculating $\mathcal{B}(B_{c}^{-}\rightarrow D_{s}^{*-}(\rightarrow D_{s}^{-}\pi^{0}) \ell^{+}\ell^{-})$ from expressions of $\mathcal{B}(B_{c}^{-}\rightarrow D_{s}^{*-}\ell^{+}\ell^{-})$ in terms of Helicity amplitudes and discussed in section \ref{subsubsection:decay width theory}, and then multiplying it by $\mathcal{B}(D_{s}^{*-}\rightarrow D_{s}^{-}\pi^{0})$, in each $q^{2}$ bin. Additionally, we also check for lepton flavor universality by calculating the ratio between heavy and light lepton branching fractions in the fourth $q^{2}$ bin, defined as
\begin{equation}
	R^{\tau\mu}=\frac{\mathcal{B}(B_{c}^{-}\rightarrow D_{s}^{*-}(\rightarrow D_{s}^{-}\pi^{0}) \tau^{+}\tau^{-})}{\mathcal{B}(B_{c}^{-}\rightarrow D_{s}^{*-}(\rightarrow D_{s}^{-}\pi^{0}) \ell^{+}\ell^{-})},
\end{equation}
and is found to be $0.335(5)$ in this work.

\begin{figure}[htb!]
	\centering
	\subfloat[\centering]{\includegraphics[width=7.0cm]{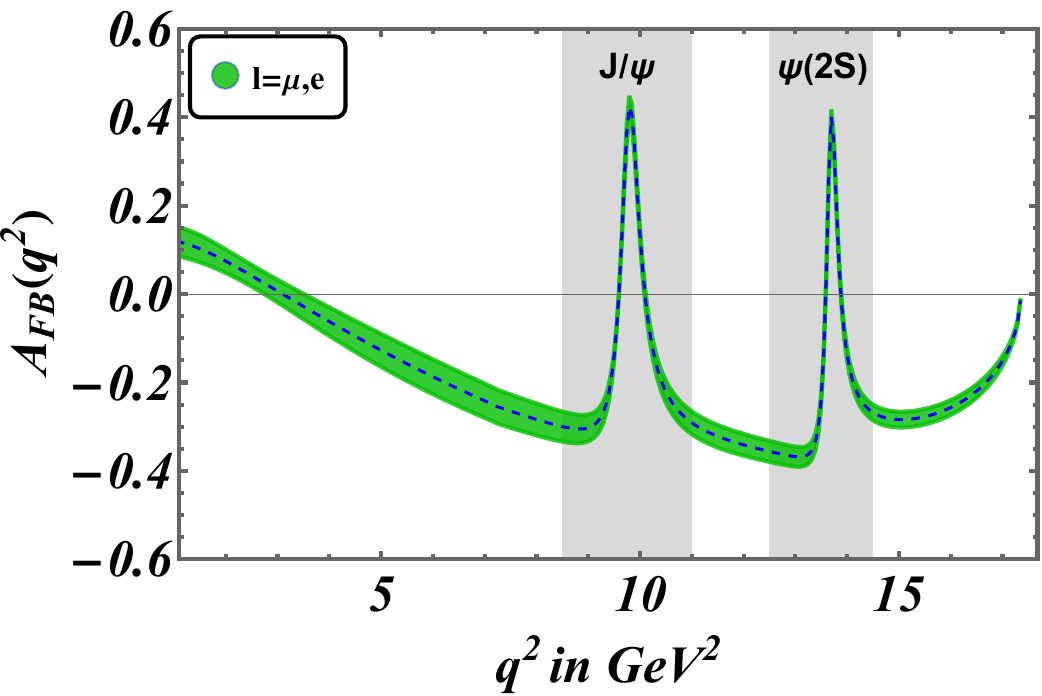}}
	\qquad
	\subfloat[\centering]{\includegraphics[width=7.0cm]{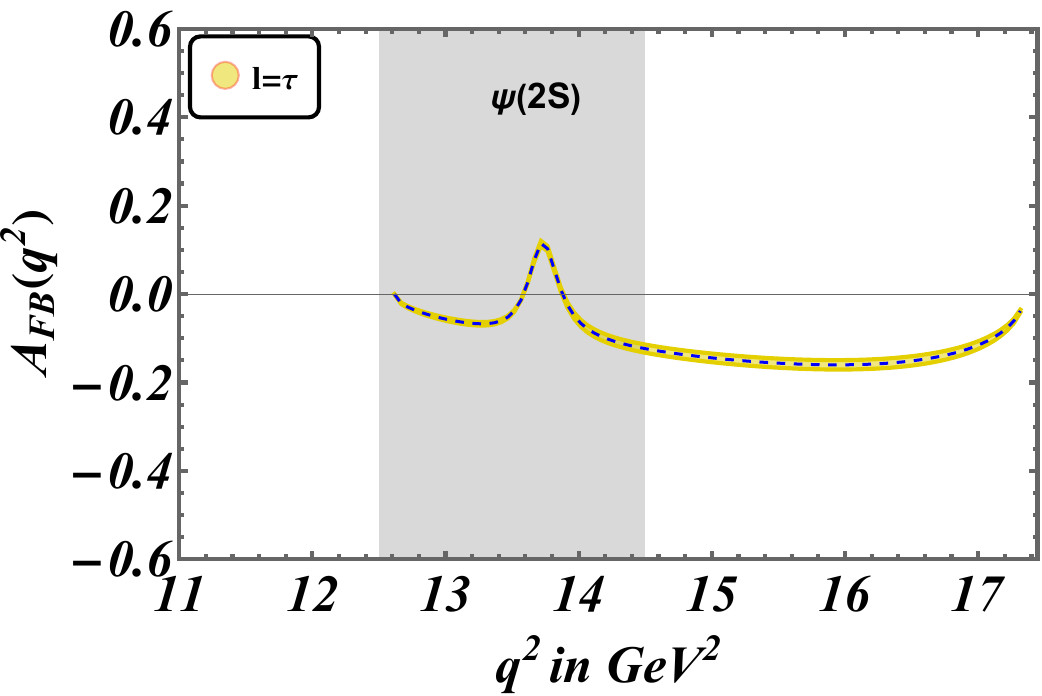}}
	\qquad
	\subfloat[\centering]{\includegraphics[width=7.0cm]{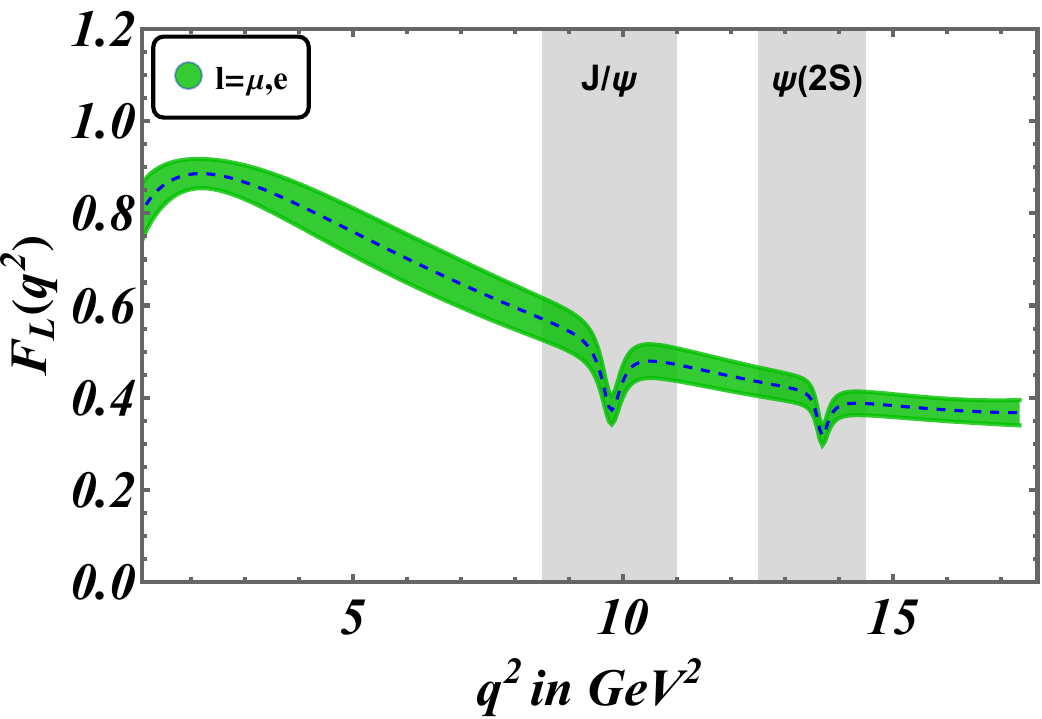}}
	\qquad
	\subfloat[\centering]{\includegraphics[width=7.0cm]{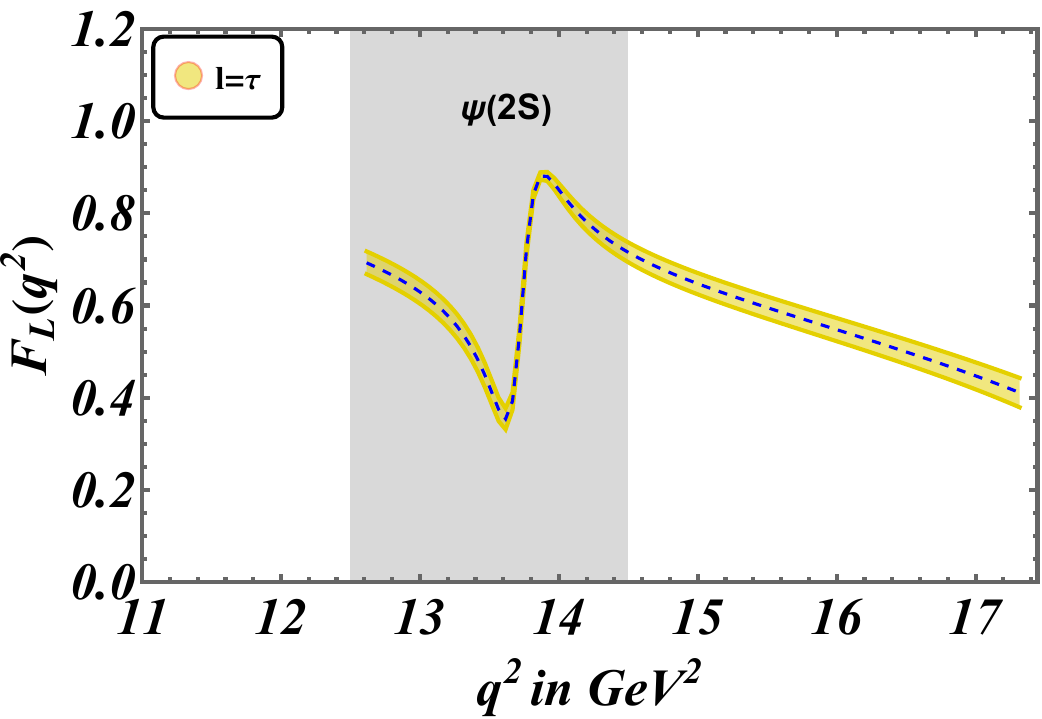}}
	\qquad
	\subfloat[\centering]{\includegraphics[width=7.0cm]{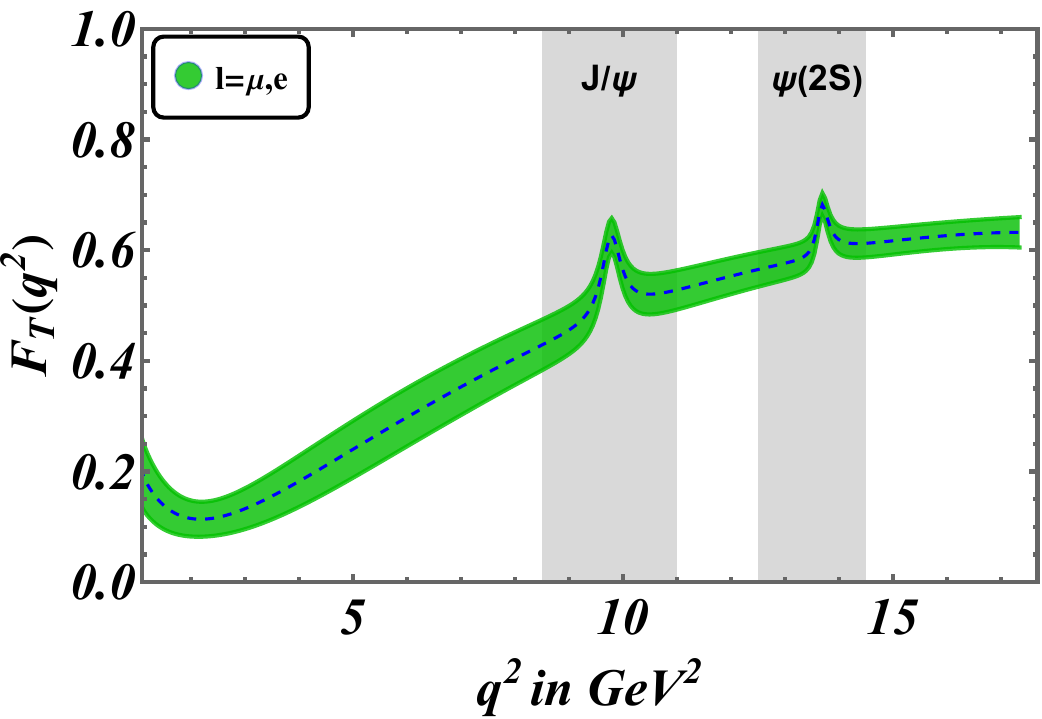}}
	\qquad
	\subfloat[\centering]{\includegraphics[width=7.0cm]{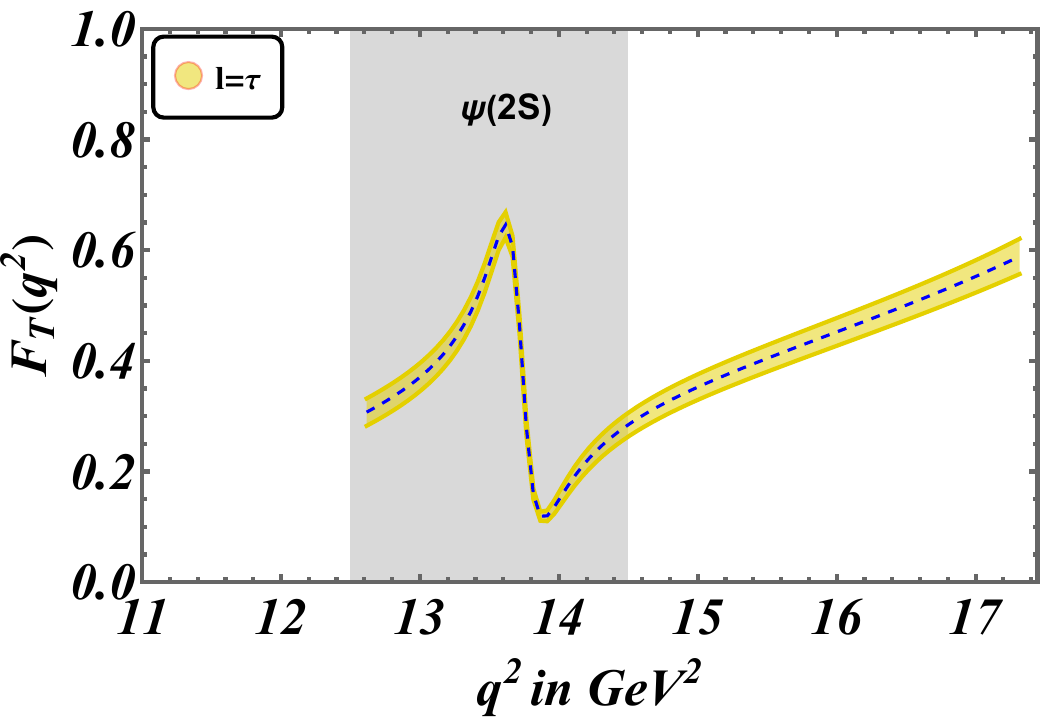}}
	\caption{The $q^{2}$ dependence of lepton forward-backward asymmetry parameter $A_{FB}$, $D_{s}^{*}$ longitudinal and transverse polarization fractions $F_{L}$ and $F_{T}$ in $B_{c}\rightarrow D_{s}^{*-}(\rightarrow D_{s}^{-}\pi^{0}) l^{+}l^{-}$ channel. The green and yellow plots denote the results for light lepton and tau lepton modes, respectively and the gray bands denote the region of $J/\psi$ and $\psi(2S)$ resonances.}
	\label{fig:FL and FT in terms of angular observables}
\end{figure}

After the CP averaged observables and branching fractions, we next focus on some of the more established observables. These are the forward backward asymmetry $A_{FB}$, and the $D_{s}^{*}$ longitudinal and transverse polarization fractions $F_{L}$ and $F_{T}$ respectively. We take the expressions previously discussed in eqs.\eqref{eqn:forward backward asymmetry} and \eqref{eqn:lepton polarization} in subsection \ref{subsubsection:Angular analysis theory}. We present $q^{2}$ distribution of these observables in figure \ref{fig:FL and FT in terms of angular observables}.

We also compute the $q^{2}$ averaged values of these observables by calculating
\begin{equation}
	\langle \mathcal{A}\rangle=\frac{\int_{q_{min}^{2}}^{q_{max}^{2}}\mathcal{A}(q^{2})\left(\frac{d\Gamma}{dq^{2}}+\frac{d\bar{\Gamma}}{dq^{2}}\right)dq^{2}}{\int_{q_{min}^{2}}^{q_{max}^{2}}\left(\frac{d\Gamma}{dq^{2}}+\frac{d\bar{\Gamma}}{dq^{2}}\right)dq^{2}},
\end{equation}
where $\mathcal{A}=A_{FB}$, $F_{L}$ or $F_{T}$. We perform our calculations in four separate $q^{2}$ bins, and present them in table \ref{table:average of physical observables}.

\begin{table}[htb!]
	\renewcommand{\arraystretch}{1.7}
	\centering
	\begin{tabular}{|c|cc|cc|cc|}
		\hline
		$\boldsymbol{q^{2}}$ \textbf{bins} $\boldsymbol{(\text{GeV}^{2})}$&\multicolumn{2}{c}{$\boldsymbol{\langle A_{\text{FB}}\rangle}$}&\multicolumn{2}{|c|}{$\boldsymbol{\langle F_{L}\rangle}$}&\multicolumn{2}{c|}{$\boldsymbol{\langle F_{T}\rangle}$}\\
		\hline
		\textbf{Lepton Mode:}&$\boldsymbol{l=\mu,e}$&$\boldsymbol{l=\tau}$&$\boldsymbol{l=\mu,e}$&$\boldsymbol{l=\tau}$&$\boldsymbol{l=\mu,e}$&$\boldsymbol{l=\tau}$\\
		\hline
		$[1.1,6.0]$&-0.059(30)&-&0.805(46)&-&0.195(46)&-\\
		$[6.0,8.0]$&-0.238(38)&-&0.646(51)&-&0.354(51)&-\\
		$[11.0,12.5]$&-0.330(25)&-&0.452(33)&-&0.547(33)&-\\
		$[15.0,17.0]$&-0.250(16)&-0.152(10)&0.374(23)&0.552(24)&0.626(23)&0.448(24)\\
		\hline
	\end{tabular}
	\caption{$q^{2}$ averaged estimates of forward backward asymmetry, longitudinal and transverse polarization fractions in separate $q^{2}$ bins.}
	\label{table:average of physical observables}
\end{table}

\begin{figure}[htb!]
	\centering
	\subfloat[\centering]{\includegraphics[width=4.5cm]{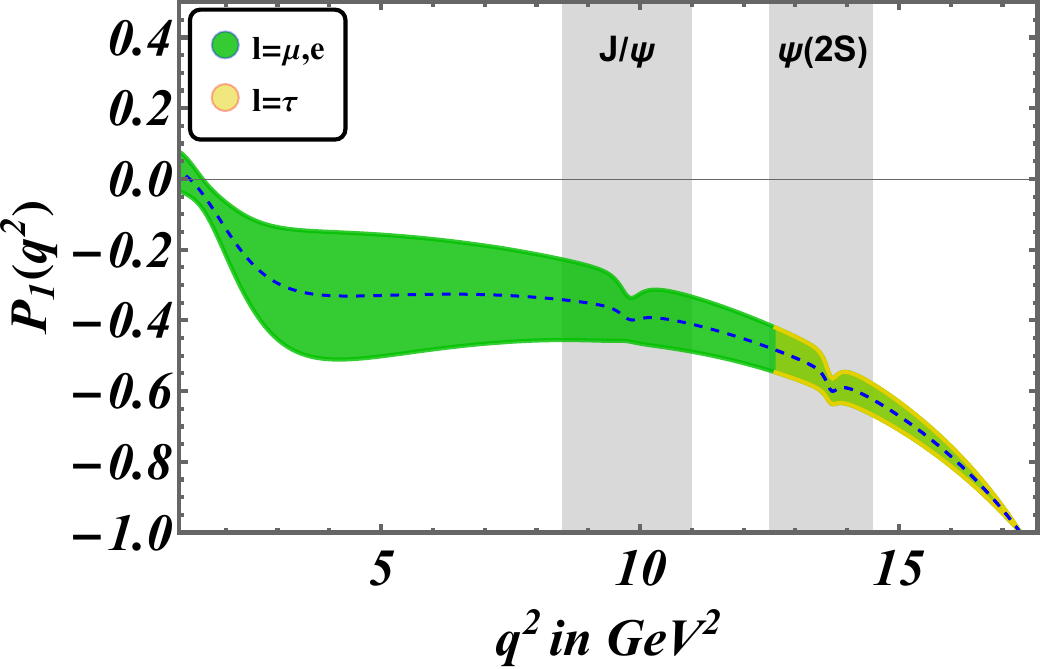}}
	\qquad
	\subfloat[\centering]{\includegraphics[width=4.5cm]{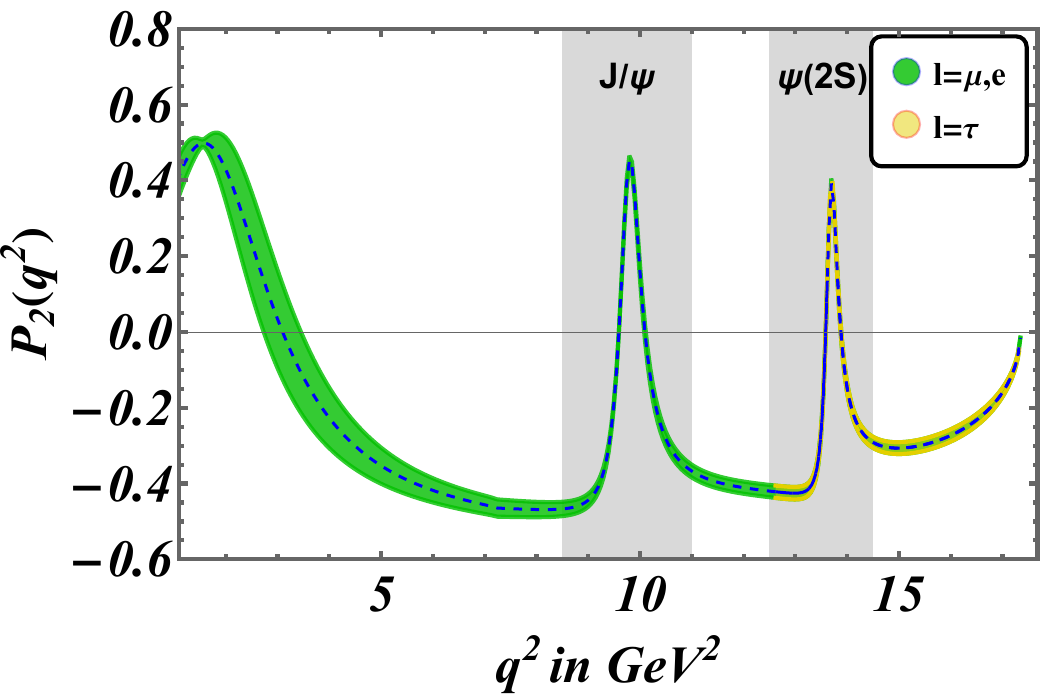}}
	\qquad
	\subfloat[\centering]{\includegraphics[width=4.5cm]{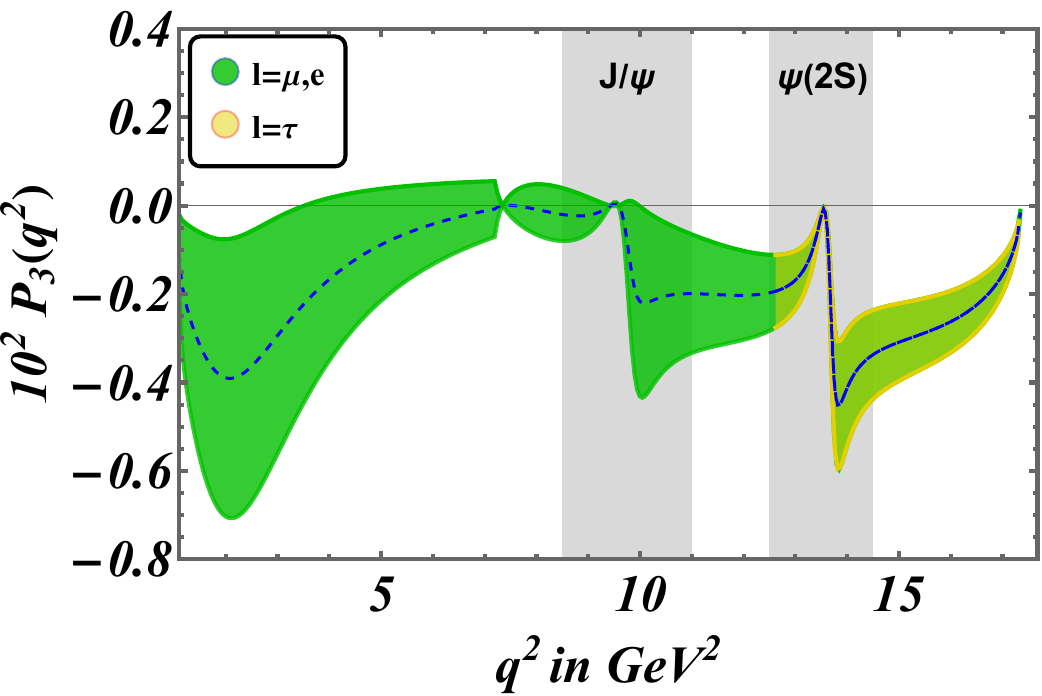}}
	\qquad
	\subfloat[\centering]{\includegraphics[width=4.5cm]{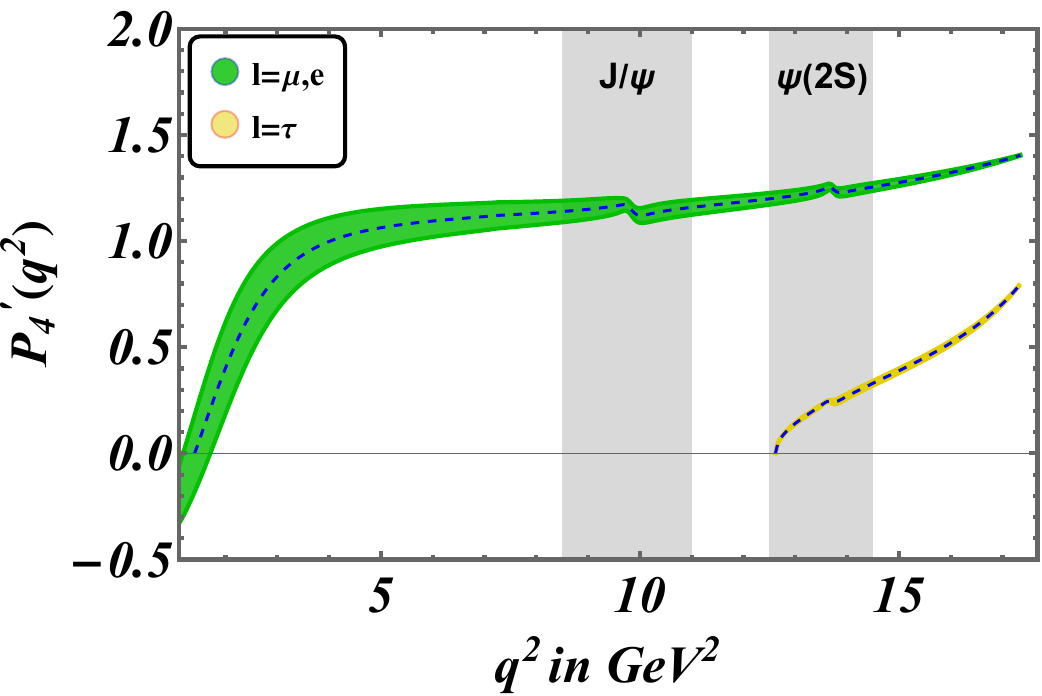}}
	\qquad
	\subfloat[\centering]{\includegraphics[width=4.5cm]{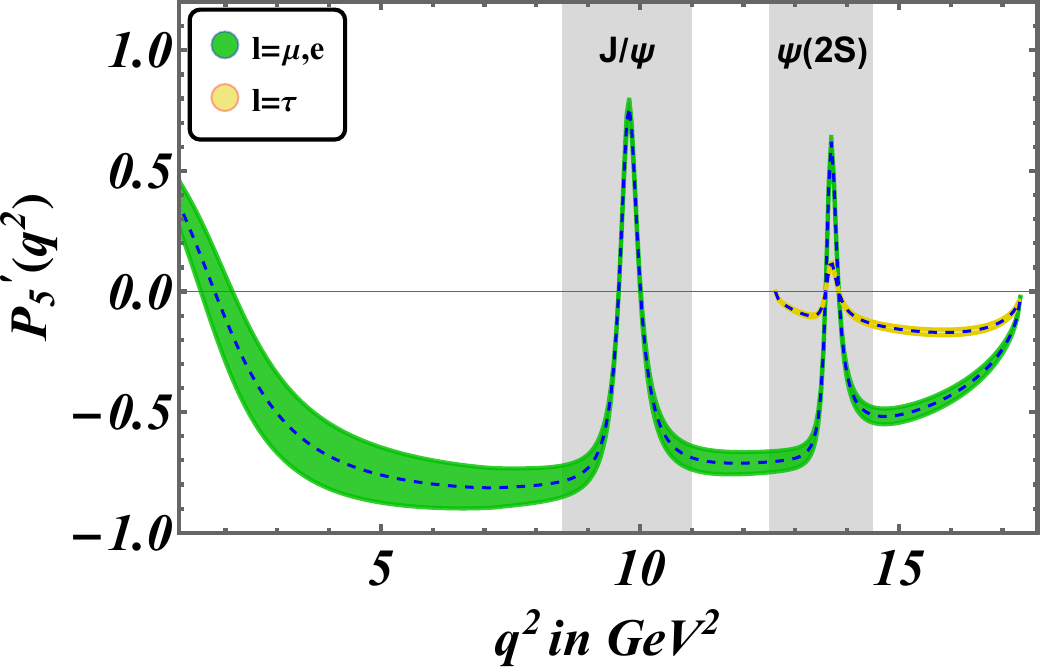}}
	\qquad
	\subfloat[\centering]{\includegraphics[width=4.5cm]{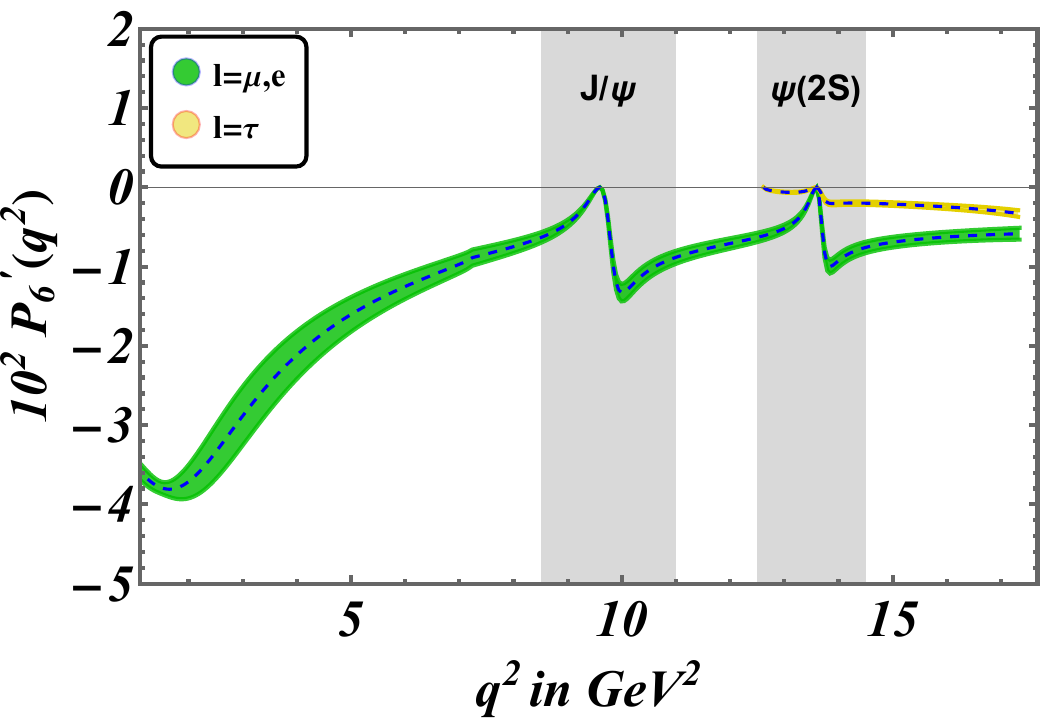}}
	\qquad
	\subfloat[\centering]{\includegraphics[width=4.5cm]{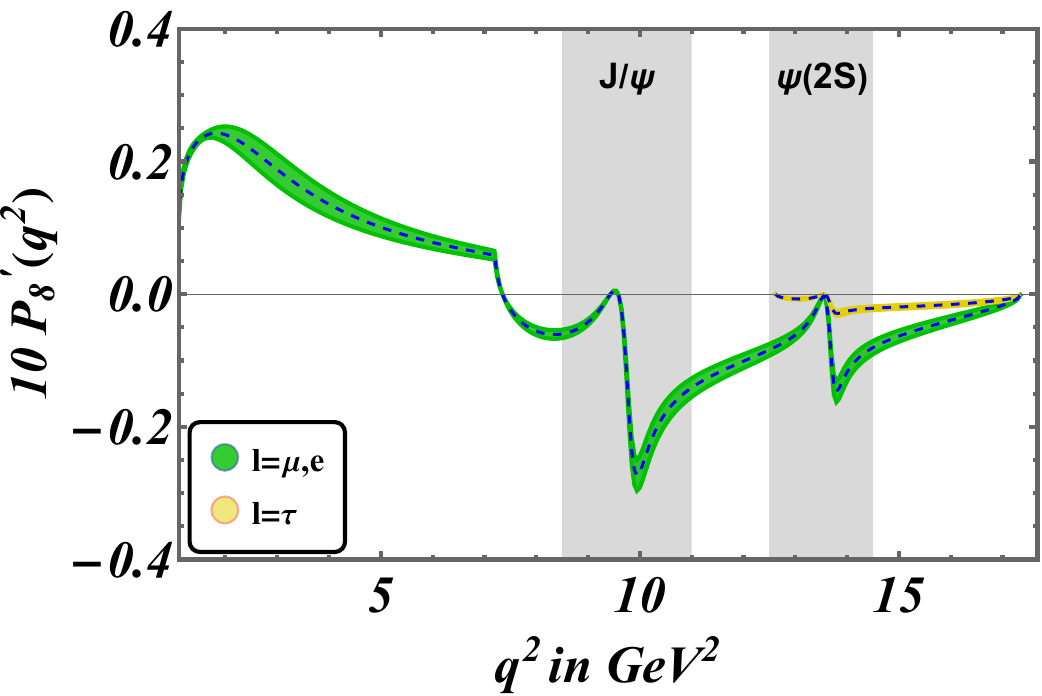}}
	\caption{The $q^{2}$ dependence of clean angular observables $P_{1,2,3}$ and $P_{4,5,6,8}^{'}$. The green and yellow plots depict our results for the light lepton and $\tau$ lepton modes, respectively and the gray bands denote the region of $J/\psi$ and $\psi(2S)$ resonances.}
	\label{fig:CP averaged clean angular observables}
\end{figure}

Further, we also present predictions for the clean angular observables $P_{1,2,3}$ and $P_{4,5,6,8}^{\prime}$, as defined in eq.\eqref{eqn:Clean observables}. These observables are of particular interest due to their reduced sensitivity to the hadronic form factors, rendering them comparatively cleaner and less affected by hadronic uncertainties. As a result, they serve as powerful probes for exploring potential new physics (NP) effects. The $q^{2}$ distribution of these observables is presented in figure \ref{fig:CP averaged clean angular observables}.

Additionally, we also present predictions for the average values of these observables in the four $q^{2}$ bins in table \ref{table:clean observables}. We have verified two of the observables, particularly $P_{2}$ and $P_{5}^{\prime}$ using the relations \cite{LHCb:2020gog}
\begin{equation}
	P_{2}=\frac{2}{3}\cdot\frac{A_{FB}}{1-F_{L}}, \qquad P_{5}^{\prime}=\frac{S_{5}}{\sqrt{F_{L}(1-F_{L})}},
\end{equation}
and have found the results in the table to be in good agreement with each other. Further, from table \ref{table:clean observables} we observe that $P_{3}$ exhibits the smallest magnitude across all four $q^{2}$ bins. This behaviour is expected, since $P_{3}$ is proportional to the angular observable $S_{9}$, which is defined through the imaginary part of the interference term $(\mathcal{A}{\parallel}^{L*}\mathcal{A}{\perp}^{L} - L \leftrightarrow R)$. In the Standard Model, this term is strongly suppressed due to the small relative strong phases between the transversity amplitudes, resulting in a negligibly small value of $S_{9}$, and consequently of $P_{3}$.

\begin{table}[htb!]
	\renewcommand{\arraystretch}{1.7}
	\centering
	\scriptsize
    \resizebox{\textwidth}{!}{%
	\begin{tabular}{|c|cc|cc|cc|cc|}
		\hline
		$\boldsymbol{q^{2}}$ \textbf{bins} $\boldsymbol{(\text{GeV}^{2})}$& \multicolumn{2}{c|}{$\boldsymbol{\langle P_{1}\rangle}$} &\multicolumn{2}{c|}{$\boldsymbol{\langle P_{2}\rangle}$ }&\multicolumn{2}{c|}{$\boldsymbol{10^{2}\times\langle P_{3}\rangle}$}&\multicolumn{2}{c|}{}\\
		\hline
		\textbf{Lepton Mode:}&$\boldsymbol{l=\mu,e}$&$\boldsymbol{l=\tau}$&$\boldsymbol{l=\mu,e}$&$\boldsymbol{l=\tau}$&$\boldsymbol{l=\mu,e}$&$\boldsymbol{l=\tau}$&&\\
		\hline
		$[1.1,6.0]$&-0.278(142)&-&-0.159(86)&-&-0.150(165)&-&&\\
		$[6.0,8.0]$&-0.327(132)&-&-0.450(25)&-&-0.012(19)&-&&\\
		$[11.0,12.5]$&-0.444(67)&-&-0.402(16)&-&-0.202(98)&-&&\\
		$[15.0,17.0]$&-0.773(29)&-0.788(28)&-0.270(16)&-0.263(16)&-0.264(68)&-0.256(66)&&\\
		\hline
		\hline
		$\boldsymbol{q^{2}}$ \textbf{bins} $\boldsymbol{(\text{GeV}^{2})}$& \multicolumn{2}{c|}{$\boldsymbol{\langle P_{4}^{'}\rangle}$}& \multicolumn{2}{c|}{$\boldsymbol{\langle P_{5}^{'}\rangle}$} &\multicolumn{2}{c|}{$\boldsymbol{10^{2}\times\langle P_{6}^{'}\rangle}$ }&\multicolumn{2}{c|}{$\boldsymbol{10 \times \langle P_{8}^{'}\rangle}$}\\
		\hline
		\textbf{Lepton Mode:}&$\boldsymbol{l=\mu,e}$&$\boldsymbol{l=\tau}$&$\boldsymbol{l=\mu,e}$&$\boldsymbol{l=\tau}$&$\boldsymbol{l=\mu,e}$&$\boldsymbol{l=\tau}$&$\boldsymbol{l=\mu,e}$&$\boldsymbol{l=\tau}$\\
		\hline
		$[1.1,6.0]$&0.839(135)&-&-0.543(141)&-&-4.488(570)&-&0.1413(174)&-\\
		$[6.0,8.0]$&1.114(60)&-&-0.805(81)&-&-1.906(225)&-&0.0192(25)&-\\
		$[11.0,12.5]$&1.180(29)&-&-0.707(45)&-&-1.476(142)&-&-0.1079(113)&-\\
		$[15.0,17.0]$&1.318(12)&0.524(11)&-0.431(29)&-0.162(11)&-1.267(138)&-0.494(57)&-0.0418(67)&-0.0150(26)\\
		\hline
	\end{tabular}
    }
	\caption{$q^{2}$ averaged estimates of clean observables $P_{i}^{(')}$ in separate $q^{2}$ bins.}
	\label{table:clean observables}
\end{table}

In table \ref{table:clean observables}, it is worth noting that the observables $P_{1},~P_{2}\text{ and }P_{3}$ exhibit about 2-3\% difference between light and heavy lepton modes. This suggests that the transversity amplitudes governing these observables are largely insensitive to lepton mass effects in the high $q^{2}$ region. In contrast, the primed observables $P_{4}^{\prime},~P_{5}^{\prime},~P_{6}^{\prime}\text{ and }P_{8}^{\prime}$ exhibit significant deviation between light and heavy lepton modes. Furthermore, we also observe that the uncertainties of $P_{1,2,3}$ are larger compared to $P_{4,5,6,8}^{\prime}$, especially in the low $q^{2}$ bins. This is due to the large error estimate that was obtained for $S_{3,6s,9}$ in table \ref{table:angular observables in bins}. The reported uncertainties for these observables are to be viewed as conservative estimates.

The results presented in this work provide a comprehensive analysis of the angular observables associated with the $B_{c}^{-}\rightarrow D_{s}^{*-}(\rightarrow D_{s}^{-}\pi^{0}) \ell^{+}\ell^{-}$ decay within the SM framework, for both light and heavy lepton modes. Recently, LHCb performed a detailed angular analysis of a related decay mode $B\rightarrow K^{*}\ell^{+}\ell^{-}$, where they reported a deviation of about $3.0~\sigma$ from SM predictions in $P_{2}$ and $P_{5}^{\prime}$ \cite{LHCb:2020gog} in the third $q^{2}$ bin. It is thus motivating to establish precise SM predictions of these observables for $B_{c}$ decay modes. These predictions will serve as benchmarks once future experimental measurements on the $B_{c}^{-}\rightarrow D_{s}^{*-}\ell^{+}\ell^{-}$ become available. Any significant deviation from the SM predictions could hint towards possible NP contributions.

	\section{Summary and conclusions}
	\label{section:Summary and conclusions}
	
	In this work, we have studied some rare decay modes of the $B_{c}$ meson in the SM. In the framework of pQCD, we have utilized the available lattice inputs on $B_{s}\rightarrow D_{s}^{(*)}$ and $B_{c}\rightarrow D_{s}$ form factors and extracted the shape parameters of the $B_c$ and $D_{s}^{(*)}$ meson wave functions, which in turn define the pQCD form factors. We have obtained the following values of the shape parameters associated with $B_c$ and $B_{s}$ meson wave functions:
	\begin{equation}
	\omega_{B_{c}} = 1.019(28)~\text{GeV},\qquad  \omega_{B_{s}} = 0.496(36)~\text{GeV}.
	\end{equation}
	Similarly, for the $D_{s}$ and $D_{s}^*$ meson wave functions we obtain
	\begin{equation}
	C_{D_{s}} = 0.495(15),\qquad C_{D_{s}^{*}} = 0.505(39), 
\end{equation}
    and
    \begin{equation}
     \omega_{D_{s}}= 0.104(12),\qquad \omega_{D_{s}^*}=0.098(15).
\end{equation}
	We find our estimate of $\omega_{B_{c}}$ to be in good agreement with our previous estimates presented in \cite{Dey:2025xdx,Dey:2025xjg}, thereby reinforcing the consistency of our methodology. Using these shape parameters, we can parameterize the wavefunction of the participating mesons. Furthermore, using these wave functions and the HQSS relations between $B_c\to D_{s}$ and $B_c\to D_{s}^*$ form factors, we have extracted the $q^2$ shapes of the $B_c\to D_{s}^*$ form factors. With this information we have then predicted a number of several observables, like branching fractions, lepton flavor universality ratios, and various angular observables, associated with the $B_c^- \rightarrow D_{s}^{*-}\ell^+\ell^-$ and $B_c^- \rightarrow D_{s}^{*-}\nu\bar{\nu}$ FCNC decays.

	\appendix
	
\section{Expressions of various tranversity amplitudes in $B_{c}^{-}\rightarrow D_{s}^{*-}\ell^{+}\ell^{-}$ }
\label{sec:apptranversity}
In this section, we present the detailed mathematical expression of the transversity amplitudes on which the angular coefficients defined in eq.\eqref{eqn:angular observables} are dependent. 
\begin{equation}
		\footnotesize
		\begin{split}
			\mathcal{A}_{\perp}^{L,R}=&-N_{\ell}\sqrt{2N_{D_{s}^{*}}}\sqrt{\lambda(m_{B_{c}}^{2},m_{D_{s}^{*}}^{2},q^{2})}\left[(C_{9}^{\text{eff}}\mp C_{10})\frac{V(q^{2})}{m_{B_{c}}+m_{D_{s}^{*}}}+2\hat{m_{b}}C_{7}^{\text{eff}}T_{1}(q^{2})\right],\\
			\mathcal{A}_{\parallel}^{L,R}=&N_{\ell}\sqrt{2N_{D_{s}^{*}}}\left[(C_{9}^{\text{eff}}\mp C_{10})(m_{B_{c}}+m_{D_{s}^{*}})A_{1}(q^{2})+2\hat{m_{b}}C_{7}^{\text{eff}}(m_{B_{c}}^{2}-m_{D_{s}^{*}}^{2})T_{2}(q^{2})\right],\\
			\mathcal{A}_{0}^{L,R}=&\frac{N_{\ell}\sqrt{N_{D_{s}^{*}}}}{2m_{D_{s}^{*}}\sqrt{q^{2}}}\biggl[(C_{9}^{\text{eff}}\mp C_{10})\left\lbrace (m_{B_{c}}^{2}-m_{D_{s}^{*}}^{2}-q^{2})(m_{B_{c}}+m_{D_{s}^{*}})A_{1}(q^{2})-\frac{\lambda(m_{B_{c}}^{2},m_{D_{s}^{*}}^{2},q^{2})}{m_{B_{c}}+m_{D_{s}^{*}}}A_{2}(q^{2})\right\rbrace\\&+2m_{b}C_{7}^{\text{eff}}\left\lbrace (m_{B_{c}}^{2}+3m_{D_{s}^{*}}^{2}-q^{2})T_{2}(q^{2})-\frac{\lambda(m_{B_{c}}^{2},m_{D_{s}^{*}}^{2},q^{2})}{m_{B_{c}}^{2}-m_{D_{s}^{*}}^{2}}T_{3}(q^{2})\right\rbrace\biggr],\\
			\mathcal{A}_{t}(q^{2})=&2N_{\ell}\sqrt{N_{D_{s}^{*}}}\sqrt{\frac{\lambda(m_{B_{c}}^{2},m_{D_{s}^{*}}^{2},q^{2})}{q^{2}}}C_{10}A_{0}(q^{2}),
		\end{split}
		\label{eqn:transversity amplitudes}
	\end{equation}
	with $\hat{m_{b}}=m_{b}/q^{2}$, $\beta_{\ell}=\sqrt{1-4m_{\ell}^{2}/q^{2}}$, and the normalization factors are expressed as
	\begin{equation}
		\begin{split}
			&N_{\ell}=\frac{i\alpha_{e}G_{F}}{4\sqrt{2}\pi}V_{tb}V_{ts}^{*},\\
			&N_{D_{s}^{*}}=\frac{8\sqrt{\lambda}q^{2}}{3\times 256 \pi^{3}m_{B_{c}}^{3}}\sqrt{1-\frac{4m_{\ell}^{2}}{q^{2}}}~\mathcal{B}(D_{s}^{*-}\rightarrow D_{s}^{-}\pi^{0}),
		\end{split}
	\end{equation}
	with $\mathcal{B}(D_{s}^{*-}\rightarrow D_{s}^{-}\pi^{0})=5.77(35)\%$ \cite{ParticleDataGroup:2024cfk}.
	These transversity amplitudes depend on the WCs and the QCD form factors. The Wilson coefficients $C_{7,9}^{\text{eff}}(\mu)$ in eq. \eqref{eqn:transversity amplitudes} have the form \cite{Chen:2001zc,Soni:2020bvu}\begin{equation}
		\begin{split}
			&C_{7}^{\text{eff}}(\mu)=C_{7}(\mu)+i \alpha_{s}(\mu)\left[\frac{2}{9}\eta^{14/23}\left\lbrace\frac{x_{t}(x_{t}^{2}-5x_{t}-2)}{8(x_{t}-1)^{3}}+\frac{3x_{t}^{2}\ln(x_{t})}{4(x_{t}-1)^{4}}-0.1687\right\rbrace-0.03C_{2}(\mu)\right],\\
			&C_{9}^{\text{eff}}(q^{2},\mu)=C_{9}(\mu)+Y_{\text{pert}}(q^{2},\mu)+Y_{\text{res}}(q^{2},\mu),
		\end{split}
		\label{eqn:Wilson Coefficients}
	\end{equation}	 
	with $x_{t}=m_{t}^{2}/m_{W}^{2}$ and $\eta=\alpha_{s}(m_{W})/\alpha_{s}(\mu)$ with $\alpha_{s}(\mu)$ being calculated at $\mu=m_{b}$. The short-distance contributions from the soft gluon emission and the one-loop contribution from the four Fermi operators $O_{1-6}$ are collected in the $Y_{\text{pert}}(q^{2},\mu)$ part of $C_{9}^{\text{eff}}(\mu)$ and can be written as \cite{Jin:2020jtu}
	\begin{equation}
		\begin{split}
			Y_{\text{pert}}(\hat{s},\mu)=&0.124\omega(\hat{s})+g(\hat{m_{c}},\hat{s})C(\mu)+\lambda_{u}\left[g(\hat{m_{c}},\hat{s})-g(0,\hat{s})\right](3C_{1}(\mu)+C_{2}(\mu))\\
			&-\frac{1}{2}g(0,\hat{s})(C_{3}(\mu)+3C_{4}(\mu))-\frac{1}{2}g(1,\hat{s})(4C_{3}(\mu)+4C_{4}(\mu)+3C_{5}(\mu)+C_{6}(\mu))\\
			&+\frac{2}{9}(3C_{3}(\mu)+C_{4}(\mu)+3C_{5}(\mu)+C_{6}(\mu)),
		\end{split}
		\label{eqn:Ypert}
	\end{equation}
	where $\hat{s}=q^{2}/m_{b}^{2}$, $\hat{m_{c}}=m_{c}/m_{b}$, $\lambda_{u}=(V_{ub}V_{us}^{*})/(V_{tb}V_{ts}^{*})$ and $C(\mu)=3C_{1}(\mu)+C_{2}(\mu)+3C_{3}(\mu)+C_{4}(\mu)+3C_{5}(\mu)+C_{6}(\mu)$, with the Wilson Coefficients $C_{1-10}(\mu)$ being evaluated upto the next to leading order correction at $\mu=m_{b}$ scale. The numerical values of the Wilson coefficients at $m_{b}$ scale \cite{Buchalla:1995vs,Soni:2020bvu} that have been used in this work are tabulated in table \ref{table:tabWilson}.
	
	\begin{table}[t!]
		\centering
		\renewcommand{\arraystretch}{1.3}
		\begin{tabular}{|cccccccccc|}
			\hline
			$C_{1}$&$C_{2}$&$C_{3}$&$C_{4}$&$C_{5}$&$C_{6}$&$C_{7}$&$C_{8}$&$C_{9}$&$C_{10}$\\
			\hline
			-0.175&1.076&0.01258&-0.03279&0.01112&-0.03634&-0.302&-0.148&4.232&-4.410\\
			\hline
		\end{tabular}
		\caption{Numerical values of Wilson coefficients $C_{1-10}$ evaluated at $m_{b}$ scale \cite{Buchalla:1995vs}.}
		\label{table:tabWilson}
	\end{table}
	
	In addition, in eq.\eqref{eqn:Ypert}, the term $\omega(\hat{s})$ representing the one gluon correction to the matrix element of the operator $O_{9}$ is represented as \cite{Jin:2020jtu,Soni:2020bvu,Wang:2012ab}
	\begin{equation}
		\begin{split}
			\omega({\hat{s}})=&-\frac{2}{9}\pi^{2}+\frac{4}{3}\int_{0}^{\hat{s}}\frac{\ln(1-u)}{u}du-\frac{2}{3}\ln(\hat{s})\ln(1-\hat{s})-\frac{5+4\hat{s}}{3(1+2\hat{s})}\ln(1-\hat{s})\\
			&-\frac{2\hat{s}(1+\hat{s})(1-2\hat{s})}{3(1-\hat{s})^{2}(1+2\hat{s})}\ln(\hat{s})+\frac{5+9\hat{s}-6\hat{s}^{2}}{6(1-\hat{s})(1+2\hat{s})}.
		\end{split}
	\end{equation}
	The functions $g(z,\hat{s})$ and $g(0,\hat{s})$ in eq.\eqref{eqn:Ypert} representing the one loop contributions of the $O_{1-6}$ are represented as \cite{Jin:2020jtu}
	
	\begin{equation}
		\begin{split}
			g(z,\hat{s})=-\frac{8}{9}\ln(z)+\frac{8}{27}+\frac{4}{9}x-\frac{2}{9}(2+x)\sqrt{|1-x|}
			\times \begin{cases}
				\ln\left|\frac{1+\sqrt{1-x}}{1-\sqrt{1-x}}\right|-i\pi \quad \text{for }x\equiv \frac{4z^{2}}{\hat{s}}< 1\\
				2 \arctan\left(\frac{1}{\sqrt{x-1}}\right) \quad \text{for }x\equiv \frac{4z^{2}}{\hat{s}}> 1,
			\end{cases}
		\end{split}
	\end{equation}
	and
	\begin{equation}
		g(0,\hat{s})=\frac{8}{27}-\frac{8}{9}\ln\left(\frac{m_{b}}{\mu}\right)-\frac{4}{9}\ln\left(\hat{s}\right)+\frac{4}{9}i\pi,
	\end{equation}
	and $\lambda_{u}=V_{ub}V_{us}^{*}/V_{tb}V_{ts}^{*}$. In addition to $Y_{\text{pert}}(q^{2},\mu)$, the third term $Y_{\text{res}}(q^{2},\mu)$ in eq.\eqref{eqn:Wilson Coefficients} describes the lolong-distanceontributions to $C_{9}^{\text{eff}}(\mu)$ and is associated with intermediate light vector mesons, i.e., $\rho$, $\omega$ and $\phi$ mesons, and vector charmonium mesons, i.e., $J/\psi\text{ and }\psi(2S)$ is expressed as \cite{Nayek:2018rcq,Jin:2020jtu}
	\begin{equation}
		\begin{split}
			Y_{\text{res}}(q^{2},\mu)=&-\frac{3\pi}{\alpha_{e}^{2}}\biggl\{C(\mu)\sum_{V_{i}=J/\psi,\psi(2S)}\frac{m_{V_{i}}\mathcal{B}\left(V_{i}\rightarrow \ell^{+}\ell^{-}\right)\Gamma_{V_{i}}}{q^{2}-m_{V_{i}}^{2}+im_{V_{i}}\Gamma_{V_{i}}}\\
			&-\lambda_{u}g(0,\hat{s})\left[3C_{1}(\mu)+C_{2}(\mu)\right]\sum_{V_{i}=\rho,\phi,\omega}\frac{m_{V_{i}}\mathcal{B}(V_{i}\rightarrow \ell^{+}\ell^{-})\Gamma_{V_{i}}}{q^{2}-m_{V_{i}}^{2}+im_{V_{i}}\Gamma_{V_{i}}}\biggr\},
		\end{split}
	\end{equation}
	where $m_{V_{i}}$ and $\Gamma_{V_{i}}$ represent the mass and the total decay width of the vector meson $V_{i}$ respectively, and $\mathcal{B}(V_{i}\rightarrow \ell^{+}\ell^{-})$ is the braching ratio of each of the dilepton decay mode. We have used the numerical values of these parameters as input in this work \cite{ParticleDataGroup:2024cfk} and present them in table \ref{table:tabresinp}.
	\begin{table}[htb!]
		\renewcommand{\arraystretch}{1.5}
		\centering
		\begin{tabular}{|c|c|c|c|c|}
			\hline
			Intermediate Meson& $m_{V_{i}}$  & $\Gamma_{V_{i}}$ & \multicolumn{2}{c|} {$\mathcal{B}(V_{i}\rightarrow \ell^{+}\ell^{-})$ with}\\
			\cline{4-5}
			$V_{i}$&in MeV& in MeV& $l=\mu$&$l=\tau$\\
			\hline
			$\rho$&775.26(23)&147.4(8)&4.55(28)$\times$ $10^{-5}$&-\\
			$\omega$&782.66(13)&8.68(13)&7.4(1.8)$\times$ $10^{-5}$&-\\
			$\phi$&1019.461(16)&4.249(13)&2.85(19)$\times$ $10^{-4}$&-\\
			$J/\psi$&3096.900(6)&0.0926(17)&5.961(33)$\times$ $10^{-2}$&-\\
			$\psi(2S)$&3686.10(6)&0.294(8)&8.0(6)$\times$ $10^{-3}$&3.1(4)$\times$ $10^{-3}$\\
			$\psi(3770)$&3773.7(7)&27.2(1.0)&9.6(7)$\times$ $10^{-6}$&-\\
			$\psi(4040)$&4040.0(4.0)&84.0(12.0)&1.02(17)$\times$ $10^{-5}$&-\\
			\hline
		\end{tabular}
		\caption{Masses, total decay widths and dilepton branching fractions of the intermediate vector mesons.}
		\label{table:tabresinp}
	\end{table}
	It is to be noted that the last two rows in this equation represent the dilepton branching fractions of $\psi(3770)$ and $\psi(4040)$ inspite of lying in the physical $q^{2}$ range, offer negligible contribution to $Y_{\text{res}}$, and their effects are smeared out in $q^{2}$ distribution of the observables, something that we will showcase in the results section. This is primarily because their dilepton decay widths are highly suppressed compared to that of $J/\psi$ and $\psi(2S)$.

	\section{Expressions of form factors in pQCD approach}
	\label{section:appendix pQCD form factor expressions}
	In this appendix, we present the analytical expressions of the form factors already discussed in section \ref{subsection:form factors}.
	\begin{itemize}
		\item For $B_{c,s}\rightarrow D_{s}$ transition the auxiliary form factors $f_{1}(q^{2})$ and $f_{2}(q^{2})$ have the analytic forms \cite{Hu:2019bdf}
		\begin{align}
				f_{1}(q^{2})=&16\pi M^{2}C_{F}\int_{0}^{1} dx_{1}dx_{2}\int_{0}^{b_{c}} b_{1}db_{1}b_{2}db_{2}\phi_{B_{c,s}}(x_{1},b_{1})\phi_{D_{s}}(x_{2},b_{2})\nonumber\\
				&\Biggl\{[2r - r^{2}x_{2} - r_{b}r]\cdot H_{1}(t_{1})+\biggl[r(2r_{q}-r)\nonumber\\
				&+\frac{x_{1}r}{2}\left(-2+2\eta+\sqrt{\eta^{2}-1}-\frac{2\eta}{\sqrt{\eta^{2}-1}}
				+\frac{\eta^{2}}{\sqrt{\eta^{2}-1}}\right)\biggr]\cdot H_{2}(t_{2})\Biggr\},\nonumber\\[0.4em]
				f_{2}(q^{2})=&16\pi M^{2}C_{F}\int_{0}^{1} dx_{1}dx_{2}\int_{0}^{b_{c}} b_{1}db_{1}b_{2}db_{2}\phi_{B_{c,s}}(x_{1},b_{1})\phi_{D_{s}}(x_{2},b_{2})\nonumber\\
				&\Biggl\{[2r_{b} - 1 - 2rx_{2}(1 -\eta)]\cdot H_{1}(t_{1})+\left[2r-r_{q}-\frac{x_{1}}{2}+\frac{x_{1}}{2\sqrt{\eta^{2}-1}}(2-\eta)\right]\cdot H_{2}(t_{2})\Biggr\},\\[0.4em]
				F_{T}(q^{2})=&8\pi M^{2}(1+r) C_{F}\int_{0}^{1}dx_{1}dx_{2}\int_{0}^{b_{c}}b_{1}db_{1}b_{2}db_{2}\phi_{B_{c,s}}(x_{1},b_{1})\phi_{D_{s}}(x_{2},b_{2})\nonumber\\
				&\times \biggl\lbrace [2 - r_{b} - rx_{2}]\cdot H_{1}(t_{1})+[2r - r_{q} + \frac{2x_{1} - x_{1}(\eta + \sqrt{\eta^{2} - 1})}{
 2 \sqrt{\eta^2 - 1}}]\cdot H_{2}(t_{2})\biggr\rbrace.\nonumber
            \label{eqn: BqDs form factors pQCD}
		\end{align}
		\item For $B_{c,s}\rightarrow D_{s}^{*}$ transition the axial-vector and vector form factors $A_{0,1,2}(q^{2})$ and $V(q^{2})$ respectively have the analytic forms \cite{Hu:2019bdf}
		\begin{align}
				A_{0}(q^{2})=&8\pi M^{2}C_{F}\int_{0}^{1} dx_{1}dx_{2}\int_{0}^{b_{c}} b_{1}db_{1}b_{2}db_{2}\phi_{B_{c,s}}(x_{1},b_{1})\phi_{D_{s}^{*}}(x_{2},b_{2})\nonumber\\
				&\Biggl\{[2r_{b} - 1 + r(2 - r_{b}) - rx_{2}(r+2-2\eta)]\cdot H_{1}(t_{1})\nonumber\\
				&+\left[r^{2}+r_{q}+\frac{x_{1}}{2}+\frac{\eta x_{1}}{2\sqrt{\eta^{2}-1}}+\frac{rx_{1}}{2\sqrt{\eta^{2}-1}}(1-2\eta(\eta+\sqrt{\eta^{2}-1}))\right]\cdot H_{2}(t_{2})\Biggr\},\nonumber\\[0.2em]
				A_{1}(q^{2})=&8\pi M^{2}C_{F}\int_{0}^{1} dx_{1}dx_{2}\int_{0}^{b_{c}} b_{1}db_{1}b_{2}db_{2}\phi_{B_{c,s}}(x_{1},b_{1})\phi_{D_{s}^{*}}(x_{2},b_{2})\frac{r}{1+r}\nonumber\\
				&\Biggl\{2[\eta(2-r_{b})+2r_{b}-1+x_{2}r\eta-2rx_{2}]\cdot H_{1}(t_{1})+\left[2r_{q}+2\eta r-x_{1}\right]\cdot H_{2}(t_{2})\Biggr\},\\[0.2em]
				A_{2}(q^{2})=&\frac{(1+r)^{2}(\eta-r)}{2r(\eta^{2}-1)}A_{1}(q^{2})\nonumber\\
				&-8\pi M^{2}C_{F}\int_{0}^{1}dx_{1}dx_{2}\int_{0}^{b_{c}}b_{1}db_{1}b_{2}db_{2}\phi_{B_{c,s}}(x_{1},b_{1})\phi_{D_{s}^{*}}(x_{2,b_{2}})\frac{1+r}{\eta^{2}-1}\nonumber\\
				&\times \biggl\lbrace[(2-r_{b})(1-r\eta)-(2r_{b}-1)(r-\eta)-rx_{2}(1-2r+\eta(2+r-2\eta))]\cdot H_{1}(t_{1})\nonumber\\
				&+\left[r+r_{q}(\eta-r)-\eta r^{2}+rx_{1}\eta^{2}-\frac{x_{1}}{2}(\eta+r)+x_{1}(\eta r-\frac{1}{2})\sqrt{\eta^{2}-1}\right]\cdot H_{2}(t_{2})\biggr\rbrace,\nonumber\\[0.2em]
				V(q^{2})=&8\pi M^{2}C_{F}\int_{0}^{1}dx_{1}dx_{2}\int_{0}^{b_{c}}b_{1}db_{1}b_{2}db_{2}\phi_{B_{c,s}}(x_{1},b_{1})\phi_{D_{s}^{*}}(x_{2,b_{2}})(1+r)\nonumber\\
				&\left\lbrace[2-r_{b}-rx_{2}]\cdot H_{1}(t_{1})+\left[r+\frac{x_{1}}{2\sqrt{\eta^{2}-1}}\right]\cdot H_{2}(t_{2})\right\rbrace.\nonumber
			\label{eqn: BqDsStar form factors pQCD}
		\end{align}
		and the tensor form factors $T_{1,2,3}$ have the analytic forms 
		\begin{align}
				T_{1}(q^{2})=&8\pi  M^{2}C_{F}\int_{0}^{1}dx_{1}dx_{2}\int_{0}^{b_{c}}b_{1}db_{1}b_{2}db_{2}\phi_{B_{c,s}}(x_{1},b_{1})\phi_{D_{s}^{*}}(x_{2,b_{2}})\nonumber\\
				&\biggl\lbrace[2-r_{b}+r(2-r_{b})-rx_{2}(2+r-2\eta)]\cdot H_{1}(t_{1})\nonumber\\&+[r(1+r_{q})-\frac{x_{1}(r\eta+r\sqrt{\eta^{2}-1}-1)}{2\sqrt{\eta^{2}-1}}]\cdot H_{2}(t_{2})\biggr\rbrace,\nonumber\\[0.3 em]
				T_{2}(q^{2})=&16\pi  M^{2}C_{F}\frac{r}{1-r^{2}}\int_{0}^{1}dx_{1}dx_{2}\int_{0}^{b_{c}}b_{1}db_{1}b_{2}db_{2}\phi_{B_{c,s}}(x_{1},b_{1})\phi_{D_{s}^{*}}(x_{2,b_{2}})\nonumber\\
				&\biggl\lbrace[(1-\eta r)(2-r_{b})+(\eta-r)(2r_{b}-1)+rx_{2}(2\eta^{2}+r(2-\eta)-(2\eta+1))]\cdot H_{1}(t_{1})\nonumber\\
				&+[r_{q}+r\eta(1-r_{q})-\frac{x_{1}}{2}-r^{2}+\frac{x_{1}r}{2}(\eta+\sqrt{\eta^{2}-1})]\cdot H_{2}(t_{2})\biggr\rbrace,\\[0.3 em]
				T_{3}(q^{2})=& 8\pi  M^{2}C_{F}\int_{0}^{1}dx_{1}dx_{2}\int_{0}^{b_{c}}b_{1}db_{1}b_{2}db_{2}\phi_{B_{c,s}}(x_{1},b_{1})\phi_{D_{s}^{*}}(x_{2,b_{2}})\nonumber\\
				&\times \biggl\lbrace[(2r_{b}-1)-r(2-r_{b})+rx_{2}(r+2\eta-2)]\cdot H_{1}(t_{1})\nonumber\\
                &+\left[r(1-r_{q})+\frac{x_{1}}{2\sqrt{\eta^{2}-1}}(1+r(\eta+\sqrt{\eta^{2}-1}))\right]\cdot H_{2}(t_{2})\biggr\rbrace.\nonumber
			\label{eqn: BqDsStar Tensor form factors pQCD}
		\end{align}
	\end{itemize}
	
In all the above expressions $r=m/M$, $r_{b}=m_{b}/M$ and $r_{q}=m_{q}/M$ with $q=c$ and $q=s$ for $b\to c$ and $b\to s$ decay channels, respectively, and 
	\begin{equation}
		\label{eqn:Evolution function}
		H_{i}(t_{i})=\alpha_{s}(t_{i})h_{i}(x_{1},x_{2},b_{1},b_{2})\exp[-S_{ab}(t_{i})],
	\end{equation}
	with $\alpha_{s}(t)$, $h_{i}(x_{1},x_{2},b_{1},b_{2})$ and $S_{ab}(t)$  representing the strong coupling constant evaluated at scale $t$, the hard kernel and the Sudakov factor respectively. The detailed expressions for these terms are presented in the following appendix. It is worth noting that by taking the limit $r_b \to 1$ and $r_s \to 0$, the expressions derived here explicitly reduce to those presented in \cite{PhysRevD.90.094018}. The form factors defined in the above equations are functions of initial and final state meson LCDAs, $\phi_{B_{c}}(x_{1},b_{1})$ and $\phi_{D_{s}^{*}}(x_{2},b_{2})$, whose explicit forms of these meson LCDAs have been already discussed in section \ref{subsection:LCDAs}. 
	The LCDAs $\phi_{B_s}(x_1,b_1)$ and $\phi_{B_c}(x_1,b_1)$ depend on the parameters $\omega_{B_s}$ and $\omega_{B_c}$, respectively, while $\phi_{D_s^{(*)}}(x_2,b_2)$ depend on $C_{D_s^{(*)}}$ and $\omega_{D_s^{(*)}}$. These parameters are treated as free parameters and are determined through a $\chi^2$ minimization. In addition, the heavy-quark masses $m_b$ and $m_c$ are included as nuisance parameters, with the corresponding values quoted in the main text. 
	
	\section{Scales and relevant functions in the hard kernel}
	\label{section:appendix hard functions}

	In this appendix, we present analytic expressions for the hard functions and scales. The hard kernel $h_{i}$ comes from the Fourier transform of virtual quark and gluon propagators
	\begin{equation}
		\begin{split}
			h_{1}(x_{1},x_{2},b_{1},b_{2})=&K_{0}(\beta_{1}b_{1})\biggl[\theta(b_{1}-b_{2})I_{0}(\alpha_{1}b_{2})K_{0}(\alpha_{1}b_{1})\\&
			+\theta(b_{2}-b_{1})I_{0}(\alpha_{1}b_{1})K_{0}(\alpha_{1}b_{2})\biggr]S_{t}(x_{2}),\\[0.5em]
			h_{2}(x_{1},x_{2},b_{1},b_{2})=&K_{0}(\beta_{2}b_{2})\biggl[\theta(b_{1}-b_{2})I_{0}(\alpha_{2}b_{2})K_{0}(\alpha_{2}b_{1})\\&+\theta(b_{2}-b_{1})I_{0}(\alpha_{2}b_{1})K_{0}(\alpha_{2}b_{2})\biggr]S_{t}(x_{1}),\\
		\end{split}
	\end{equation}
	where $K_{0}$ and $I_{0}$ are the modified Bessel functions. The functions $\alpha_{1},~\alpha_{2}\text{ and }\beta_{1,2}$ represent maximum of the virtualities of the internal particles and are expressed as
	\begin{equation}
		\begin{split}
			\alpha_{1}=&m_{B_{s}}\sqrt{x_{2}r\eta^{+}},\\[1em]
			\alpha_{2}=&m_{B_{s}}\sqrt{x_{1}r\eta^{+}-r^{2}+r_{c}^{2}},\\[1em]
			\beta_{1,2}=&m_{B_{s}}\sqrt{x_{1}x_{2}r\eta^{+}},
		\end{split}
	\end{equation}
	for $B_{s}\rightarrow D_{s}^{(*)}$ form factors \cite{Hu:2019bdf,Wang:2012lrc}, and
	\begin{equation}
		\begin{split}
			\alpha_{1}=&m_{B_{c}}\sqrt{2rx_{2}\eta+r_{b}^{2}-1-r^{2}x_{2}^{2}},\\[1em]
			\alpha_{2}=&m_{B_{c}}\sqrt{rx_{1}\eta^{+}+r_{s}^{2}-r^{2}},\\[1em]
			\beta_{1,2}=&m_{B_{c}}\sqrt{x_{1}x_{2}r\eta^{+}-r^{2}x_{2}^{2}},
		\end{split}
	\end{equation}
	for $B_{c}\rightarrow D_{s}^{(*)}$ form factors \cite{Hu:2019qcn}. In addition to the hard kernels, in eq.\eqref{eqn:Evolution function}, the Sudakov factors $S_{ab}(t)$ evaluated in the modified pQCD framework have been taken from \cite{PhysRevD.97.113001}. The hard scale $t$ is chosen to be the maximum of the virtuality of internal momentum transition in the hard amplitudes
	\begin{equation}
		\begin{split}
		t_{1}=\text{max}(\alpha_{1},1/b_{1},1/b_{2}),\\[1em] t_{2}=\text{max}(\alpha_{2},1/b_{1},1/b_{2}),
		\end{split}
	\end{equation}
	and the jet function $S_{t}(x)$ has the same form as eq. \eqref{eqn:jet function}.
	\section{Inputs utilized in this work}
	\label{section:appendix synthetic data of form factors}
	
	In this appendix we present the inputs of $B_{s}\rightarrow D_{s}^{(*)}$ and $B_{c}\rightarrow D_{s}$ form factors at discrete $w$ values, or more fundamentally $q^{2}$ values, which has been used in the chi-square minimizations we performed in this work.
	
	\begin{itemize}
		\item In table \ref{table:input BcDs}, we present lattice inputs for $B_{c}\rightarrow D_{s}$ form factors at $w=1.0,~1.15$ and $1.3$. These have been used to extract the soft function coefficients in table \ref{table:Soft function parameters}.
        
		\begin{table}[htb!]
			\renewcommand{\arraystretch}{2.0}
			\centering
			\resizebox{\textwidth}{!}{
			\begin{tabular}{|c|c|ccccccccc|}
				\hline
				\textbf{Form Factors} & \textbf{Value} &  &  &  & \textbf{Correlation}&  & &&& \\ 
				\cline{3-11}
				\textbf{at} $\boldsymbol{w}$ & \textbf{from HPQCD}&$F_{+}(1.0)$ &$F_{+}(1.15)$ & $F_{+}(1.3)$&$F_{0}(1.0)$ & $F_{0}(1.15)$& $F_{0}(1.3)$&$F_{T}(1.0)$ & $F_{T}(1.15)$& $F_{T}(1.3)$\\
				\hline 
				$F_{+}(1.0)$ & 1.464(128)  &1.0 & 0.945 & 0.664 & -0.062 & 0.234 & 0.228 & 0.061 & 0.053 & 0.021\\ 
				
				$F_{+}(1.15)$ & 0.830(50)  & & 1.0 & 0.862 & -0.077 & 0.295 & 0.340 & 0.067 & 0.068 & 0.035\\ 
				
				$F_{+}(1.3)$ & 0.527(23)  & &  & 1.0 & -0.075 & 0.322 & 0.525 & 0.063 & 0.084 & 0.066\\ 
				
				$F_{0}(1.0)$ & 0.737(11)  & &  &  & 1.0 & 0.659 & 0.338& 0.095 & 0.093 & 0.043\\ 
				
				$F_{0}(1.15)$ & 0.512(8)  & &  &  &  & 1.0 & 0.799 & 0.061 & 0.090 & 0.063\\ 
				
				$F_{0}(1.3)$ & 0.385(8)  & &  &  &  &  & 1.0 & 0.038 & 0.092 &0.102\\
				
				$F_{T}(1.0)$ & 2.032(64)  & &  &  &  &  & &1.0&0.511& 0.103\\ 
				
				$F_{T}(1.15)$ & 1.208(31)  & &  &  &  & & && 1.0&0.829\\ 
				
				$F_{T}(1.3)$ & 0.787(31)  & &  &  &  &  & &&& 1.0\\ 
				\hline
			\end{tabular} 
		}
			\captionof{table}{HPQCD inputs for $B_{c}\rightarrow D_{s}$ form factors, along-with their correlation.}	
			\label{table:input BcDs}
		\end{table}
        
		\item In tables \ref{table: inputs BcDStar A012V} and \ref{table: inputs BcDStar T123},  we present inputs for rest of the $B_{c}\rightarrow D_{s}^{*}$ form factors generated using the soft function parameters extracted in table \ref{table:Soft function parameters}. We have treated each form factor as uncorrelated with the others. This was done because all the form factors were derived from the same universal functions $\Sigma_{1}$ and $\Sigma_{2}$, which made the full correlation matrix positive semidefinite and led to numerical instability during chi-square minimization in table \ref{table:BGL Bc to Ds*}. So to avoid this we have considered all the form factors to be uncorrelated.  
        
		\begin{table}[htb!]
			\renewcommand{\arraystretch}{2.5}
			\centering
			\scriptsize
            \resizebox{\textwidth}{!}{%
			\begin{tabular}{|c|c|ccc|c|c|ccc|}
				\hline
				\textbf{Form}&\textbf{Value}&&\textbf{Correlation}&&\textbf{Form}&\textbf{Value}&&\textbf{Correlation}&\\
				\cline{3-5} \cline{8-10}
				\textbf{Factors}&\textbf{Obtained}&$A_{0}(1.0)$&$A_{0}(1.15)$&$A_{0}(1.3)$&\textbf{Factors}&\textbf{Obtained}&$A_{1}(1.0)$&$A_{1}(1.15)$&$A_{1}(1.3)$\\
				\hline
				$A_{0}(1.0)$&2.554(39)&1.0&0.619&0.273&$A_{1}(1.0)$&0.746(11)&1.0&0.694&0.405\\
				$A_{0}(1.15)$&1.367(19)&&1.0&0.755&$A_{1}(1.15)$&0.496(9)&&1.0&0.833\\
				$A_{0}(1.3)$&0.816(18)&&&1.0&$A_{1}(1.3)$&0.363(9)&&&1.0\\
				\hline
				\textbf{Form}&\textbf{Value}&&\textbf{Correlation}&&\textbf{Form}&\textbf{Value}&&\textbf{Correlation}&\\
				\cline{3-5} \cline{8-10}
				\textbf{Factors}&\textbf{Obtained}&$A_{12}(1.0)$&$A_{12}(1.15)$&$A_{12}(1.3)$&\textbf{Factors}&\textbf{Obtained}&$V(1.0)$&$V(1.15)$&$V(1.3)$\\
				\hline
				$A_{12}(1.0)$&0.246(4)&1.0&0.652&0.355&$V(1.0)$&2.165(60)&1.0&0.682&0.196\\
				$A_{12}(1.15)$&0.176(4)&&1.0&0.818&$V(1.15)$&1.243(30)&&1.0&0.833\\
				$A_{12}(1.3)$&0.140(5)&&&1.0&$V(1.3)$&0.788(27)&&&1.0\\
				\hline
			\end{tabular}
            }
			\caption{Inputs for $B_{c}\rightarrow D_{s}^{*}$ axial-vector and vector form factors $A_{0,1,12}$ and $V$ obtained using soft function parameters extracted in this work.}
			\label{table: inputs BcDStar A012V}
		\end{table}

        \begin{table}[htb!]
			\renewcommand{\arraystretch}{2.0}
			\centering
			\resizebox{\textwidth}{!}{
			\begin{tabular}{|c|c|ccccccccc|}
				\hline
				\textbf{Form Factors} & \textbf{Value} &  &  &  & \textbf{Correlation}&  & &&& \\ 
				\cline{3-11}
				\textbf{at} $\boldsymbol{w}$ & \textbf{from HPQCD}&$T_{1}(1.0)$ &$T_{1}(1.15)$ & $T_{1}(1.3)$&$T_{2}(1.0)$ & $T_{2}(1.15)$& $T_{2}(1.3)$&$T_{23}(1.0)$ & $T_{23}(1.15)$& $T_{23}(1.3)$\\
				\hline 
				$T_{1}(1.0)$ & 1.571(32)  &1.0 & 0.714 & 0.259 & 0.0 & 0.0 & 0.0 & 0.0 & 0.0 & 0.0\\ 
				
				$T_{1}(1.15)$ & 0.901(15)  & & 1.0 & 0.813 & 0.0 & 0.0 & 0.0 & 0.0 & 0.0 & 0.0\\ 
				
				$T_{1}(1.3)$ & 0.574(13)  & &  & 1.0 & 0.0 & 0.0 & 0.0 & 0.0 & 0.0 & 0.0\\ 
				
				$T_{2}(1.0)$ & 0.747(11)  & &  &  & 1.0 & 0.683 & 0.365& 0.0 & 0.0 & 0.0\\ 
				
				$T_{2}(1.15)$ & 0.535(8)  & &  &  &  & 1.0 & 0.799 & 0.0 & 0.0 & 0.0\\ 
				
				$T_{2}(1.3)$ & 0.412(8)  & &  &  &  &  & 1.0 & 0.0 & 0.0 &0.0\\
				
				$T_{23}(1.0)$ & 0.990(14)  & &  &  &  &  & &1.0&0.658& 0.363\\ 
				
				$T_{23}(1.15)$ & 0.557(14)  & &  &  &  & & && 1.0&0.803\\ 
				
				$T_{23}(1.3)$ & 0.355(19)  & &  &  &  &  & &&& 1.0\\ 
				\hline
			\end{tabular} 
		}
			\captionof{table}{Inputs for $B_{c}\rightarrow D_{s}^{*}$ tensor form factors $T_{1,2,23}$ obtained using the soft function parameters extracted in this work.}	
			\label{table: inputs BcDStar T123}
		\end{table}        
	\end{itemize}

	\section{Correlation matrices}
	\label{section:appendix correlation matrices}
	In this appendix, we present the correlation matrices obtained between the extracted parameters in each of the chi-square optimizations.
	
	\begin{itemize}
        \item In table \ref{table:Extraction of LCDA shape parameters input correlations}, we present the correlation matrix between lattice HPQCD and LCSR values of $B_{s}\to D_{s}^{(*)}$ and $B_{c}\to D_{s}$ form factors used as inputs in constructing the $\chi^{2}$-function of section \ref{section:Extraction of LCDA parameters}.
        \begin{table}[htb!]
            \scriptsize
    \centering
    \setlength{\tabcolsep}{5.0pt} 
    \renewcommand{\arraystretch}{2.0} 
            \begin{tabular}{|c|c|cccccccccc|}
            \hline
            &\rotatebox{90}{Channels}&&\rotatebox{90}{$B_{s}\to D_{s}^{*}$}&&\rotatebox{90}{$B_{s}\to D_{s}$}&&\rotatebox{90}{$B_{s}\to D_{s}^{*}$}&&\rotatebox{90}{$B_{s}\to D_{s}$}&\rotatebox{90}{$B_{c}\to D_{s}$}&\\
            \hline
                 \raisebox{1em}{Channels}&\raisebox{1 em}{Form Factors}& \rotatebox{90}{\textcolor{red}{$A_{0}(0)$}}&\rotatebox{90}{\textcolor{red}{$A_{1}(0)$}}&\rotatebox{90}{\textcolor{red}{$V(0)$}}&\rotatebox{90}{\textcolor{blue}{$F_{+}(0)$}}&\rotatebox{90}{\textcolor{blue}{$A_{0}(0)$}}&\rotatebox{90}{\textcolor{blue}{$A_{1}(0)$}}&\rotatebox{90}{\textcolor{blue}{$V(0)$}}&\rotatebox{90}{\textcolor{red}{$F_{+}(0)$}}& \rotatebox{90}{\textcolor{red}{$F_{+}(0)$}}&\rotatebox{90}{\textcolor{red}{$F_{T}(0)$}}\\ 
            \hline
            &\textcolor{red}{$A_{0}(0)$}&1.0&0.353&-0.010&0.0&0.0&0.0&0.0&0.0&0.0&0.0\\
            $B_{s}\to D_{s}^{*}$&\textcolor{red}{$A_{1}(0)$}&&1.0&0.375&0.0&0.0&0.0&0.0&0.0&0.0&0.0\\
            &\textcolor{red}{$V(0)$}&&&1.0&0.0&0.0&0.0&0.0&0.0&0.0&0.0\\
            $B_{s}\to D_{s}$&\textcolor{blue}{$F_{+}(0)$}&&&&1.0&0.104&0.126&0.146&0.0&0.0&0.0\\
            &\textcolor{blue}{$A_{0}(0)$}&&&&&1.0&0.205&0.219&0.0&0.0&0.0\\
            $B_{s}\to D_{s}^{*}$&\textcolor{blue}{$A_{1}(0)$}&&&&&&1.0&0.247&0.0&0.0&0.0\\
            &\textcolor{blue}{$V(0)$}&&&&&&&1.0&0.0&0.0&0.0\\
            $B_{s}\to D_{s}$&\textcolor{red}{$F_{+}(0)$}&&&&&&&&1.0&0.0&0.0\\
            $B_{c}\to D_{s}$&\textcolor{red}{$F_{+}(0)$}&&&&&&&&&1.0&-0.008\\
            &\textcolor{red}{$F_{T}(0)$}&&&&&&&&&&1.0\\
            \hline
            \end{tabular}
            \caption{Correlation matrix between lattice HPQCD and LCSR values of $B_{s}\to D_{s}^{(*)}$ and $B_{c}\to D_{s}$ form factors used as inputs in constructing the $\chi^{2}$-function of section \ref{section:Extraction of LCDA parameters}. The red and blue headings represent form factors taken from lattice HPQCD and LCSR papers, respectively.}
            \label{table:Extraction of LCDA shape parameters input correlations}
        \end{table}
		\item In table \ref{table: correlation LCDA BcD}, we present the correlation matrix between the LCDA shape parameters of $B_{s}$, $B_{c}$ and $D_{s}^{(*)}$ mesons, whose values we had extracted and showcased in table \ref{table:extracted value of Bc and D shape parameters}.
	\begin{table}[htb!]
		\scriptsize
    \centering
    \setlength{\tabcolsep}{2.0pt} 
    \renewcommand{\arraystretch}{2.5} 
    \resizebox{\textwidth}{!}{ 
    \begin{tabular}{|c|cccccccccccccccc|}
        \hline
         & \rotatebox{90}{$\omega_{B_{c}}$} & \rotatebox{90}{$\omega_{B_{s}}$} & \rotatebox{90}{$C_{D_{s}}$} & \rotatebox{90}{$C_{D_{s}^{*}}$} & \rotatebox{90}{$\omega_{D_{s}}$} & \rotatebox{90}{$\omega_{D_{s}^{*}}$} & \rotatebox{90}{$m_{b}$} & \rotatebox{90}{$m_{c}$} & \rotatebox{90}{$\delta_{f_{1}}^{B_{s}\to D_{s}}$} & \rotatebox{90}{$\delta_{f_{2}}^{B_{s}\to D_{s}}$} & \rotatebox{90}{$\delta_{A_{0}}^{B_{s}\to D_{s}^{*}}$} & \rotatebox{90}{$\delta_{A_{1}}^{B_{s}\to D_{s}^{*}}$} & \rotatebox{90}{$\delta_{V}^{B_{s}\to D_{s}^{*}}$} & \rotatebox{90}{$\delta_{f_{1}}^{B_{c}\to D_{s}}$} & \rotatebox{90}{$\delta_{f_{2}}^{B_{c}\to D_{s}}$} & \rotatebox{90}{$\delta_{F_{T}}^{B_{c}\to D_{s}}$} \\
        \hline
        $\omega_{B_{c}}$ & 1.0 & -0.145 & -0.201 & -0.141 & -0.014 & -0.015 & 0.011 & -0.017 & -0.014 & 0.004 & -0.021 & 0.014 & -0.023 & -0.022 & 0.080 & -0.029 \\
        $\omega_{B_{s}}$ & & 1.0 & 0.755 & 0.283 & -0.076 & 0.041 & -0.029 & 0.012 & 0.044 & -0.026 & 0.044 & -0.037 & 0.150 & 0.064 & -0.224 & 0.060 \\
        $C_{D_{s}}$ & & & 1.0 & 0.122 & -0.110 & 0.335 & -0.262 & -0.238 & 0.300 & -0.253 & 0.290 & -0.271 & 0.041 & 0.359 & -0.410 & 0.269 \\
        $C_{D_{s}^{*}}$ & & & & 1.0 & 0.416 & -0.092 & -0.012 & 0.181 & -0.029 & 0.034 & -0.151 & -0.014 & -0.082 & -0.087 & -0.143 & 0.087 \\
        $\omega_{D_{s}}$ & & & & & 1.0 & 0.138 & -0.175 & 0.018 & 0.200 & -0.184 & 0.179 & -0.191 & -0.629 & 0.170 & -0.168 & 0.304 \\
        $\omega_{D_{s}^{*}}$ & & & & & & 1.0 & -0.912 & -0.799 & 0.898 & -0.799 & 0.623 & -0.923 & -0.204 & 0.989 & -0.853 & 0.892 \\
        $m_{b}$ & & & & & & & 1.0 & 0.830 & -0.933 & 0.888 & -0.316 & 0.987 & 0.026 & -0.885 & 0.940 & -0.922 \\
        $m_{c}$ & & & & & & & & 1.0 & -0.927 & 0.940 & -0.224 & 0.862 & -0.160 & -0.814 & 0.856 & -0.776 \\
        $\delta_{f_{1}}^{B_{s}\to D_{s}}$ & & & & & & & & & 1.0 & -0.980 & 0.399 & -0.943 & -0.119 & 0.913 & -0.948 & 0.872 \\
        $\delta_{f_{2}}^{B_{s}\to D_{s}}$ & & & & & & & & & & 1.0 & -0.259 & 0.897 & 0.030 & -0.821 & 0.928 & -0.800 \\
        $\delta_{A_{0}}^{B_{s}\to D_{s}^{*}}$ & & & & & & & & & & & 1.0 & -0.304 & -0.667 & 0.676 & -0.259 & 0.396 \\
        $\delta_{A_{1}}^{B_{s}\to D_{s}^{*}}$ & & & & & & & & & & & & 1.0 & 0.031 & -0.893 & 0.953 & -0.907 \\
        $\delta_{V}^{B_{s}\to D_{s}^{*}}$ & & & & & & & & & & & & & 1.0 & -0.245 & 0.000 & -0.074 \\
        $\delta_{f_{1}}^{B_{c}\to D_{s}}$ & & & & & & & & & & & & & & 1.0 & -0.843 & 0.887 \\
        $\delta_{f_{2}}^{B_{c}\to D_{s}}$ & & & & & & & & & & & & & & & 1.0 & -0.844 \\
        $\delta_{F_{T}}^{B_{c}\to D_{s}}$ & & & & & & & & & & & & & & & & 1.0 \\
        \hline
    \end{tabular}
    }
    \caption{Correlation Matrix between the extracted LCDA parameters and nuisance parameters.}
    \label{table: correlation LCDA BcD}
\end{table}

		\item In table \ref{table:correlation form factors BcDStar}, we present the correlation matrix between the seven $B_{c}\rightarrow D_{s}^{*}$ semileptonic form factors at $q^{2}=0$ predicted in table \ref{table:form factor predictions BcDStar}.
		\begin{table}[htb!]
\renewcommand{\arraystretch}{2.3}
    \footnotesize
    \centering
    \begin{tabular}{|c|ccccc|}
        \hline
        &$A_{0}(0)$&$A_{1}(0)$&$V(0)$&$T_{1}(0)$&$T_{3}(0)$\\
        \hline
        $A_{0}(0)$ & 1.0 & 0.105 & -0.134 & 0.140 & 0.123 \\
        $A_{1}(0)$ & & 1.0 & -0.193 & 0.210 & 0.175  \\
        $V(0)$ & & & 1.0 & -0.261 & -0.231 \\
        $T_{1}(0)$ & & & & 1.0 & 0.966 \\
        $T_{3}(0)$ & & & & & 1.0  \\
        \hline
    \end{tabular}
    \caption{Correlation matrix between the $B_{c}\rightarrow D_{s}^{*}$ form factors calculated at $q^2=0$.}
    \label{table:correlation form factors BcDStar}
\end{table}

		\item In table \ref{table:correlation BGL parameters BcDsStar}, we present the correlation matrix between the BGL coefficients of $B_{c}\rightarrow D_{s}^{*}$ form factors extracted in table \ref{table:BGL Bc to Ds*}.
    \begin{table}
    \scriptsize
    \centering
    \setlength{\tabcolsep}{2.0pt} 
    \renewcommand{\arraystretch}{2.5} 
    \resizebox{\textwidth}{!}{
        \begin{tabular}{|c|ccccccccccccccccccc|}
            \hline
             & $a_{0}^{A_{0}}$ & $a_{1}^{A_{0}}$ & $a_{2}^{A_{0}}$ & $a_{0}^{A_{1}}$ & $a_{1}^{A_{1}}$ & $a_{2}^{A_{1}}$ & $a_{1}^{A_{12}}$ & $a_{2}^{A_{12}}$ & $a_{0}^{V}$ & $a_{1}^{V}$ & $a_{2}^{V}$ & $a_{0}^{T_{1}}$ & $a_{1}^{T_{1}}$ & $a_{2}^{T_{1}}$ & $a_{1}^{T_{2}}$ & $a_{2}^{T_{2}}$ & $a_{0}^{T_{23}}$ & $a_{1}^{T_{23}}$ & $a_{2}^{T_{23}}$ \\
            \hline
$a_{0}^{A_{0}}$ & 1.0 & 0.343 & -0.149 & 0.002 & 0.003 & 0.002 & 0.46 & 0.344 & -0.003 & -0.003 & -0.0005 & 0.001 & 0.002 & 0.001 & 0.002 & 0.002 & -0.002 & -0.003 & -0.002 \\
$a_{1}^{A_{0}}$ &  & 1.0 & 0.546 & 0.003 & 0.005 & 0.004 & 0.879 & 0.657 & -0.006 & -0.006 & -0.001 & 0.003 & 0.004 & 0.002 & 0.004 & 0.003 & -0.003 & -0.007 & -0.004 \\
$a_{2}^{A_{0}}$ &  &  & 1.0 & 0.002 & 0.004 & 0.003 & 0.687 & 0.514 & -0.005 & -0.005 & -0.0008 & 0.002 & 0.003 & 0.002 & 0.004 & 0.003 & -0.003 & -0.005 & -0.003 \\
$a_{0}^{A_{1}}$ &  &  &  & 1.0 & 0.451 & 0.016 & 0.003 & 0.002 & -0.004 & -0.004 & -0.0007 & 0.002 & 0.003 & 0.002 & 0.004 & 0.003 & -0.005 & -0.009 & -0.006 \\
$a_{1}^{A_{1}}$ &  &  &  &  & 1.0 & 0.734 & 0.005 & 0.004 & -0.006 & -0.007 & -0.001 & 0.004 & 0.005 & 0.003 & 0.006 & 0.005 & -0.008 & -0.015 & -0.01 \\
$a_{2}^{A_{1}}$ &  &  &  &  &  & 1.0 & 0.004 & 0.003 & -0.005 & -0.006 & -0.0009 & 0.003 & 0.004 & 0.002 & 0.005 & 0.004 & -0.006 & -0.013 & -0.008 \\
$a_{1}^{A_{12}}$ &  &  &  &  &  &  & 1.0 & 0.692 & -0.006 & -0.007 & -0.001 & 0.003 & 0.004 & 0.002 & 0.005 & 0.003 & -0.003 & -0.007 & -0.005 \\
$a_{2}^{A_{12}}$ &  &  &  &  &  &  &  & 1.0 & -0.005 & -0.005 & -0.0008 & 0.002 & 0.003 & 0.002 & 0.004 & 0.003 & -0.003 & -0.005 & -0.003 \\
$a_{0}^{V}$ &  &  &  &  &  &  &  &  & 1.0 & 0.517 & -0.211 & -0.003 & -0.005 & -0.003 & -0.006 & -0.004 & 0.004 & 0.008 & 0.005 \\
$a_{1}^{V}$ &  &  &  &  &  &  &  &  &  & 1.0 & -0.057 & -0.004 & -0.006 & -0.003 & -0.007 & -0.005 & 0.004 & 0.008 & 0.006 \\
$a_{2}^{V}$ &  &  &  &  &  &  &  &  &  &  & 1.0 & -0.0006 & -0.0009 & -0.0005 & -0.001 & -0.0008 & 0.0007 & 0.001 & 0.0009 \\
$a_{0}^{T_{1}}$ &  &  &  &  &  &  &  &  &  &  &  & 1.0 & 0.181 & -0.308 & 0.449 & 0.331 & 0.093 & 0.189 & 0.125 \\
$a_{1}^{T_{1}}$ &  &  &  &  &  &  &  &  &  &  &  &  & 1.0 & 0.047 & 0.678 & 0.5 & 0.141 & 0.285 & 0.189 \\
$a_{2}^{T_{1}}$ &  &  &  &  &  &  &  &  &  &  &  &  &  & 1.0 & 0.352 & 0.259 & 0.073 & 0.148 & 0.098 \\
$a_{1}^{T_{2}}$ &  &  &  &  &  &  &  &  &  &  &  &  &  &  & 1.0 & 0.501 & 0.161 & 0.325 & 0.216 \\
$a_{2}^{T_{2}}$ &  &  &  &  &  &  &  &  &  &  &  &  &  &  &  & 1.0 & 0.119 & 0.24 & 0.159 \\
$a_{0}^{T_{23}}$ &  &  &  &  &  &  &  &  &  &  &  &  &  &  &  &  & 1.0 & 0.306 & -0.334 \\
$a_{1}^{T_{23}}$ &  &  &  &  &  &  &  &  &  &  &  &  &  &  &  &  &  & 1.0 & 0.611 \\
$a_{2}^{T_{23}}$ &  &  &  &  &  &  &  &  &  &  &  &  &  &  &  &  &  &  & 1.0 \\
            \hline
        \end{tabular}
        }
    \captionof{table}{Correlation between extracted BGL parameters of $B_{c}\rightarrow D_{s}^{*}$ form factors (with $A_{12}$ and $T_{23}$).}
    \label{table:correlation BGL parameters BcDsStar}
\end{table}

	\end{itemize}

	\bibliographystyle{JHEP}
	\bibliography{biblio}
	
\end{document}